\pdfoutput=1
\documentclass[preprint,12pt]{elsarticle}

\journal{Computer-Aided Design}

\usepackage{amsfonts}
\usepackage{amssymb}
\usepackage{amsbsy}
\usepackage{amsmath}
\usepackage{dsfont}
\usepackage{enumerate}
\usepackage{stackengine}
\usepackage{subfig}
\usepackage{algorithm}
\usepackage{makeidx}
\usepackage{graphicx}
\usepackage{multicol}
\usepackage{xargs}
\usepackage[svgnames]{xcolor}
\usepackage{algpseudocode}

\newcommand{\argmin}{\textrm{argmin}}

\newcommand{\zR}{\mathbb{R}}
\newcommand{\zdim}{d}
\newcommand{\zmetric}{\textbf{M}}
\newcommand{\zvep}{\textbf{U}}
\newcommand{\zvepT}{\textbf{U}^\mathrm{T}}
\newcommand{\zvap}{\textbf{D}}

\newcommand{\ztr}{\mathrm{tr}}
\newcommand{\zmaster}{E^M}
\newcommand{\zbmaster}{E^{\hat{M}}}

\newcommand{\zequilater}{E^{\triangle}}

\newcommand{\zphysical}{E^P}
\newcommand{\zbphysical}{E^{\hat{P}}}

\newcommand{\zequilaterphysicalmap}{\zphi_E}

\newcommand{\zequilatermap}{\zphi_{\triangle}}

\newcommand{\zphysicalmap}{\zphi_P}

\newcommand{\zbphysicalmap}{\zphi_{\hat{P}}}

\newcommand{\zJacobianequilaterphysical}{\textbf{D}\zequilaterphysicalmap}

\newcommand{\zJacobianequilater}{\textbf{D}\zequilatermap}

\newcommand{\zJacobianphysical}{\textbf{D}\zphysicalmap}

\newcommand{\zu}{\textbf{u}}

\newcommand{\zx}{\textbf{x}}
\newcommand{\zxi}{\boldsymbol{\xi}}
\newcommand{\zbxi}{\hat{\boldsymbol{\xi}}}
\newcommand{\zy}{\textbf{y}}
\newcommand{\zp}{\textbf{p}}
\newcommand{\zbp}{\hat{\textbf{p}}}

\newcommand{\zphi}{\boldsymbol{\phi}}

\newcommand{\zbpsi}{\hat{N}}
\newcommand{\zallbpsi}{\hat{\textbf{N}}}

\newcommand{\zmesh}{\altmathcal{M}}
\newcommand{\zbmesh}{\hat{\zmesh}}

\DeclareMathAlphabet{\altmathcal}{OMS}{cmsy}{m}{n}
\newcommand{\loc}{\psi}

\newcommand{\zmodel}{\Lambda}
\newcommand{\znurbs}{\Gamma}
\newcommand{\zchull}{\text{CH}}

\newcommand{\ztrimmednurbs}{\overline{\Gamma}}
\newcommand{\zmatrix}{\mathbb{N}}
\newcommand{\zmat}{\zmatrix}
\newcommand{\zadj}{\adj{\zmatrix\left(\zx\right)}}
\newcommand{\zadjj}{\adj{\zmatrix}}
\newcommand{\adj}[1]{\text{adj}\left(#1\right)}
\newcommand{\gfun}{\gamma}

\newcommand{\gnfun}{\hat{\gamma}}
\newcommand{\gnfunn}{\gnfunn_{\znurbs}}
\newcommand{\tr}{\text{tr}}

\graphicspath{{figures/}}
\usepackage{xargs}
\begin{document}

\begin{frontmatter}
\title{Combining high-order metric interpolation and geometry implicitization for curved $r$-adaption}



\author{Guillermo Aparicio-Estrems}
\ead{guillermo.aparicio@bsc.es}
\author{Abel Gargallo-Peir\'{o}}
\ead{abel.gargallo@bsc.es}
\author{Xevi Roca\corref{mycorrespondingauthor}}
\cortext[mycorrespondingauthor]{Corresponding author}
\ead{xevi.roca@bsc.es}
\address{Barcelona Supercomputing Center, 08034 Barcelona, Spain}




\begin{abstract}
We detail how to use Newton's method for distortion-based curved $r$-adaption to a discrete high-order metric field while matching a target geometry. Specifically, we combine two terms: a distortion measuring the deviation from the target metric; and a penalty term measuring the deviation from the target boundary. For this combination, we consider four ingredients. First, to represent the metric field, we detail a log-Euclidean high-order metric interpolation on a curved (straight-edged) mesh. Second, for this metric interpolation, we detail the first and second derivatives in physical coordinates. Third, to represent the domain boundaries, we propose an implicit representation for 2D and 3D NURBS models. Fourth, for this implicit representation, we obtain the first and second derivatives. The derivatives of the metric interpolation and the implicit representation allow minimizing the objective function with Newton's method. For this second-order minimization, the resulting meshes simultaneously match the curved features of the target metric and boundary. Matching the metric and the geometry using second-order optimization is an unprecedented capability in curved (straight-edged) $r$-adaption. This capability will be critical in global and cavity-based curved (straight-edged) high-order mesh adaption.
\end{abstract}

\begin{keyword}
$r$-adaption, mesh optimization, curved high-order meshes
\end{keyword}

\end{frontmatter}


\section{Introduction}
\label{sec:intro}

The capability to relocate mesh nodes without changing the mesh topology, referred to as $r$-adaptivity, is a key ingredient in many adaptive PDE-based applications \cite{yano2012,loseille2011continuous,coupez2015implicit}. In these applications, to improve the solution accuracy, an error indicator or estimator determines the target stretching and alignment of the mesh. Then, to match these target features, an $r$-adaption procedure modifies the whole mesh (global) \cite{huang:AdaptMovingMesh, knupp:algebraicQuality} or a previously re-meshed cavity (local) \cite{alauzet:AnisotropicMeshAdaptation,gruau20053d,frey2005anisotropic}.

In either case, $r$-adaptivity contributes to increasing the solution accuracy for a fixed number of degrees of freedom supported on a straight-edged mesh \cite{coupez2015implicit,huang:AdaptMovingMesh,alauzet:AnisotropicMeshAdaptation,hecht1998bamg,coupez2011metric}. However, straight-edged meshes might not be an efficient support in many applications. Especially in applications where additional straight-edged mesh elements are artificially required to match highly curved solution features \cite{fidkowski2011review}.

To efficiently match curved solution features, many practitioners have recently started to exploit curved high-order meshes. These meshes can be stretched and aligned in a pointwise varying fashion through anisotropic procedures \cite{coupez:BasisFrameworkHighOrderAnisotropicMeshAdaptation}, geodesic approaches for curved edges \cite{johnen2021quality,zhang2018curvilinear}, shock-tracking methods \cite{zahr2018optimization,zahr2020implicit,zahr2020r}, and deformation analogies \cite{marcona2017variational,marcon2020rp}. Alternatively, the curved $r$-adaption can be driven, as for straight-edged elements \cite{huang:AdaptMovingMesh, knupp:algebraicQuality}, by distortion measures. These measures are defined point-wise and are aware of either a target deformation matrix \cite{dobrev2019target} or a target metric \cite{aparicio2018defining}.

In adaptivity applications, the target deformations and metrics are not known a priori. These target fields are reconstructed a posteriori from the solution on the last mesh. Specifically, this mesh supports the resulting discrete representation of the target field. This discrete representation is key to interpolate the required field values in the adaptive procedure. However, to also preserve the geometric accuracy, the mesh adaption procedures have to be devised to simultaneously match the target curved boundaries. Hence, to enable high-order adaptivity, we need the capability to interpolate target fields on a high-order mesh while matching a target boundary.

Considering the previous issues, we aim to use Newton's optimization for distortion-based curved $r$-adaption to a discrete high-order metric field and a geometry model. This work extends our previous work \cite{aparicio2022metricinterpolation}. In this extension, we also detail how to compatibly combine an optimization based $r$-adaption with a valid-to-valid mesh curving approach. To this end, our contribution is to propose an implicit model representation that measures the deviations of the mesh to the target geometry.

For the optimization based $r$-adaption, we need three existent ingredients. First, to minimize the distortion, we use the specific-purpose solver in \cite{aparicio2019imr,aparicio2021icosahom}. Second, we represent the metric field as a log-Euclidean high-order metric interpolation \cite{rochery2021p2} on a curved high-order mesh. Third, we locate physical points in the curved background mesh similar to the approach in \cite{dobrev2018towards}. We also need to extend to discrete metric fields a distortion-based curved $r$-adaption framework \cite{aparicio2018defining}.

To match the curved boundaries, we also need three existing ingredients. First, a non-interpolative approach to match the target curved geometry \cite{ruiz2016generation,ruiz2017augmented}. Second, an implicit CAD geometry representation method for 2D NURBS curves and a 3D NURBS surfaces \cite{upreti2014algebraic} or for embedded NURBS entities \cite{laurent2014implicit} such as 3D curves. Third, a series of conjuction and trimming operations to assemble the implicit representations of the individual entities \cite{upreti2014algebraic,biswas2004approximate}.

To compatibly combine the optimization based $r$-adaption with the mesh curving, the main novelty is twofold. First, for the non-interpolative mesh curving approach, we propose a model implicitization \cite{upreti2014algebraic,laurent2014implicit,biswas2004approximate}. Second, we also provide the first and second-order derivatives of the implicit representation of the model. As in \cite{aparicio2022metricinterpolation}, we also provide the first and second derivatives in physical coordinates for the log-Euclidean high-order metric interpolation. The model implicitation derivatives and the metric interpolation derivatives are critical to use Newton's method for distortion minimization while targeting a curved geometry. This minimization leads to unprecedented second-order optimization results for curved $r$-adaption for a discrete high-order metric representation on a curved (straight-edged) mesh while targeting a curved (straight-edged) geometry.

This paper focuses on enabling Newton's method for $r$-adaption, but it is focused neither on $r$-adaption nor $h$-adaption cycles.
Specifically, we detail how to optimize the high-order mesh coordinates to match a target metric and a curved boundary.
Then, to verify the methodology and the corresponding derivatives, we optimize initial isotropic and anisotropic straight-edged meshes.
These results do not consider any adaptivity cycles because we want to demonstrate if Newton’s method can be used.

The rest of the paper is organized as follows.
In Section \ref{sec:relatedwork}, we overview the related work.
In Section \ref{sec:preliminaries}, we introduce the preliminaries on metric-aware measures for high-order elements.
In Section \ref{sec:interpolation}, we detail the high-order metric interpolation and its derivatives.
In Section \ref{sec:implicit}, we propose an implicit representation for NURBS models, and we obtain the first and second derivatives of this representation.
Moreover, we detail the objective function that accounts for the metric and geometry deviations.
In Section \ref{sec:results}, we show Newton's method results for different geometries, meshes, and metrics.
Finally, we present the concluding remarks. 

\section{Related work}\label{sec:relatedwork}
Next, we overview the work related to matching a target discrete field and a target geometry model. Regarding matching discrete fields, we overview works on target deformations, target metrics, and discrete field representations. For matching geometry models, the related work is about non-interpolative mesh curving, surface fitting methods, and implicit geometry representations.

To match a deformation matrix, distortion optimization for curved $r$-adaption to a discrete target field is detailed in \cite{dobrev2019target}. The method is really well-suited for simulation-driven $r$-adaption \cite{dobrev2018towards,dobrev2020simulation}. It evaluates the distortion in a physical point by interpolating the target matrix on a discrete field.
Although the derivatives of the target matrices are not zero, the method assumes they are zero. Moreover, the second derivatives are also assumed to be zero. Since non-null derivatives are assumed to be zero although the approach implements Newton's method, the curved $r$-adaption minimization corresponds to a quasi-Newton method.

To match a metric, distortion-based curved $r$-adaption to an analytic field can be performed with Newton's minimization \cite{aparicio2018defining,aparicio2019imr,aparicio2021icosahom}. The formulation for an analytic metric is derived in \cite{aparicio2018defining}, while a specific-purpose globalization and a pre-conditioned Netwon-CG method are proposed in \cite{aparicio2019imr,aparicio2021icosahom} to minimize the mesh distortion. Since the method deals with an analytic metric, it does not specify the derivatives for a metric represented by a discrete high-order field.

Regarding a discrete field representation, a convenient approach is to use a log-Euclidean \cite{arsigny:Log-EuclideanMetrics} high-order metric interpolation \cite{rochery2021p2}. This metric interpolation drives a cavity-based adaption approach, where the remeshed cavities are improved by locally smoothing the curved quadratic edges. To smooth these edges, the method optimizes the mid-node position. The optimization only uses the first derivatives of the log-Euclidean metric interpolation in terms of the curved edge coordinates. Accordingly, the method does not provide the first and second derivatives of the discrete metric field in physical coordinates.

High-order mesh curving methods that approximate the target geometry in a non-interpolative manner are presented in \cite{ruiz2016generation,ruiz2017augmented}. Specifically, a new methodology to optimize a curved high-order mesh in terms of both element quality and a distance-based geometric approximation is developed. For this, a penalty method is proposed to solve the constrained minimization problem.

Previous surface fitting methods based in field interpolation are presented in \cite{knupp2021adaptive}.
They are specially designed for dynamically changing geometry according to a solution.
For this, a background mesh is required to interpolate the solution.
Moreover, the resolution of the background mesh determines the precision of the dynamic geometry. Hence, for CAD models, the background mesh resolution controls the geometry accuracy.

In contrast to previous methods, implicit CAD geometry representation methods provide a field for geometric approximation without using a background mesh \cite{upreti2014algebraic,laurent2014implicit,biswas2004approximate}. Specifically, one first computes the implicit representation of each NURBS entity. This is the case of a 2D NURBS curve and 3D NURBS surface \cite{upreti2014algebraic} or a generally embedded NURBS entity \cite{laurent2014implicit}. Then, one applies convex-hull conjunction and normalization, convex-hull trimming, and NURBS conjunction to assemble the representations of the individual entities \cite{upreti2014algebraic,biswas2004approximate}. Even if they are not a full representation of the model they provide a useful tool for representing the model in a entity-wise fashion.

\section{Preliminaries: metric-aware measures for high-order elements}\label{sec:preliminaries}

\begin{figure}[t]
	\centering
	\includegraphics[width=0.7\textwidth]{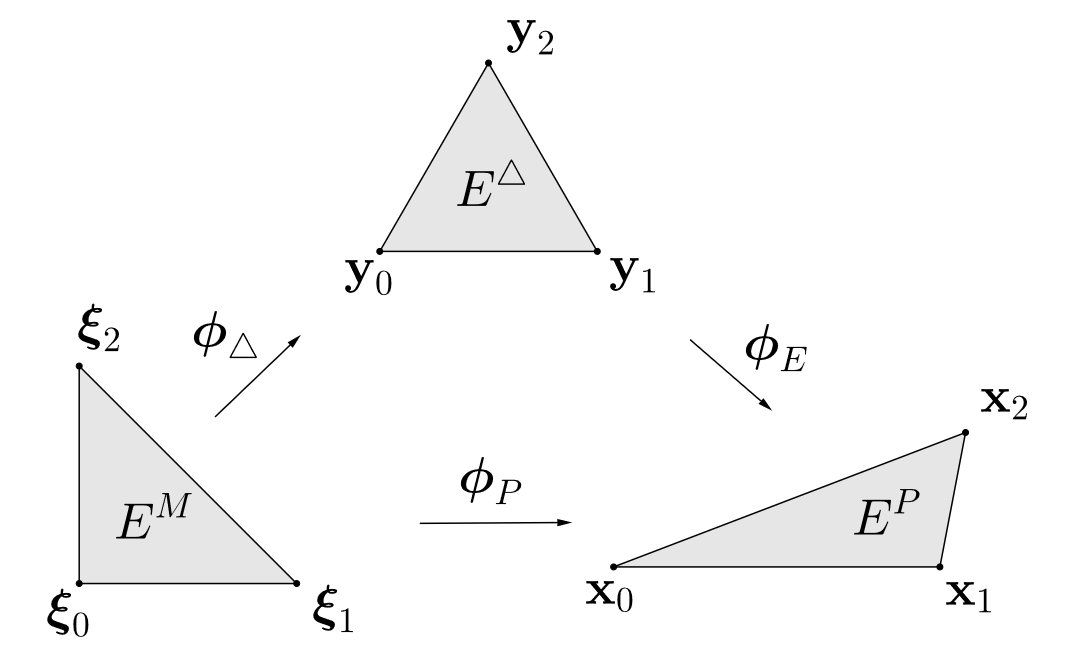}
	\caption{Mappings between the master, the ideal, and the physical elements in the linear case.}
	\label{fig:mappingRefIdealPhysical}
\end{figure}

In this section, we review the definition of the Jacobian-based quality measure for high-order elements equipped with a metric, presented in \cite{aparicio2018defining}.
To define and compute a Jacobian-based measure for simplices \cite{knupp:algebraicQuality}, three elements are required: the master, the ideal, and the physical, see Figure \ref{fig:mappingRefIdealPhysical} for the linear triangle case. The master $(\zmaster)$ is the element from which the iso-parametric mapping is defined. The equilateral element $\left(\zequilater\right)$ represents the target configuration in the isotropic case. The physical $(\zphysical)$ is the element to be measured.

To summarize the results in \cite{aparicio2018defining}, we present the expression of the metric distortion measure in terms of the equilateral element $\zequilater$.
First, we need to compute a mapping from the master to the equilateral and physical elements, denoted as $\zequilatermap$ and $\zphysicalmap$, respectively.
By means of these mappings, we determine a mapping between the equilateral and physical elements by the composition
\begin{equation*}
\zequilaterphysicalmap :\zequilater\xrightarrow{\zequilatermap^{-1}} \zmaster \xrightarrow{\zphysicalmap} \zphysical.
\end{equation*}

As detailed in \cite{aparicio2018defining}, we define the point-wise distortion measure for a high-order element $\zphysical$ equipped with a point-wise metric $\zmetric$, at a point $\zy\in\zequilater$ as
\begin{equation}\label{eq:distortionoperator}
\altmathcal{N}\zequilaterphysicalmap(\zy) = \frac{\ztr\left( {\zJacobianequilaterphysical(\zy)}^{\text{T}}\cdot \zmetric\left(\zequilaterphysicalmap(\zy)\right) \cdot\zJacobianequilaterphysical(\zy) \right)}{\zdim\left(\det\left({\zJacobianequilaterphysical(\zy)}^{\text{T}}\cdot \zmetric\left(\zequilaterphysicalmap(\zy)\right) \cdot\zJacobianequilaterphysical(\zy)\right)\right)^{1/\zdim}},
\end{equation}
where the Jacobian of the map $\zequilaterphysicalmap$ is given by
\begin{equation*}
	\zJacobianequilaterphysical(\zy) := \zJacobianphysical(\zequilatermap^{-1}\left(\zy\right))\cdot \zJacobianequilater^{-1}(\zy).
\end{equation*}
Herein, $\zJacobianphysical$ and $\zJacobianequilater$ denote the Jacobian of the physical and equilateral transformation, respectively.
Specifically, the physical mapping can be expressed in terms of the $d$-simplex shape functions $N_i$, that is
\begin{equation*}
	\zphysicalmap\left(\zxi\right) = \sum_{i=1}^n N_i(\zxi)\zx_i,
\end{equation*}
where $n = \binom{d+p}{p}$ is the number of nodes, $\zxi$ are the master coordinates, and $\zx_i$ denotes the physical coordinates of the high-order nodes. 
In addition, the equilateral mapping can be expressed in terms of the linear shape functions $N_i$, that is
\begin{equation*}
	\zequilatermap\left(\zxi\right) = \sum_{i=1}^{d+1} N_i(\zxi)\zy_i,
\end{equation*}
where $\zy_i$ are the coordinates of an equilateral $d$-simplex.

Note that $\altmathcal{N}$ is a non-linear operator that transforms a mapping between the equilateral and physical elements to a mapping from an point to a scalar.
In this work, for operators, we use the standard notation without parentheses.

Note that the distortion measure is independent of the computation of the metric $\zmetric\left(\zequilaterphysicalmap(\zy)\right)$, either using an analytical or a discretized representation.

We regularize the determinant in the denominator of Equation \eqref{eq:distortionoperator} in order to detect inverted elements \cite{garanzha:regularization, storti:globalSmoothUntangl,escobar:untanglingSmoothing,gargallo:generation3Doptimization}. In particular, we define
\begin{equation*}
\sigma_0 = \frac{1}{2}(\sigma + |\sigma|),
\end{equation*}
where
\begin{equation*}
\sigma = \det\left({\zJacobianequilaterphysical(\zy)}\right)\sqrt{\det\left(\zmetric\left(\zequilaterphysicalmap(\zy)\right)\right)}.
\end{equation*}
Then, we define the point-wise regularized distortion measure of a physical element $\zphysical$ at a point $\zy\in\zequilater$ as
\begin{equation}\label{eq:distortion}
\altmathcal{N}_0\zequilaterphysicalmap(\zy) := \frac{\ztr(\zJacobianequilaterphysical(\zy)^\mathrm{T}\cdot \zmetric\left(\zequilaterphysicalmap(\zy)\right)\cdot \zJacobianequilaterphysical(\zy))}{d\sigma_0^{2/d}},
\end{equation}
where we introduce the sub-script 0 to distinguish the regularized operator from the non-regularized one.
In addition, we define the corresponding point-wise quality measure
\begin{equation}\label{eq:pointwisequality}
\altmathcal{Q}\zequilaterphysicalmap(\zy) = \frac{1}{\altmathcal{N}_0\zequilaterphysicalmap(\zy)}.
\end{equation}
Finally, we define the regularized elemental distortion by
\begin{equation*}\label{eq:distortionReg}
\eta_{\left(\zphysical, \zmetric \right)} := \frac{\int_{\zequilater}\altmathcal{N}_0\zequilaterphysicalmap(\zy)\ d\zy}{\int_{\zequilater} 1\ d\zy},
\end{equation*}
and its corresponding quality
\begin{equation}\label{eq:qualityreg}
\text{q}_{\left(\zphysical, \zmetric \right)} = \frac{1}{\eta_{\left(\zphysical, \zmetric \right)}}.
\end{equation}

We can improve the mesh configuration by means of relocating the nodes of the mesh according to a given distortion measure \cite{aparicio2018defining,aparicio2019imr,aparicio2021icosahom,gargallo:PhDdissertation}.
In \cite{aparicio2018defining} it is proposed an optimization of the distortion (quality) of a mesh $\zmesh$ equipped with a target metric $\zmetric$ that describes the desired alignment and stretching of the mesh elements. 
To optimize a given mesh $\zmesh$, first it is defined the mesh distortion by
\begin{equation}\label{eq:minfun}
\altmathcal{F}\left(\zmesh \right) := \sum_{\zphysical\in\zmesh} \int_{\zequilater} \left(\altmathcal{N}_0\zequilaterphysicalmap(\zy)\right)^2\ d\zy,
\end{equation}
which allows to pose the following global minimization problem
\begin{equation}\label{eq:argmin}
\zmesh^* := \argmin_{\zmesh} \altmathcal{F}\left(\zmesh\right),
\end{equation}
to improve the mesh configuration according to $\altmathcal{F}$.
In particular, herein, the degrees of freedom of the minimization problem in Equation \eqref{eq:argmin} correspond to the spatial coordinates of the mesh nodes.

To evaluate the distortion minimization formulation presented in Equation \eqref{eq:argmin}, an input metric is required.
The reviewed $r$-adaption procedure has been applied for analytic metrics in \cite{aparicio2018defining}. 
In the following section, we detail the interpolation process that is required to extend the presented framework to dicrete metrics.
\section{Log-Euclidean metric interpolation}\label{sec:interpolation}
In this section, we formulate a metric interpolation process that allows both the distortion evaluation, Equation \eqref{eq:distortion}, and its optimization, Equation \eqref{eq:argmin}.
In Section \ref{sec:meshInterp} we detail the log-Euclidean metric interpolation for  linear and high-order elements first presented in \cite{arsigny:Log-EuclideanMetrics} and \cite{rochery2021p2,ekelschot2019parallel}, respectively. 
Then, in Section \ref{sec:interpDeriv} we present, as a contribution of this work, the gradient and the Hessian of the log-Euclidean interpolation. Their computation will be used for the distortion minimization problem.

\subsection{Metric Interpolation}\label{sec:meshInterp}
\begin{figure*}[t]
	\centering
	\begin{tabular}{ccc}
		\subfloat[]{\label{fig:physicalmesh}
			\includegraphics[width=0.25\textwidth]{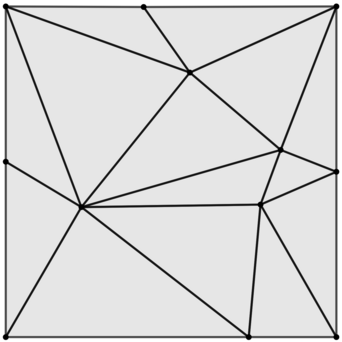}}
		&
		\subfloat[]{\label{fig:backgroundmesh}
			\includegraphics[width=0.25\textwidth]{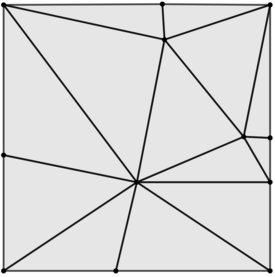}}
		&
		\subfloat[]{\label{fig:intersection}
			\includegraphics[width=0.25\textwidth]{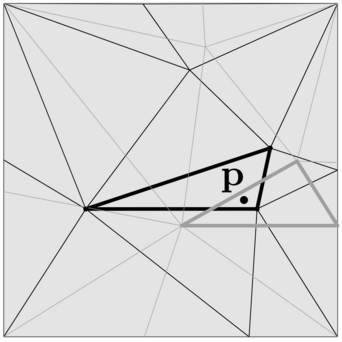}}
		\\
	\end{tabular}
	\caption{Point localization: (a) physical mesh, (b) background mesh, and (c) a point $\textbf{p}$ in the corresponding physical and background element (bold edges).}
	\label{fig:backgroundintersection}
\end{figure*}

\begin{figure}[t]
	\centering
	\includegraphics[width=0.6\textwidth]{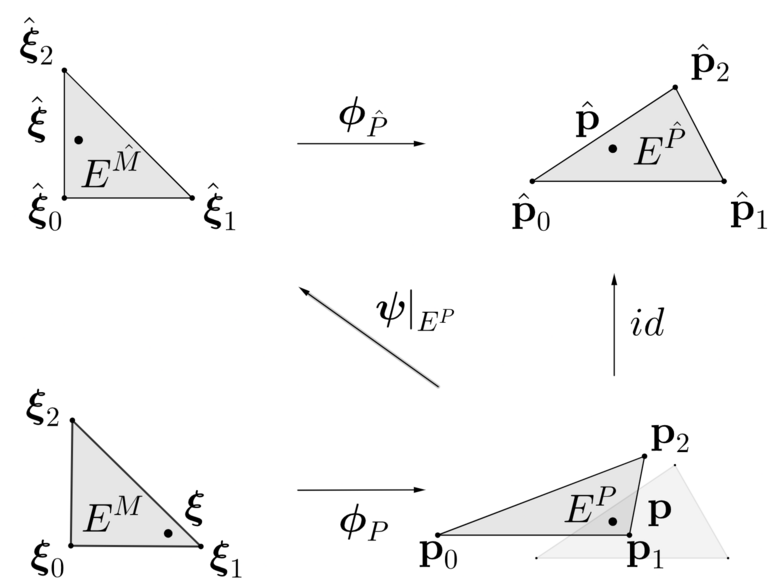}
	\caption{Mappings between the master and the physical elements (below) and their background analogs (above).}
	\label{fig:diagrambackground}
\end{figure}
In this section, we introduce the definition of the log-Euclidean metric interpolation at the background mesh. First, we introduce the required notation of the mappings and their parameters with the corresponding diagram. Secondly, we detail the interpolation procedure. 

To evaluate the metric-aware distortion measure in Equation \eqref{eq:distortion} featuring discrete metrics, two meshes are required. 
On the one hand, the \textit{physical} mesh $\zmesh$, Figure \ref{fig:backgroundintersection}\subref{fig:physicalmesh}, is the domain where the elements are deformed in order to solve the problem presented in Equation \eqref{eq:argmin}. 
%
%
On the other hand, the background mesh $\zbmesh$, Figure \ref{fig:backgroundintersection}\subref{fig:backgroundmesh}, is a mesh that stores discrete metric values as a nodal field.
%
%
%

To evaluate the point-wise metric-aware distortion measure, we need to compute the interpolation of the point-wise metric values. For this, the localization between both meshes is required \cite{dobrev2018towards,mittal2019nonconforming,sitaraman2010parallel}.
In particular, a physical point $\zp\in\zmesh$ is located at the background mesh $\zbmesh$ where the metric is interpolated, see Figure \ref{fig:backgroundintersection}\subref{fig:intersection}. In what follows, we introduce the elements and the mappings required for this localization procedure.
%


We integrate the distortion measure presented in Equation \eqref{eq:distortion} over the equilateral element via the master element $\zmaster$.
In particular, for the metric evaluation, we map via $\zphysicalmap$, each integration point $\zxi\in\zmaster$ to a point $\zp$ of the physical element $\zphysical$, see Figure \ref{fig:diagrambackground}.
To compute the metric at $\zp$ we need to locate $\zp$ in the background mesh, where the values of the metric are stored, see the intersection between $\zphysical$ and the background element $\zbphysical$ in Figure \ref{fig:diagrambackground}.
In addition, Figure \ref{fig:diagrambackground} shows the procedure to obtain the coordinate to interpolate the metric from the quadrature points. In particular, we map a reference point $\zxi\in\zmaster$ to a physical point $\zp = \zphysicalmap\left(\zxi\right)\in\zphysical$, which we identify it with a point $\zbp\in\zbphysical$ of the background mesh and its preimage is the background reference point $\zbxi = {\zbphysicalmap}^{-1}\left(\zbp\right)\in\zbmaster$.



Given a physical point $\zp$, we find it convenient to denote by $\psi$ any mapping from a background element containing $\zp$ that provides the coordinates in the background master element $\zbmaster$. Using this notation, we understand that any projection of a physical point $\zp$ onto a point $\zbxi$ of the background master element $\zbmaster$ corresponds to the evaluation of the non-linear function $\zbxi = \loc(\zp)$.

To evaluate this non-linear function, we exploit that the expression of $\psi|_{\zphysical}$, defined in the intersection of a physical element $\zphysical$ and a fixed background element $\zbphysical$, is given by
\begin{equation}\label{eq:loc}
\begin{array}{cccc}
\loc|_{\zphysical} : & \zphysical \cap \zbphysical& \rightarrow & \zbmaster \\
&\zp &\mapsto &	\zbphysicalmap^{-1}\left(\zp\right).
\end{array}
\end{equation}
Specifically, we solve the non-linear inverse expression in the image term, Equation \eqref{eq:loc}, by applying Newton's minimization to the squared distance.
That is, as in Section 2.3 of \cite{mittal2019nonconforming}, we solve
\begin{equation*}
\zbxi = \argmin\lim_{\hat{\zeta}} \bigg|\bigg|\zbphysicalmap\left(\hat{\zeta}\right) - \zp\bigg|\bigg|^2.
\end{equation*}
The result is a numerical approximation of the point coordinates in the background master element. An alternative approach \cite{dobrev2018towards} is to seek the zeros of the vector equation
\begin{equation*}
\zbphysicalmap\left(\zbxi\right) - \zp = \textbf{0}.
\end{equation*}

Once the background master coordinates associated to a given physical point have been computed, it is necessary to interpolate the metric supported by the background mesh at the corresponding master coordinate.
To do so, we use the log-Euclidean interpolation proposed in \cite{arsigny:Log-EuclideanMetrics,rochery2021p2}:
\begin{equation}\label{eq:metricInterpolationHO}
\zmetric\left(\zallbpsi\right) := \exp\left(\textbf{L}(\zallbpsi)\right),\ \ \textbf{L}(\zallbpsi) := \sum_{j=1}^{\hat{n}} \zbpsi_j \log \hat{\zmetric}_j,
\end{equation}
where for the $j$-th node of the master element $\zbmaster$, $\hat{\zmetric}_j$, and $\zbpsi_j$ are the corresponding metric value and shape function, respectively. In addition, $\zallbpsi$ denotes all the shape functions, $\hat{n} = \binom{d+\hat{p}}{\hat{p}}$ is the number of nodes, and where $\hat{p}$ is the interpolation degree which corresponds to the polynomial degree of the master element $\zbmaster$. Finally, $\textbf{M}(\zallbpsi)$ is characterized by the \textit{eigenvalue-based} matrix exponential function
\begin{equation}\label{eq:exponential}
\zmetric\left(\zallbpsi\right) = \zvep \cdot \exp \zvap \cdot \zvepT,
\end{equation}
where $\zvap,\ \zvep$ are given from the eigenvalue decomposition of the matrix $\textbf{L}(\zallbpsi) =: \zvep \cdot \zvap \cdot \zvepT$. Finally, for each physical point $\zp$ the metric interpolation is given by $\zmetric\left(\zallbpsi\left(\loc\left(\zp\right)\right)\right)$.

\subsection{Gradient and Hessian}
\label{sec:interpDeriv}

This section provides the expressions for the gradient and Hessian of the metric interpolation over a background mesh in terms of the physical coordinates.
For this, we detail first the case for the metric interpolation at a single element and then for the background mesh. In particular, our approach uses the gradient and Hessian of the eigenvalue decomposition presented in \cite{andrew1993derivatives}.

To compute the derivatives of the metric $\zmetric$ we first differentiate the eigenvalue-based exponential matrix function presented in Equation \eqref{eq:exponential} and then we differentiate the $\textbf{L}$ function presented in Equation \eqref{eq:metricInterpolationHO}.
By denoting $x_j$ the coordinates of $\zp$ and $\partial_j := \frac{\partial}{\partial x_j}$, $\partial_{jk} := \partial_j \partial_k = \frac{\partial}{\partial x_j}\frac{\partial}{\partial x_k}$ the partial derivatives in terms of the physical coordinates of $\zp$, we can compute the spatial derivatives of the metric interpolation of Equation \eqref{eq:metricInterpolationHO}. In particular, the first-order derivatives are given by
\begin{eqnarray*}\label{eq:gradient}
\partial_{j} \zmetric(\zallbpsi) = \partial_j \exp \textbf{L}(\zallbpsi) = \partial_j \left(\zvep \cdot \exp \zvap \cdot \zvepT \right) =\\
\left(\partial_j \zvep\right) \cdot \exp \zvap \cdot \zvepT + \zvep \cdot \left(\partial_j \exp \zvap\right) \cdot \zvepT + \zvep \cdot \exp \zvap \cdot \left(\partial_j\zvepT\right),
\end{eqnarray*}
and the second-order derivatives are given by
\begin{eqnarray*}\label{eq:hessian}
	\partial_{jk} \zmetric(\zallbpsi) = \partial_{jk} \exp \textbf{L}(\zallbpsi) = \partial_{jk} \left(\zvep \cdot \exp \zvap \cdot \zvepT \right) =\\
	\left(\partial_{jk} \zvep\right) \cdot \exp \zvap \cdot \zvepT + \partial_k \zvep \cdot \left(\partial_j \exp \zvap\right) \cdot \zvepT + \partial_k \zvep \cdot \exp \zvap \cdot \left(\partial_j\zvepT\right) +\\
	\left(\partial_j \zvep\right) \cdot \partial_k \exp \zvap \cdot \zvepT + \zvep \cdot \left(\partial_{jk} \exp \zvap\right) \cdot \zvepT + \zvep \cdot \partial_k \exp \zvap \cdot \left(\partial_j\zvepT\right) + \\
	\left(\partial_j \zvep\right) \cdot \exp \zvap \cdot \partial_k \zvepT + \zvep \cdot \left(\partial_j \exp \zvap\right) \cdot \partial_k \zvepT + \zvep \cdot \exp \zvap \cdot \left(\partial_{jk}\zvepT\right).
\end{eqnarray*}

Note that, since the matrix $\zvap$ is diagonal, we have
\begin{eqnarray*}
	\partial_j \exp \zvap &=& \exp\left(\zvap\right)\cdot \partial_j\zvap,\\
	\partial_{jk} \exp \zvap &=&  \exp\left(\zvap\right) \cdot \left( \partial_k \zvap \cdot \partial_j\zvap + \partial_{jk}\zvap\right).
\end{eqnarray*}


The presented first and second-order derivatives of the metric require the first and second-order spatial derivatives of the eigenvalue decomposition (eigenvalues and eigenvectors), respectively. 
Their computation is appended in Section \ref{sec:appendix}.

In addition, the derivatives of the eigenvalues and eigenvectors depend on the derivatives of the $\textbf{L}$ function presented in Equation \eqref{eq:metricInterpolationHO}. In particular, they are given by
\begin{equation*}
\nabla_{} \textbf{L} = \sum_{j} \left(\log{\hat{\zmetric}_j}\right) \nabla_{}\zbpsi_j,\ \ \nabla_{}^2 \textbf{L} = \sum_{j} \left(\log{\hat{\zmetric}_j}\right) \nabla_{}^2\zbpsi_j,
\end{equation*}
where $\nabla_{}$ is the gradient with respect to physical coordinates.
Therefore, to differentiate the metric interpolation $\zmetric\left(\zallbpsi\left(\loc\left(\zp\right)\right)\right)$ at a physical point $\zp$, the derivatives of the map $\loc$ presented in Equation \eqref{eq:loc} and of the shape functions $\zallbpsi$ are required.

The derivatives of $\loc|_{\zphysical}$ are given, at each patch $\zphysical\cap\zbphysical$, by the ones of the inverse of the physical map $\zbphysicalmap^{-1}$ corresponding to the background mesh. To obtain the derivatives of the shape functions $\zallbpsi$ in terms of the physical coordinates $\zp$, we consider the chain rule for the composition $\zallbpsi\circ\loc|_{\zphysical}$ and the restriction of the map $\loc|_{\zphysical}$ at each patch $\zphysical\cap\zbphysical$. We finally obtain the gradient
\begin{equation}\label{eq:derpsi}
\nabla \zallbpsi = \nabla_{\zbxi} \zallbpsi \cdot \nabla \zbphysicalmap^{-1},
\end{equation}
where $\nabla_{\zbxi}$ is the gradient with respect to $\zbxi$ coordinates, and the Hessian
\begin{equation}\label{eq:derpsi2}
\nabla^2 \zbpsi_j = \left(\nabla \zbphysicalmap^{-1}\right)^{\text{T}} \cdot \nabla_{\zbxi}^2 \zbpsi_j \cdot \nabla_{} \zbphysicalmap^{-1} + \nabla_{\zbxi} \zbpsi_j \cdot \nabla_{}^2 \zbphysicalmap^{-1},
\end{equation}
where
\begin{eqnarray*}\label{eq:derinvphis}
	\nabla_{} \zbphysicalmap^{-1} &=& \left(\nabla_{\zbxi} \zbphysicalmap\right)^{-1},
	\\
	\nabla_{}^2 \zbphysicalmap^{-1} = \nabla_{} \left(\left(\nabla_{\zbxi} \zbphysicalmap\right)^{-1}\right) &=& -\nabla_{} \zbphysicalmap^{-1} \cdot \nabla_{\zbxi}^2 \zbphysicalmap \cdot \nabla_{}\zbphysicalmap^{-1}.
\end{eqnarray*}


\section{Implicit CAD representation: metric and geometry aware optimization}\label{sec:implicit}
Herein, we propose a high-order mesh curving method by an implicitization that measures the geometric deviation.
First, in Section \ref{sec:cadimplicit}, we present a model implicitization for the mesh curving process.
Then, in Section \ref{sec:implicitders}, we detail the first and second-order derivatives for the implicit representation.
Finally, in Section \ref{sec:penalty}, we consider the penalty method to solve the corresponding constrained second-order minimization process for the curving problem.

\subsection{Implicit CAD representation}\label{sec:cadimplicit}
In this section, we present an entity-wise CAD representation for curves in 2D, and for curves, and surfaces in 3D.
For this, we consider the implicit representation of embedded NURBS \cite{laurent2014implicit}, and the Boolean algebraic operations for implicit representations \cite{upreti2014algebraic,biswas2004approximate}.
Then, we assemble these representations to obtain an implicit representation of a CAD model.
Finally, we detail the algorithm of the considered methodology.

We consider a CAD model $\zmodel$ composed of a sequence of NURBS entities.
These NURBS entities can be decomposed into a sequence of B\'{e}zier patches $\znurbs_i,\ i=1,...,n$.
In particular, we describe a $d$-dimensional B\'{e}zier patch $\znurbs\subset\zR^D$ embedded in a $D$-dimensional space in terms of a parameterization
\begin{equation*}\label{eq:param}
\varphi^{}_\znurbs:[0,1]^d\rightarrow\zR^D,\ \ \varphi^{}_\znurbs\left( u \right) \in \znurbs,\ u\in[0,1]^d.
\end{equation*}
In addition, the implicit representation of $\znurbs$ can be obtained as in \cite{laurent2014implicit}
\begin{equation*}\label{eq:implicitfun}
\gamma^{}_\znurbs:\zR^{D}\rightarrow\zR,\ \ \gamma^{}_\znurbs\left( x \right) = 0 \text{ if and only if } x \in \znurbs.
\end{equation*}
Our objective is to obtain a representation $\gamma^{}_\zmodel$ of the model $\zmodel$ that is expressed in terms of the representations $\gamma^{}_{\znurbs_i}$ of the patches $\znurbs_i$.
To combine these implicit representations we use algebraic Boolean operations between real-valued functions \cite{biswas2004approximate}.

\begin{figure*}[t!]
	\centering
	\begin{tabular}{ccc}
		\includegraphics[width=0.20\textwidth]{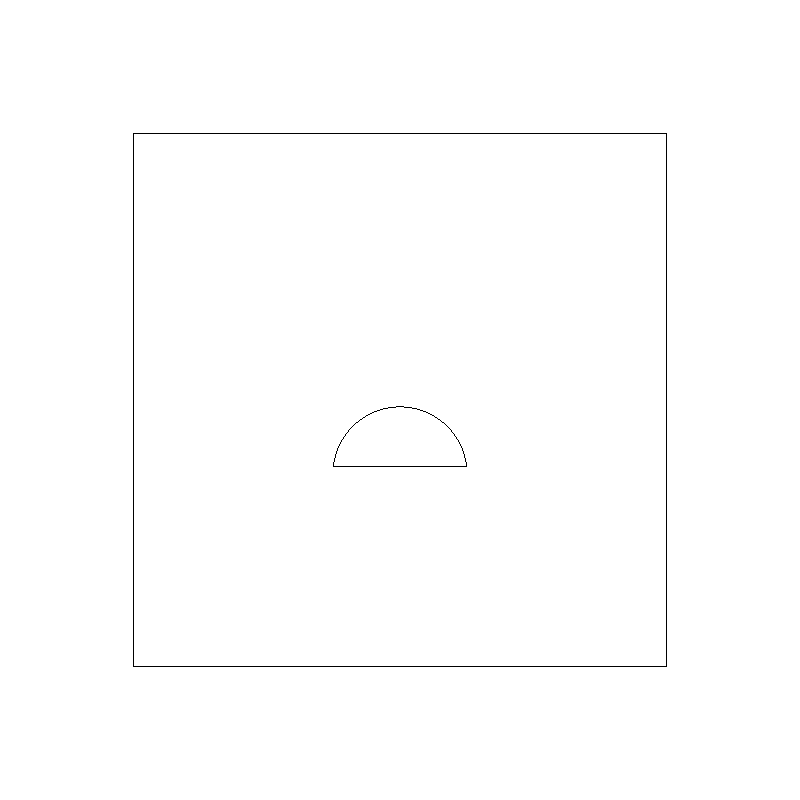}&
		\includegraphics[width=0.20\textwidth]{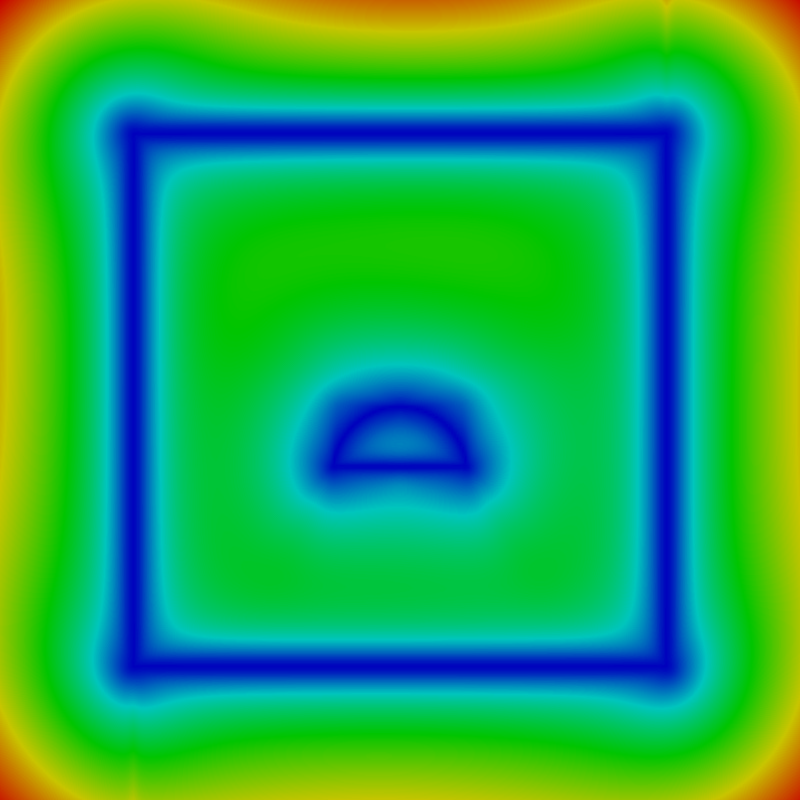}&
		\includegraphics[width=0.20\textwidth]{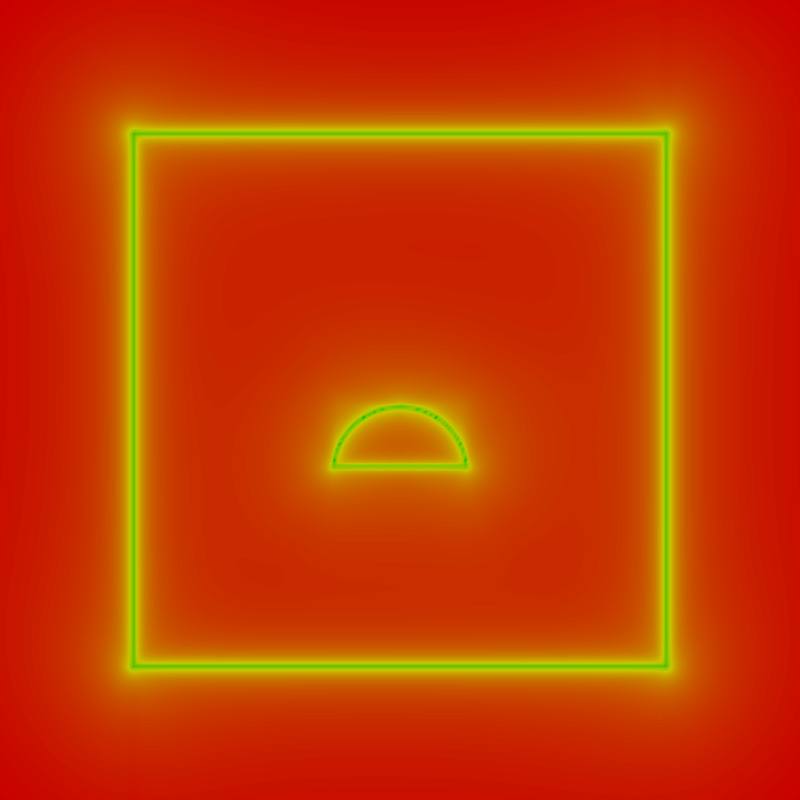}
		\\
		\includegraphics[width=0.20\textwidth]{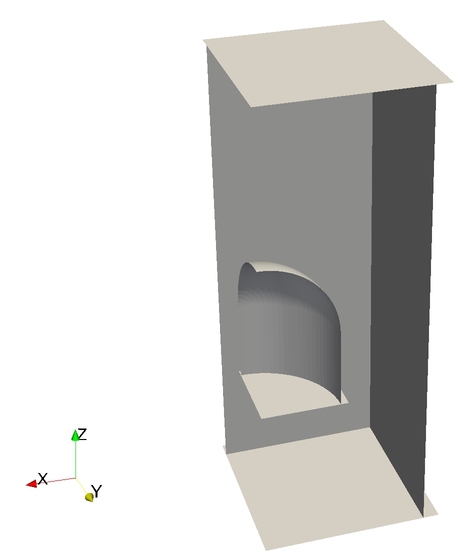}&			
		\includegraphics[width=0.26\textwidth]{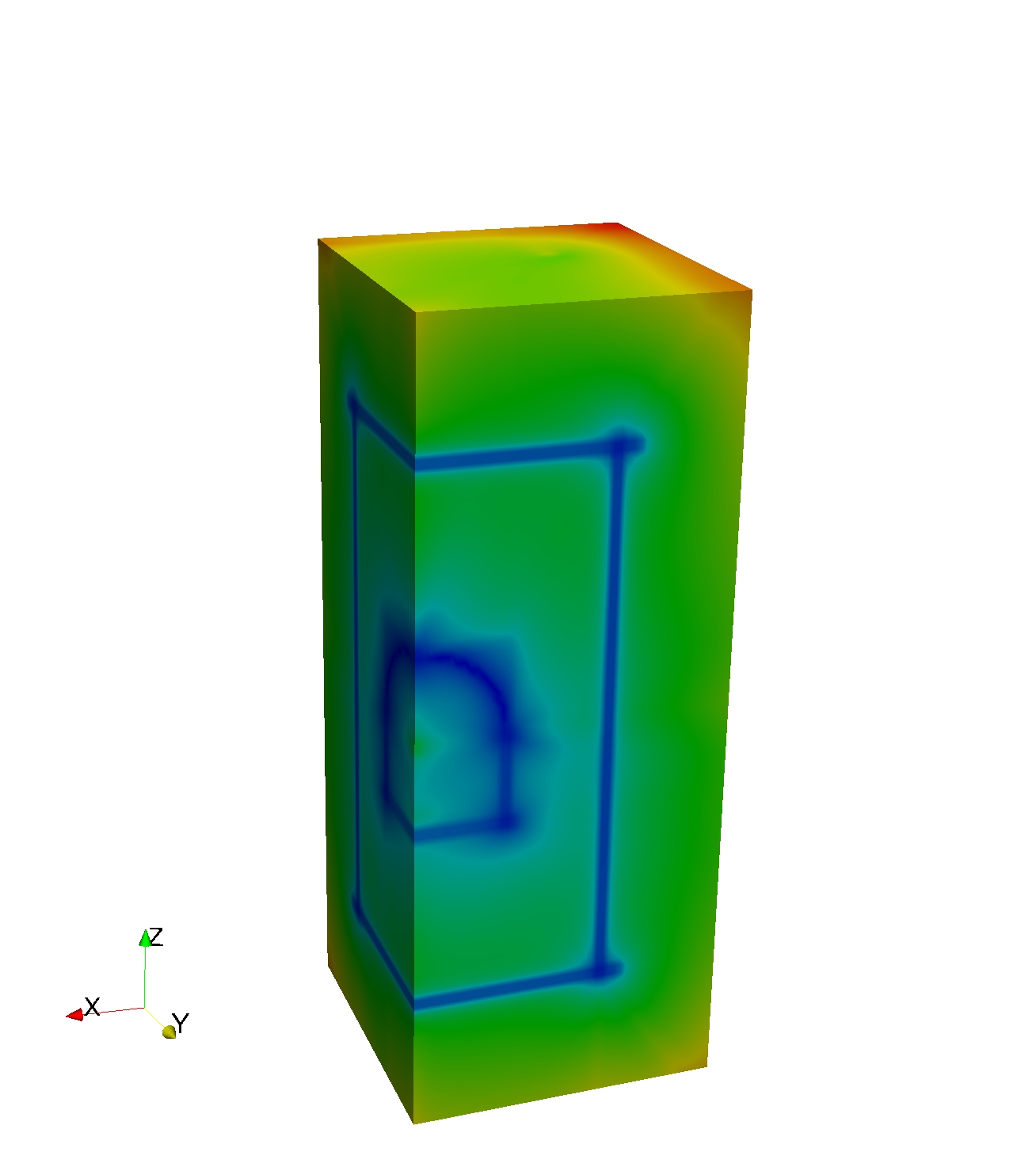}&
		\includegraphics[width=0.26\textwidth]{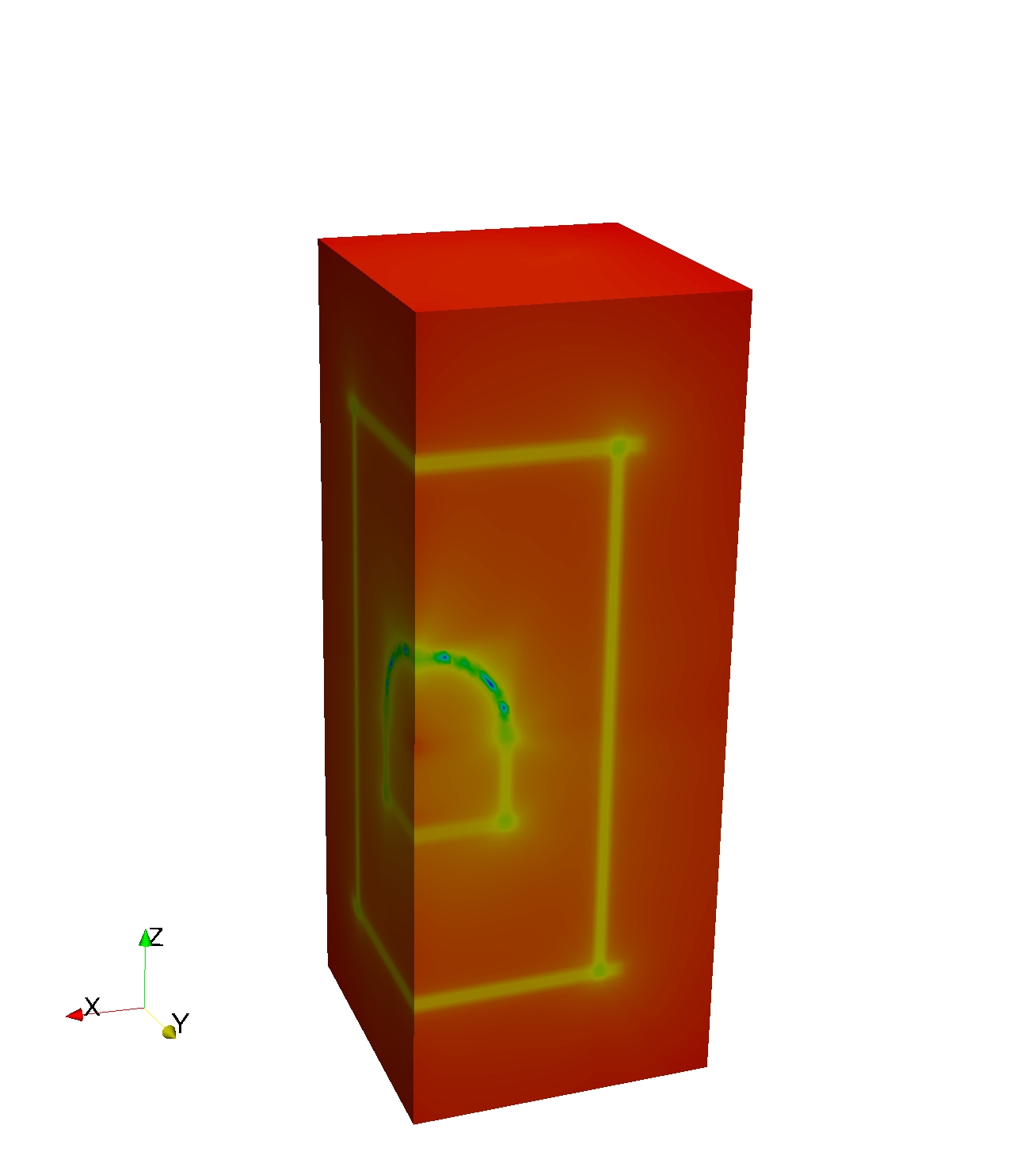}
		\\
	\end{tabular}
	\caption{Implicit representation of (first row) a 2D CAD geometry, and (second row) a 3D CAD geometry. CAD model, and implicit representation in linear, and logarithmic scale in columns.}
	\label{fig:model}
\end{figure*}
In Figure \ref{fig:model}, we show a 2D and a 3D model.
They are mapped via the B\'{e}zier parameterizations $\varphi^{}_{\znurbs_i}$ and their level-sets are represented via the implicit function $\gamma^{}_{\zmodel}$.
The level-sets are illustrated in linear and logarithmic scaling.
As we observe, the functions are numerically zero at the model.
In addition, they smoothly increase far from the model region.

The implicit representation of a CAD model requires a knot preprocessing of the NURBS entities.
Specifically, two knot insertion procedures are required \cite{upreti2014algebraic}.
The first knot insertion, is used to decompose the NURBS entity into B\'{e}zier patches.
The second one, is used to avoid auto-intersections for curves of degree $p \geq 3$.
In this case, we perform an auto-intersection detection process.
Note that the auto-intersection points are given by the equation $\|\nabla\gamma^{}_{\znurbs}\| = 0$.
Then, we detect the auto-intersections by minimizing the quantity $\|\nabla\gamma^{}_{\znurbs}\|^2$ via a one-dimensional search bisection over the parametric line.

To trim the implicit representation in its corresponding domain, we consider the convex-hull of the B\'{e}zier patch control points.
Specifically, for degenerate cases we extrude the set of control points.
To compute the extrusion directions we perform a null space computation via the singular value decomposition.
This defines a valid convex-hull.
We apply this procedure to the degenerate cases given by 2D segments, 3D planes, 3D curves, and 3D cylinders.
Furthermore, we also apply this procedure to approximately degenerate cases such as almost flat curves and surfaces.
For each B\'{e}zier patch $\znurbs$, we consider its implicit representation $\gfun$ defined in the projective space $\mathbb{P}\left(\zR^D\right)$.
In particular, the patch control points determine a vector of matrices that is, $\mathbb{M} = \left(\mathbb{M}_x,\mathbb{M}_y,\mathbb{M}_z,\mathbb{M}_w\right)$ in 3D, corresponding to the projective coordinates $x, y, z,$ and $w$.
Then, we define the implicit representation at a point $\zx\in\mathbb{P}\left(\zR^D\right)$ as in \cite{laurent2014implicit}
\begin{equation}\label{eq:implicit}
\gfun\left(\zx\right) := \det\left(\mathbb{M}\left(\zx\right)\cdot \mathbb{M}\left(\zx\right)^{\text{T}}\right),\ \ \mathbb{M}\left(\zx\right) := \mathbb{M}\ \zx.
\end{equation}
For example, in 3D we set $\zx = (x,y,z,1)$ and hence, $\mathbb{M}\ \zx = \mathbb{M}_x x + \mathbb{M}_y y + \mathbb{M}_z z + \mathbb{M}_w$.
Finally, we normalize the functions $\gfun\left(\zx\right)$ to ensure that they match during the assembly procedure \cite{upreti2014algebraic}.
Specifically, we define the normalized function $\gnfun$ by
\begin{equation}\label{eq:normalized}
\gnfun :=\frac{\gfun}{\|\nabla\gfun\|}.
\end{equation}

The implicit function of a B\'{e}zier patch described in Equation \eqref{eq:normalized} extends over an infinite parametric space.
For this reason, it is standard to trim the patch via a convex hull operation to ensure that the function does not extend beyond the patch limits \cite{upreti2014algebraic,biswas2004approximate}.
Specifically, we  first compute $\zchull\left(\znurbs\right)$, the implicit representation of the convex hull of the B\'{e}zier patch $\znurbs$.
Then, to obtain an implicit representation of $\znurbs$ trimmed by $\zchull\left(\znurbs\right)$, we use a trimming function, denoted by $\text{trim}$, proposed in \cite{biswas2004approximate}

\begin{equation}\label{eq:trim}
\gamma^{}_{\overline{\znurbs}} := \hat{\gamma}^{}_{\zchull\left(\znurbs\right)}\text{ trim }
\hat{\gamma}^{}_{\znurbs} = \sqrt{\hat{\gamma}_{\znurbs}^2 +
\left(
\frac{
\sqrt{\hat{\gamma}_{\znurbs}^4 + \hat{\gamma}_{\zchull\left(\znurbs\right)}^2
}
-
\hat{\gamma}_{\zchull\left(\znurbs\right)}
}{2}
\right)^2},
\end{equation}


where $\gfun^{}_{\Gamma}$ denotes the representation of the B\'{e}zier patch $\Gamma$, see Equation \eqref{eq:implicit}.
The trimming operation of Equation \eqref{eq:trim} is twice differentiable at all points where $\hat{\gamma}^{}_{\znurbs} \neq 0$.
Here, the function $\gamma^{}_{\overline{\znurbs}}$ is an implicit representation of the B\'{e}zier patch $\znurbs$ in its parametric domain $\text{Dom}\ \znurbs$ determined by the NURBS convex-hull $\zchull\left(\znurbs\right)$.

For a given model $\zmodel = \{ \znurbs_1,\znurbs_2,...,\znurbs_n\}$, its implicitization $\gamma^{}_{\zmodel}$ is obtained via the $r$-conjunction $\wedge$ of the implicitizations $\gamma^{}_{\overline{\znurbs_i}}$ of the B\'{e}zier patches $\znurbs_i$ \cite{upreti2014algebraic}.
In particular, for each B\'{e}zier patch $\znurbs_i$, we recursively update the model representation as follows
\begin{equation}\label{eq:rconj}
\gamma^{}_{\zmodel} \leftarrow \gamma^{}_{\zmodel} \wedge \gamma^{}_{\overline{\znurbs}_i} := \gamma^{}_{\zmodel} + \gamma^{}_{\overline{\znurbs}} -  \sqrt{{\gamma^{}_{\zmodel}}^2 + {\gamma^{}_{\overline{\znurbs}}}^2}.
\end{equation}

To obtain the convex-hull representation of a B\'{e}zier patch $\znurbs$, $\zchull\left(\znurbs\right)$, we apply $r$-conjunction to the hyperplane functions of the convex hull entities. Specifically, for each hyperplane entity $\text{H}$ of the convex hull $\zchull\left(\znurbs\right)$ we consider its unit normal component $\textbf{n}$ and its affine term $b$. Then, the implicit representation of $\text{H}$ is given by
\begin{equation}\label{eq:hyperplane}
\gamma^{}_{\text{H}}\left(\zx\right) := \textbf{n}\cdot \zx + b.
\end{equation}
In our case, the sign of the representation $\gamma^{}_{\text{H}}$ is chosen such that $\gamma^{}_{\text{H}} < 0$ outside the convex region enclosed by $\zchull\left(\znurbs\right)$ and $\gamma^{}_{\text{H}} \geq 0$ otherwise. Following, we apply the $r$-conjunction operation for each hyperplane $\text{H}$ to obtain the convex-hull representation $\gamma^{}_{\zchull\left(\znurbs\right)}$, see Equation \eqref{eq:rconj}. Finally, we obtain its normalized version $\hat{\gamma}^{}_{\zchull\left(\znurbs\right)}$ by applying Equation \eqref{eq:normalized}.

\begin{algorithm}[t!]
	\caption{Implicitization}\label{alg:implicit}
	\textbf{Input:} $\zmodel:= \{\znurbs_1,\ \znurbs_2,\ ...,\ \znurbs_n\}$\\
	\textbf{Output:} $\gamma^{}_{\zmodel}$
	\begin{algorithmic}[1]
		\For{$i = 1,...,n$}
		\State $\hat{\gamma}^{}_{\znurbs_i} = $ normalized implicitization of $\znurbs_i$\label{line:implicit}
		\State $\hat{\gamma}^{}_{\zchull\left(\znurbs_i\right)} = $ normalized implicitization of the convex hull of $\znurbs_i$, $\zchull\left(\znurbs_i\right)$\label{line:chull}
		\State $\gamma^{}_{\ztrimmednurbs_i} = $ trimming of $\hat{\gamma}^{}_{\znurbs_i}$ with $\hat{\gamma}^{}_{\zchull\left(\znurbs_i\right)}$\label{line:trim}
		\If{$i = 1$}\label{line:rconj0}
		\State $\gamma^{}_{\zmodel} \gets \gamma^{}_{\overline{\znurbs}_1}$
		\Else
		\State $\gamma^{}_{\zmodel} \gets \gamma^{}_{\zmodel} \wedge \gamma^{}_{\overline{\znurbs}_i}$ $r$-conjunction
		\EndIf\label{line:rconj1}
		\EndFor
	\end{algorithmic}
\end{algorithm}
In Algorithm \ref{alg:implicit}, we describe how to obtain the implicit representation $\gamma^{}_{\zmodel}$ of a model $\zmodel$.
In Line \ref{line:implicit}, we compute for each B\'{e}zier patch $\znurbs_i$ of $\zmodel$ its implicit function $\gamma^{}_{\znurbs_i}$ and we normalize it, see Equations \eqref{eq:implicit} and \eqref{eq:normalized}.
Then, in Line \ref{line:chull}, we consider the convex hull property of the B\'{e}zier control points for trimming \cite{upreti2014algebraic}.
We first obtain an implicit representation of the convex hull $\gamma^{}_{\zchull\left(\znurbs_i\right)}$ by applying a pair-wise $r$-conjunction to the hyperplane functions, see Equation \eqref{eq:hyperplane}.
Then, we compute its normalized representation $\hat{\gamma}^{}_{\zchull\left(\znurbs_i\right)}$, see Equation \eqref{eq:normalized}.
In Line \ref{line:trim}, we trim the B\'{e}zier patch representation $\hat{\gamma}^{}_{\znurbs_i}$ in terms of $\hat{\gamma}^{}_{\zchull\left(\znurbs_i\right)}$, see Equation \eqref{eq:trim}.
The obtained representation is denoted by $\gamma^{}_{\ztrimmednurbs_i}$.
Finally, in Lines \ref{line:rconj0}-\ref{line:rconj1}, we obtain the implicit representation of the model $\zmodel$ by pair-wise $r$-conjunction of $\gamma^{}_{\ztrimmednurbs_i}$, see Equation \eqref{eq:rconj}.
\subsection{Gradient and Hessian}\label{sec:implicitders}

%
Next, we present the gradient and Hessian of the geometry implicitization.
In Section \ref{sec:cadimplicit}, we describe the geometry implicitization in terms of the trimming and $r$-conjunction operations of the convex-hull and B\'{e}zier patch normalized representations.
Accordingly, we describe in this section the derivatives of the trimming and $r$-conjunction operations.
For completeness, we detail in \ref{sec:appendix2} the derivatives of the convex-hull and B\'{e}zier patch normalized representations.



As detailed in Section \ref{sec:cadimplicit}, we perform an $r$-conjunction operation to obtain the model representation.
We compute the derivatives in a straight-forward manner.
Lets denote by $\nabla f * \nabla g$ the matrix with coefficients $\partial_{j} f \partial_{k}g$ for $j,\ k = 1,...,d$.
Then, the derivatives of the $r$-conjunction, presented in Equation \eqref{eq:rconj}, are given by
\begin{equation}\label{eq:gradrconj}
\nabla\ f\wedge g = \nabla f + \nabla g - \nabla\ \sqrt{f^2 + g^2},
\end{equation}
and
\begin{equation}\label{eq:Hessrconj}
\nabla^2\ f\wedge g = \nabla^2 f + \nabla^2 g - \nabla^2\ \sqrt{f^2 + g^2},
\end{equation}
where
\begin{equation}\label{eq:dersqrt}
\nabla\ \sqrt{f^2 + g^2} = \frac{f\nabla f + g\nabla g}{\sqrt{f^2 + g^2}},
\end{equation}
and
\begin{equation}
\begin{split}
\nabla^2\ \sqrt{f^2 + g^2} = \frac{\nabla f * \nabla f + f\nabla^2 f + \nabla g * \nabla g + g\nabla^2 g}{\sqrt{f^2 + g^2}} - \\
\frac{\nabla \sqrt{f^2 + g^2} * \nabla \sqrt{f^2 + g^2}}{\sqrt{f^2 + g^2}}.
\end{split}
\label{eq:der2sqrt}
\end{equation}
Following Equation \eqref{eq:rconj}, we consider that $f := \gfun_{\zmodel}$ and $g := \gfun_{\overline{\znurbs}}$.

Similarly to the $r$-conjunction, we compute the derivatives of the trimming operation, presented in Equation \eqref{eq:trim}.
We simplify the computations by noticing that
\begin{equation*}
\tilde{h} := f\text{ trim } h = \sqrt{f^2 + g^2}\ \ \ \text{for}\ \ g := \frac{\sqrt{h^4 + f^2} - f}{2},
\end{equation*}
where, following Equation \eqref{eq:trim}, we consider $f := \gnfun_{\zchull\left(\znurbs\right)}$, $h := \gnfun_{\znurbs}$, and $\tilde{h} := \gfun_{\overline{\znurbs}}$.
Then, to obtain the derivatives of the trimming operation, we differentiate the term $\sqrt{f^2 + g^2}$, see Equations \eqref{eq:dersqrt} and \eqref{eq:der2sqrt}.
In this case, the derivatives of $g$ can be computed as follows
\begin{equation}\label{eq:dertrim}
\nabla g = \frac{1}{2}\left(\frac{2h^3\nabla h + f\nabla f}{\sqrt{h^4 + f^2}} - \nabla f\right),
\end{equation}
and
\begin{eqnarray*}\label{eq:der2trim}
\nabla^2 g = \frac{1}{2}\left(\frac{2h^2\left(h\nabla^2 h + 3\nabla h * \nabla h\right) + f\nabla^2 f + \nabla f * \nabla f}{\sqrt{h^4 + f^2}} - \right.\\
\left. \frac{\nabla \sqrt{h^4 + f^2} * \nabla \sqrt{h^4 + f^2}}{\sqrt{h^4 + f^2}} - \nabla^2 f\right),
\end{eqnarray*}
where the term $\nabla \sqrt{h^4 + f^2}$ can be computed from Equation \eqref{eq:dersqrt} for the functions $f$ and $h^2$.

As we observe, the derivatives of both the $r$-conjunction and the trimming operation require the derivatives of the convex-hull and B\'{e}zier patch normalized representations. For completeness, we detail these last derivatives in \ref{sec:appendix2}.

\subsection{Minimizing metric and geometry deviations}\label{sec:penalty}
In this section, we consider a modification of the methodology to generate curved high-order meshes featuring optimal mesh quality and geometric accuracy presented in \cite{ruiz2016generation,ruiz2022automatic}.
This technique combines a distortion measure and a geometric $L^2$-disparity measure into a single objective function.
While the element distortion term takes into account the mesh quality, the $L^2$-disparity term takes into account the geometric error introduced by the mesh approximation to the target geometry. Herein, the target geometry is an implicit representation.

Our input data is a CAD model, $\zmodel$, composed of several geometric entities in such manner that
\begin{equation*}
\zmodel = \bigcup_{k = 1}^n \zmodel_k,
\end{equation*}
where each geometric entity is composed of sub-entities.
These sub-entities are curves in 2D, and curves and surfaces in 3D.
In 3D, we consider that the curves are embedded directly in the containing space.

In our representation, we consider that the curves are the image of a segment.
Moreover, we consider that the surfaces are the image of a rectangular region.
For the 3D cases, we consider the implicitization of the curves and surfaces.
In this manner, we can allow the inner curve (surface) nodes to target the implicitization of the corresponding curve (surface).

In what follows, we propose an entity-wise implicit representation of the CAD model $\zmodel$.
We use it to measure the geometric deviation between the mesh and the model.
In particular, for each geometric entity $\zmodel_k$ we consider the implicit representation, see Section \ref{sec:implicit}.
This geometric entity is approximated by a set of boundary mesh entities, denoted by $\partial\zmesh\left( \zmodel_k \right)$.
Instead of measuring the geometric error, herein we account from the geometric deviation through the average of the square of the level set value.
This term is zero when on top of the target CAD entity, and the square ensures deriviability at the zero-level set.
Specifically, this deviation measure is integrated over the candidate boundary mesh entities as follows
\begin{equation}\label{eq:geoapprox}
\altmathcal{G}\left(\partial\zmesh\left( \zmodel_k \right)\right) := \int_{\partial\zmesh\left( \zmodel_k \right)} \gfun^2.
\end{equation}
Note that, the model representation $\gfun^{}_{\zmodel_k}$ is not differentiable at the zero level-set.
By considering the squared function ${\gfun^{}_{\zmodel_k}}^2$ we avoid the derivative singularity.

Our objective is to determine an optimal physical mesh, $\zmesh$, in terms of mesh quality and
geometric deviation.
First, the mesh quality deviation term, distortion, is presented in Section \ref{sec:preliminaries}.
Second, we consider Equation \eqref{eq:geoapprox} to take into account the geometric deviation.
Finally, we define the functional for the mesh quality and the geometric deviation
\begin{equation}\label{eq:curving}
\altmathcal{H}\left(\zmesh; \lambda\right) := \altmathcal{F}\left(\zmesh\right) + \lambda \altmathcal{G}\left(\partial\zmesh\right),
\end{equation}
where
\begin{equation*}\label{eq:curvingterm}
\altmathcal{G}\left(\partial\zmesh\right) := \sum_{k = 1}^{n} \altmathcal{G}\left(\partial\zmesh\left( \zmodel_k \right)\right),
\end{equation*}
and where $\lambda$ corresponds to the penalty parameter. This parameter $\lambda$ can be chosen heuristically or with an automatic procedure \cite{ruiz2022automatic}.

To deal with corners and geometric edges, we distinguish between nodes targeting points or curves of the geometry. For points, we associate the corresponding node with the incident curves. Moreover, for this node, the objective function accounts for the measure of the distance to all the incident curves. Thus, the optimal node is close to the target point because it is close to all the incident curves. For curves, we associate the corresponding nodes with the curve and the incident surfaces. Moreover, for these nodes, the objective function accounts for the measure of the distance to the curve and the two incident surfaces. Thus, the optimal nodes are close to the target curve and the two incident surfaces.

\begin{algorithm}[t!]
	\caption{Distortion minimization}\label{alg:adaption}
	\textbf{Input:} $\zmodel,\ \altmathcal{M},\ \hat{\altmathcal{M}},\ \hat{\textbf{M}},\ \varepsilon,\ \lambda$\\
	\textbf{Output:} $\altmathcal{M}^*$
	\begin{algorithmic}[1]
		\State $\textbf{X} \gets \textrm{coordinates}(\altmathcal{M})$
		\State $\partial\textbf{X} \gets \textrm{coordinates}(\partial \altmathcal{M})$
		\State $\textbf{M} := \textbf{M}\left(\hat{\altmathcal{M}},\hat{\textbf{M}},\textbf{X}\right)$ \Comment{Section \ref{sec:meshInterp}}
		\State $\nabla \textbf{M} := \nabla\textbf{M}\left(\hat{\altmathcal{M}},\hat{\textbf{M}},\textbf{X}\right)$; 
		$\nabla^2 \textbf{M} := \nabla^2\textbf{M}\left(\hat{\altmathcal{M}},\hat{\textbf{M}},\textbf{X}\right)$ \Comment{Section \ref{sec:interpDeriv}}
		\State $\gamma := \gamma\left( \zmodel, \partial\textbf{X} \right)$ \Comment{Section \ref{sec:cadimplicit}}
		\State $\nabla\gamma := \nabla\gamma\left( \zmodel, \partial\textbf{X} \right)$;
		$\nabla^2\gamma := \nabla^2\gamma\left( \zmodel, \partial\textbf{X} \right)$ \Comment{Section \ref{sec:implicitders}}
		\State $\altmathcal{F} := \altmathcal{F}\left(\textbf{X}, \textbf{M}\right)$; \Comment{Section \ref{sec:preliminaries}, Equation \eqref{eq:minfun}}
		\State $\nabla\altmathcal{F} := \nabla\altmathcal{F}\left(\textbf{X}, \textbf{M},\nabla\textbf{M}\right)$; $\nabla^2\altmathcal{F} := \nabla^2\altmathcal{F}\left(\textbf{X}, \textbf{M},\nabla\textbf{M},\nabla^2\textbf{M}\right)$
		\State $\altmathcal{G} := \altmathcal{G}\left(\partial\textbf{X},\gamma\right)$; \Comment{Section \ref{sec:implicit}, Equation \eqref{eq:geoapprox}}
		\State $\nabla\altmathcal{G} := \nabla\altmathcal{G}\left(\partial\textbf{X},\gamma,\nabla\gamma\right)$; $\nabla^2\altmathcal{G} := \nabla^2\altmathcal{G}\left(\partial\textbf{X},\gamma,\nabla\gamma,\nabla^2\gamma\right)$
		\State $\altmathcal{H} \gets \altmathcal{F} + \lambda \altmathcal{G}$ \Comment{Section \ref{sec:implicit}, Equation \eqref{eq:curving}}
		\State $\textbf{X}^* \gets \textrm{Non-linearSolver}\left(\altmathcal{H},\nabla \altmathcal{H},\nabla^2 \altmathcal{H},\textbf{X},\varepsilon\right)$ \Comment{Section \ref{sec:preliminaries}, Equation \eqref{eq:argmin}}
		\State $\altmathcal{M}^*\gets$ update coordinates of $\altmathcal{M}$ with $\textbf{X}^*$
	\end{algorithmic}
\end{algorithm}
In Algorithm \ref{alg:adaption}, we outline the structure of the distortion minimization. The algorithm inputs are: a CAD model $\zmodel$, a physical mesh $\altmathcal{M}$, a background mesh $\hat{\altmathcal{M}}$ equipped with a discrete metric $\hat{\textbf{M}}$, a residual tolerance $\varepsilon$, and a penalty parameter $\lambda$. The output is an optimized physical mesh $\altmathcal{M}^*$ with the same connectivity of $\altmathcal{M}$ and matching the metric $\hat{\textbf{M}}$ and the curved boundary $\zmodel$.
To outline the algorithm, we assign variables, and we declare the corresponding functions and their derivatives in terms of previously defined functions and derivatives.  We recall that, the implementation details of the values and derivatives of the log-Euclidean interpolation $\textbf{M}$ and the implicitation $\gamma$ are detailed in Section \ref{sec:interpolation} and Section \ref{sec:implicit}, respectively. Note that the derivatives of $\altmathcal{F}$ and $\altmathcal{G}$ depend on the corresponding derivatives of $\textbf{M}$ and $\gamma$, respectively.

Algorithm \ref{alg:adaption} proceeds as follows.
First, we assign the volume and boundary mesh coordinates to $\textbf{X}$ and $d\textbf{X}$, respectively.
From these coordinates, we declare the Log-Euclidean interpolation of the discrete metric $\hat{\textbf{M}}$ and its derivatives, $\nabla \hat{\textbf{M}}$ and $\nabla^2 \hat{\textbf{M}}$, see Section \ref{sec:interpolation}.
In addition, from the CAD model $\zmodel$, we declare the implicitization $\gamma$ and its derivatives, $\nabla\gamma$ and $\nabla^2\gamma$, in terms of $d\textbf{X}$, see Section \ref{sec:implicit}.
Then, we declare the distortion functional $\altmathcal{F}$ and the boundary functional  $\altmathcal{G}$.
For these functionals, we also declare the dependency of their derivatives in terms of the values and derivatives of the metric $\textbf{M}$ interpolation and the geometry implicitation $\gamma$. These declarations allow assigning the objective function $\altmathcal{H}$ according to the functionals, $\altmathcal{F}$ and $\altmathcal{G}$, and the penalty parameter $\lambda$, see Equation \eqref{eq:curving}.
Finally, we call a second-order non-linear solver to minimize the objective function up to a residual tolerance $\varepsilon$. This results in an adapted mesh $\altmathcal{M}^*$ with coordinates $\textbf{X}^*$ and with the same connectivity as $\altmathcal{M}$.

\section{Results}
\label{sec:results}


In this section, we present a 2D and a 3D example to illustrate the applicability of our distortion minimization framework for curved $r$-adaption to a high-order metric interpolation while preserving the implicit representation of the boundary.
First, we generate a background mesh $\zbmesh$ and we evaluate the analytical metric $\zmetric$ at the background mesh nodes.
Second, we generate an initial physical mesh $\zmesh$ and we measure its distortion (quality) by interpolating the metric.
Then, by relocating the nodes, we minimize the mesh distortion problem presented in Equation \eqref{eq:argmin} using the framework presented in this work.
Moreover, in the last examples, we consider a boundary term that takes into account the geometric deviation.
We relocate the nodes to minimize the distortion measure while preserving the curved features of the boundary.

To summarize the results, we present a statistics table for the element quality of Equation \eqref{eq:qualityreg}, and the figures for the initial and optimized meshes.
Specifically, we show the minimum quality, the maximum quality, the mean quality, and the standard deviation of the initial and optimized meshes.
We highlight that in all cases, the optimized mesh increases the minimum element quality and it does not include any inverted element.
In addition, the meshes resulting after the optimization are composed of elements aligned and stretched to match the target metric tensor.
In all figures, the meshes are colored according to the point-wise quality presented in Equation \eqref{eq:pointwisequality}.

Because our goal is to optimize the mesh distortion using the detailed derivatives, instead of including mathematical proofs of mesh validity, we detail how we numerically enforce the positiveness of the element Jacobians.
Specifically, we use a numerical valid-to-valid approach that uses four ingredients.
First, because we want numerically valid results, we enforce mesh validity on the integration points.
Second, to initialize the optimization, we start from a numerically valid mesh. Third, to penalize inverted elements, we modify the point-wise distortion, Equation \eqref{eq:pointwisequality}, to be infinity for non-positive Jacobians.
Specifically, we regularize the element Jacobians to be zero for non-positive Jacobians, so their reciprocals are infinite.
Note that these reciprocals appear in the distortion expression, and thus, they determine the infinite distortion value.
Fourth, to enforce numerically valid mesh displacements, we equip Newton's method with a backtracking line-search.
Specifically, if the mesh optimization update is invalid in any integration point, the objective function, Equation \eqref{eq:argmin}, is infinite.
In that case, the step is divided by two until it leads to a valid mesh update.

As a proof of concept, a mesh optimizer has been developed in Julia 1.6.2 \cite{bezanson2017julia}.
For this, we use the following external packages: Arpack v0.5.0, ILUZero v0.1.0, and TensorOperations v3.1.0.
In addition, we use the MATLAB PDE Toolbox \cite{MATLAB:2017} to generate the initial isotropic linear unstructured 2D and 3D meshes (the structured meshes are generated by subdivision), and the MMG algorithm \cite{dobrzynski2012mmg3d} to generate the initial anisotropic linear unstructured 2D and 3D meshes.
To construct the geometric models, we use the FreeCAD software \cite{riegel2016freecad}.
Finally, we use the Quickhull (Qhull) algorithm \cite{barber1996quickhull} for the convex-hull computations required in the geometric model's implicitization, see Section \ref{sec:implicit}.

The Julia prototyping code is sequential, it corresponds to the implementation of the method presented in this work and the one presented in \cite{aparicio2018defining,aparicio2019imr,aparicio2021icosahom}.
In all the examples, the optimization corresponds to finding a minimum of a nonlinear unconstrained multi-variable function.
In particular, the mesh optimizer uses an unconstrained line-search globalization with an iterative preconditioned conjugate gradients linear solver.
The stopping condition is set to reach an absolute root mean square residual, defined as $\frac{\| \nabla f(x)\|_{\ell^2}}{\sqrt{n}}$ for $x\in\zR^n$, smaller than $10^{-4}$ or a length-step smaller than $10^{-4}$. 
Each optimization process has been performed in a node featuring two Intel Xeon Platinum 8160 CPU with 24 cores, each at 2.10 GHz, and 96 GB of RAM memory.

Following, we first present the target domains to be meshed, and the considered metrics on the domain, Section \ref{sec:examples}.
In Section \ref{sec:meshes} we present the optimization results for a quadrilateral and a hexahedral domain.
In Section \ref{sec:avsd} we compare the proposed discrete based-interpolation procedure with the analytical one from \cite{aparicio2018defining,aparicio2019imr,aparicio2021icosahom}.
Finally, in Sections \ref{sec:iniAnisotropic} and \ref{sec:curvedboundaries}, we show the application of the discrete metric approach to optimize an anisotropic mesh adapted to a given metric generated by the MMG algorithm.
In particular, in Section \ref{sec:curvedboundaries}, we illustrate that our mesh adaption method based in the metric interpolation approach is compatible with curved boundaries.
\subsection{Domains and metrics}\label{sec:examples}
We consider the quadrilateral domain $\Omega=[-0.5,0.5]^2$ for the two-dimensional examples and the hexahedral domain $\Omega=[-0.5,0.5]^3$ for the three-dimensional ones. Each domain is equipped with a metric matching a boundary layer. 
In particular, our target metric $\textbf{M}$ is characterized by a boundary layer metric with a diagonal matrix $\textbf{D}$ and a deformation map $\varphi$ by the following expression
\begin{equation}\label{eq:metricDeformation}
\textbf{M} = \nabla \varphi^{\text{T}}\cdot \textbf{D} \cdot \nabla \varphi.
\end{equation}
In what follows, we first detail the boundary layer metric $\textbf{D}$ and then the deformation map $\varphi$.

The boundary layer aligns with the $x$-axis ($xy$-plane) in the 2D case (3D case).
It determines a constant unit element size along the $x$-direction ($xy$-directions), and a non-constant element size along the $y$-direction ($z$-direction).
This vertical element size grows linearly with the distance to the $x$-axis ($xy$-plane), with a factor $\alpha = 2$, and starts with the minimal value $h_{\min}=0.01$ ($h_{\min}=0.02$).
Thus, the stretching ratio blends from $1:100$ to $1:1$ (from $1:50$ to $1:1$) between $y=-0.5$ and $y=0.5$ (between $z=-0.5$ and $z=0.5$). 
We define the metric for the 2D case as:
\begin{equation}\label{eq:test2D}
\textbf{D} := \left(
\begin{array}{cc}
1 & 0\\
0 & 1/h(y)^2
\end{array}
\right)
\end{equation}
where the function $h$ is defined by
\begin{equation*}
h(x):= h_{\min} + \alpha |x|.
\end{equation*}
Similarly, the metric for the 3D case is
\begin{equation}\label{eq:test3D}
\textbf{D} := \left(
\begin{array}{ccc}
1 & 0 & 0\\
0 & 1 & 0\\
0 & 0 & 1/h(z)^2
\end{array}
\right).
\end{equation}

The deformation map $\varphi$ in Equation \eqref{eq:metricDeformation} aligns the stretching of $\textbf{D}$ according to a given curve in the 2D examples and at a given surface in the 3D examples.
In the 2D case, we define the map $\varphi$ by
\begin{equation*}
\varphi(x,y) = \left(x,\frac{10y - \cos(2\pi x)}{\sqrt{100 + 4\pi^2}} \right),
\end{equation*}
and, in the 3D case by
\begin{equation*}
{\varphi}(x,y,z) = \left(x,y,\frac{10z - \cos(2\pi x)\cos(2\pi y)}{\sqrt{100 + 8\pi^2}}\right).
\end{equation*}

\begin{figure}[t!]
	\centering
	\subfloat{\includegraphics[width=0.3\textwidth]{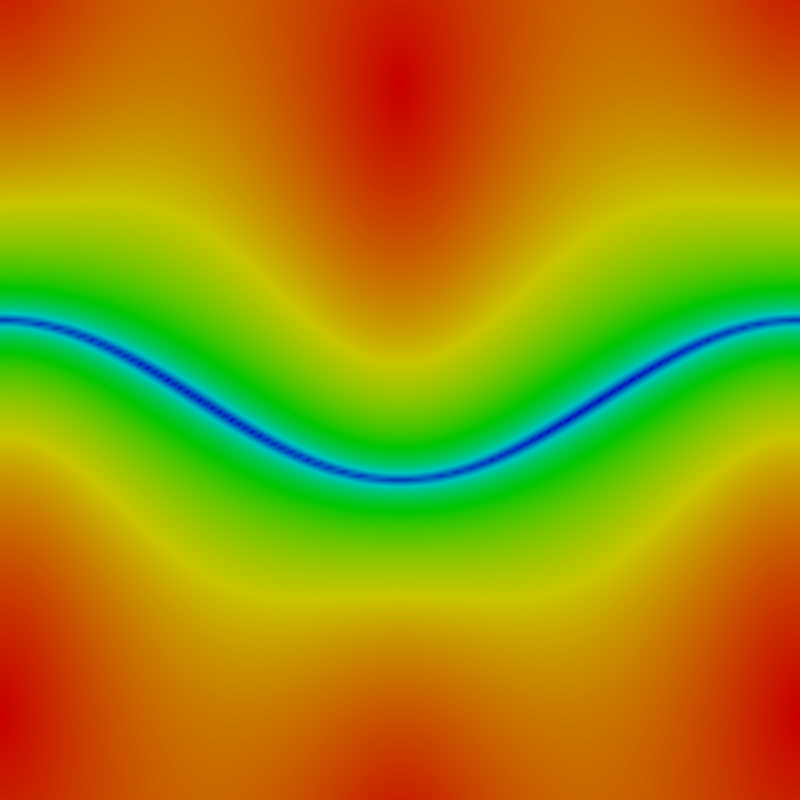}}
	 \\
	 \includegraphics[width=0.35\textwidth]{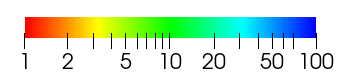} 
	 \\
	 \begin{tabular}{cc}
	 	\subfloat{\includegraphics[width=0.35\textwidth]{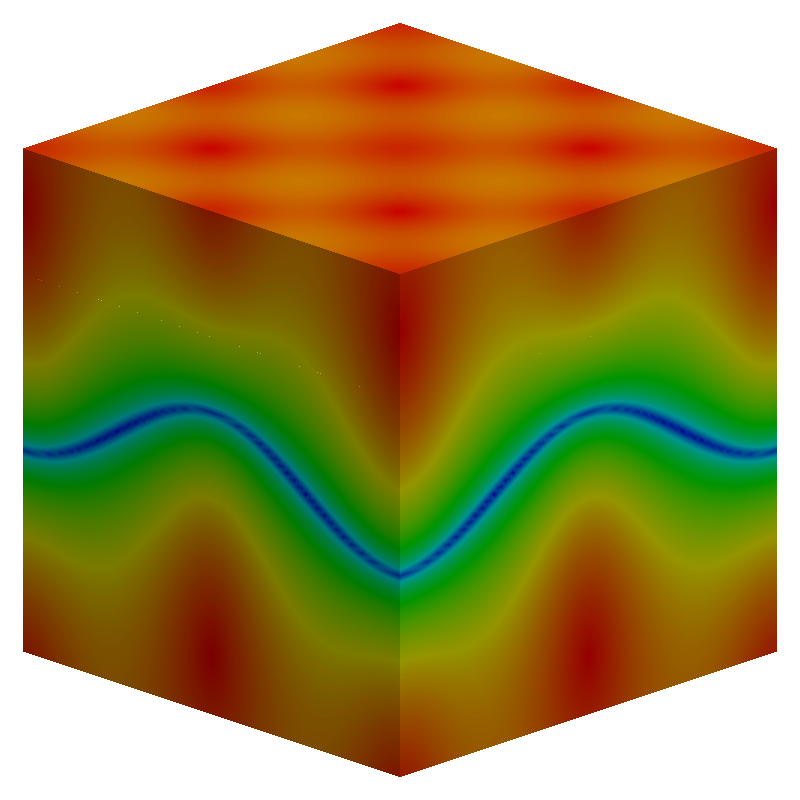}}
	 	&
	 	\subfloat{\includegraphics[width=0.35\textwidth]{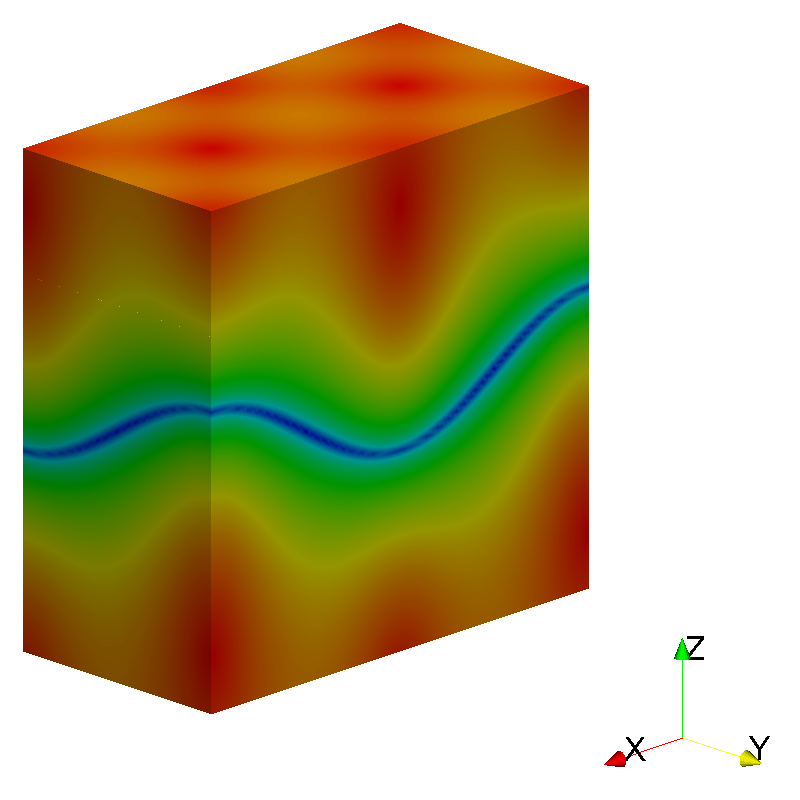}}
	 \end{tabular}
	\\
	\includegraphics[width=0.35\textwidth]{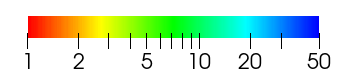}
	\caption{Anisotropic quotient values in logarithmic scale of the target metrics: (top) 2D case; (bottom left) boundaries of the 3D case; and (bottom right) solid slice of the 3D case.}
	\label{fig:stretching}	
\end{figure}
Figure \ref{fig:stretching} shows the anisotropic quotient \cite{loseille:AnisotropicAdaptiveSimulations} of the metric presented in Equations \eqref{eq:test2D} and \eqref{eq:test3D}.
Specifically, the anisotropic quotient of a metric tensor $\zmetric\in\zR^{d\times d}$ is given by
\begin{equation*}
\text{quo} = \max_{i = 1,...,d} \sqrt{\frac{\det\left(\zmetric\right)}{\lambda_i^d}}
\end{equation*}
where $\lambda_i, \ i = 1,...,d$ are the eigenvalues of $\zmetric$.
The considered metric $\textbf{M}$ attains the highest level of anisotropy, close to the curve described by the points $(x,y)\in\Omega$ such that $\varphi(x,y)=(x,0)$ in 2D, and close the surface described by the points $(x,y,z)\in\Omega$ such that $\varphi(x,y,z)=(x,y,0)$ in 3D. 
\subsection{Distortion minimization: initial isotropic straight-edged meshes}\label{sec:meshes}
In this example, we present the optimization results for initially isotropic meshes on the domain equipped with the metrics presented in Section \ref{sec:examples}. 
We describe first the initial meshes $\zmesh$ together with the background meshes $\zbmesh$ where the metric is interpolated. Next, we present the optimized meshes $\zmesh^*$ and to conclude, we present the results obtained from the optimization process. Herein, both the background and physical meshes are meshes of the same polynomial degree.

\begin{figure}[t!]
	\centering
	\setlength{\tabcolsep}{10pt}
	\renewcommand{\arraystretch}{1.5}
	\begin{tabular}{ccc}
		\subfloat[]{\label{fig:p1_0}
			\includegraphics[width=0.25\textwidth]{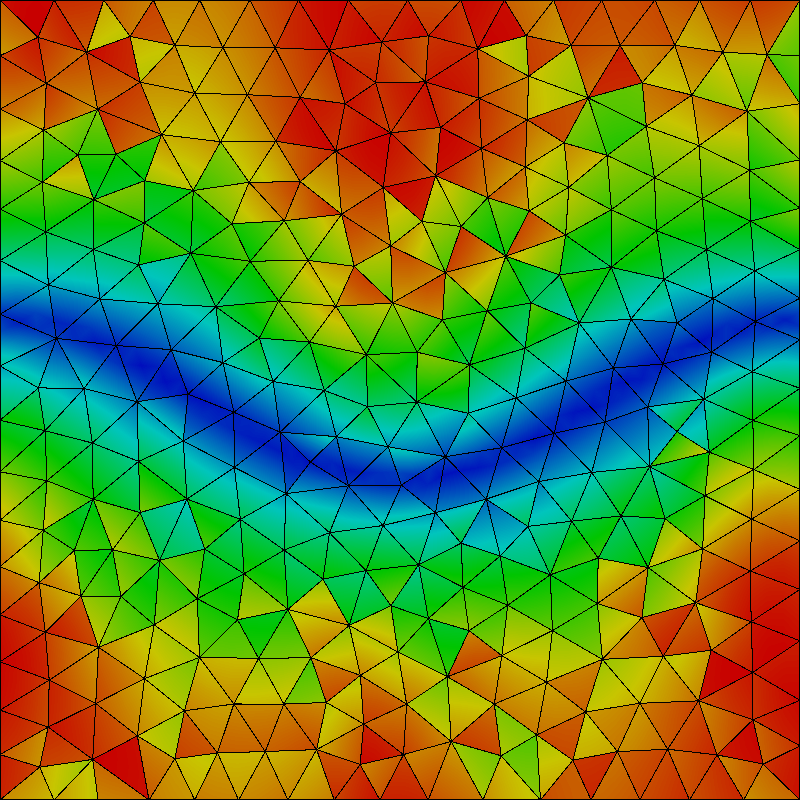}}
		&
		\subfloat[]{\label{fig:p2_0}
			\includegraphics[width=0.25\textwidth]{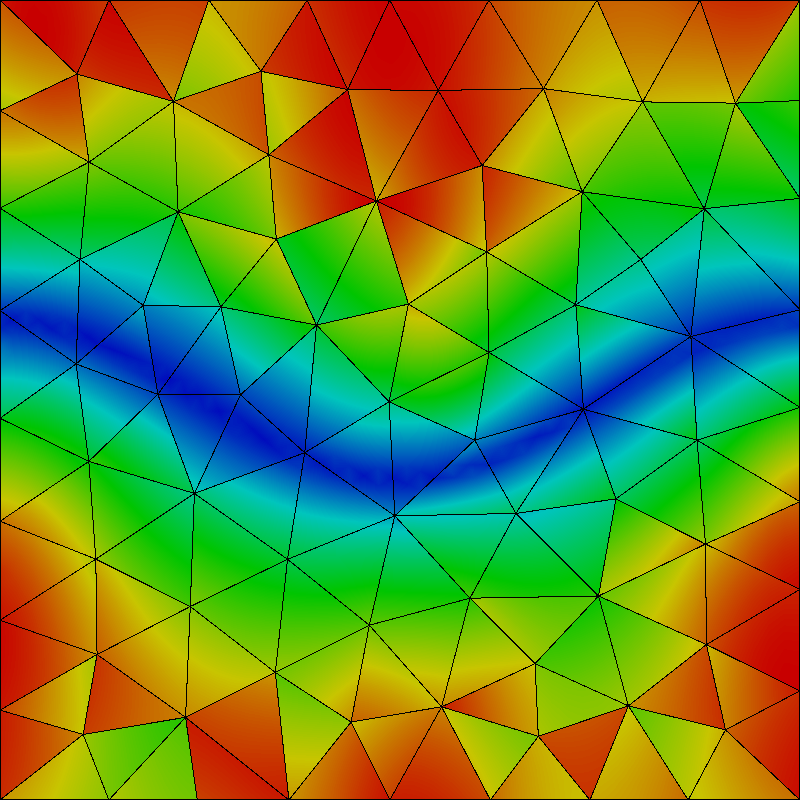}}
		&
		\subfloat[]{\label{fig:p4_0}
			\includegraphics[width=0.25\textwidth]{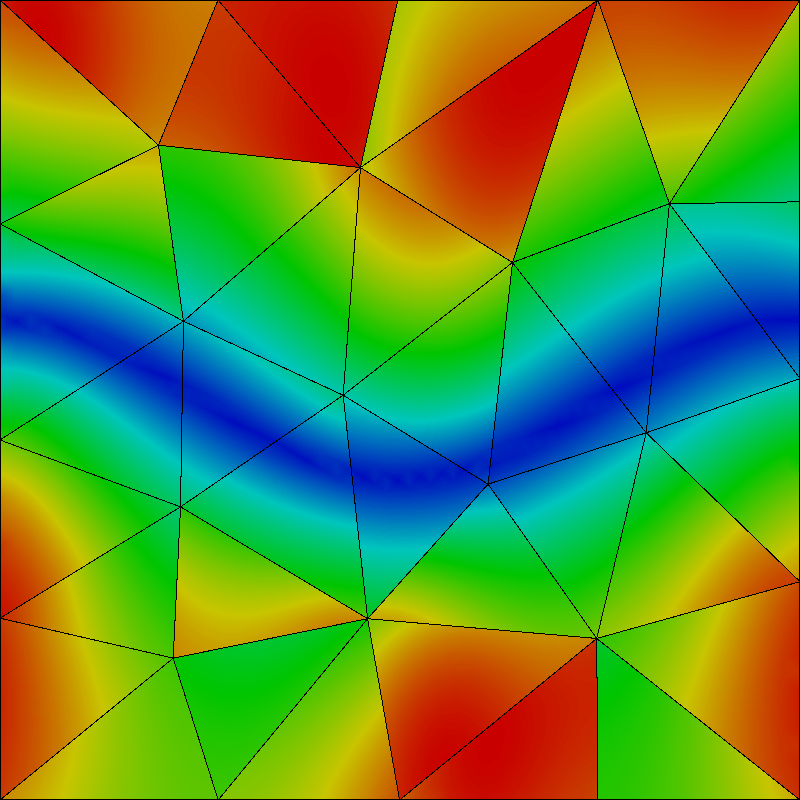}}
		\\
		\subfloat[]{\label{fig:p1_1}
			\includegraphics[width=0.25\textwidth]{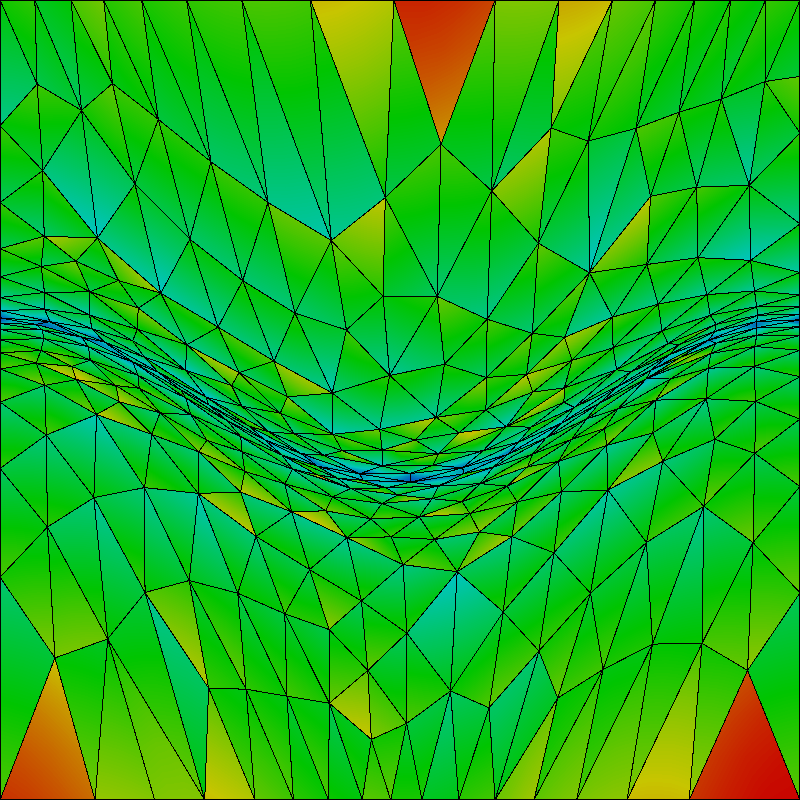}}
		&
		\subfloat[]{\label{fig:p2_1}
			\includegraphics[width=0.25\textwidth]{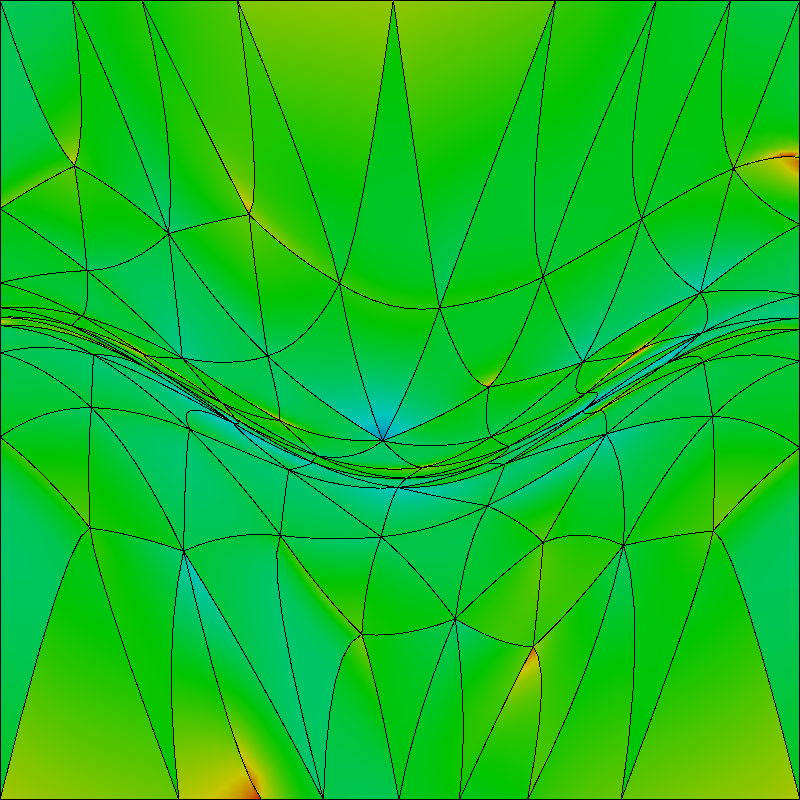}}
		&
		\subfloat[]{\label{fig:p4_1}
			\includegraphics[width=0.25\textwidth]{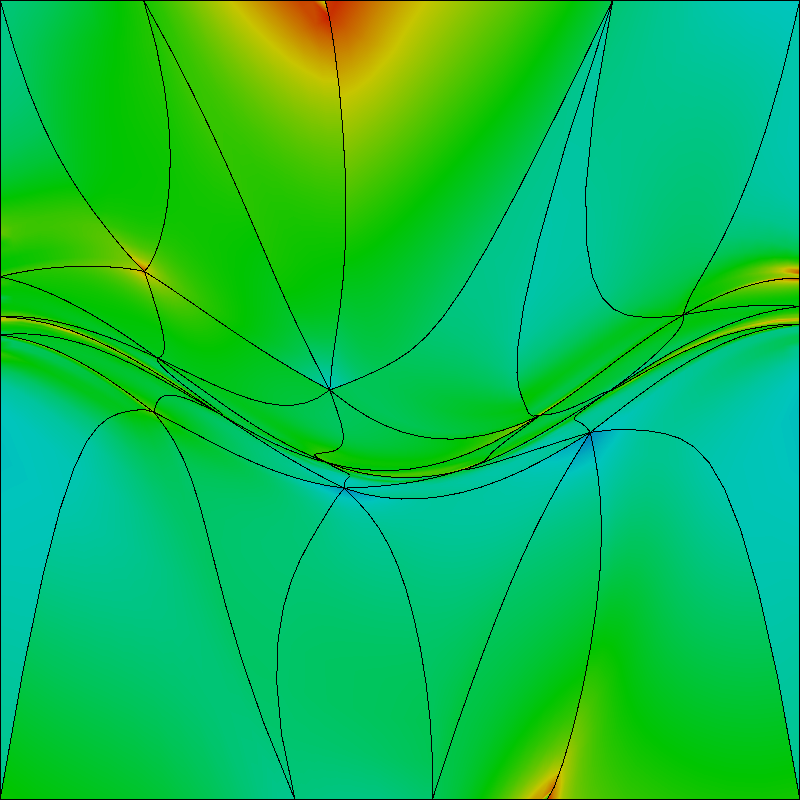}}
		\\
		\subfloat[]{\label{fig:p1_1_a}
			\includegraphics[width=0.25\textwidth]{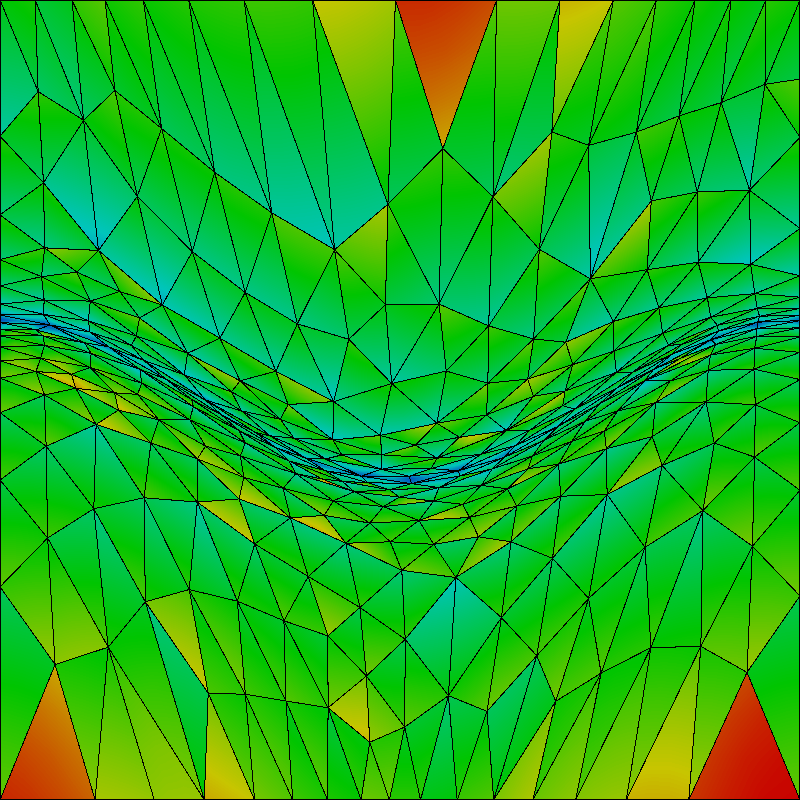}}
		&
		\subfloat[]{\label{fig:p2_1_a}
			\includegraphics[width=0.25\textwidth]{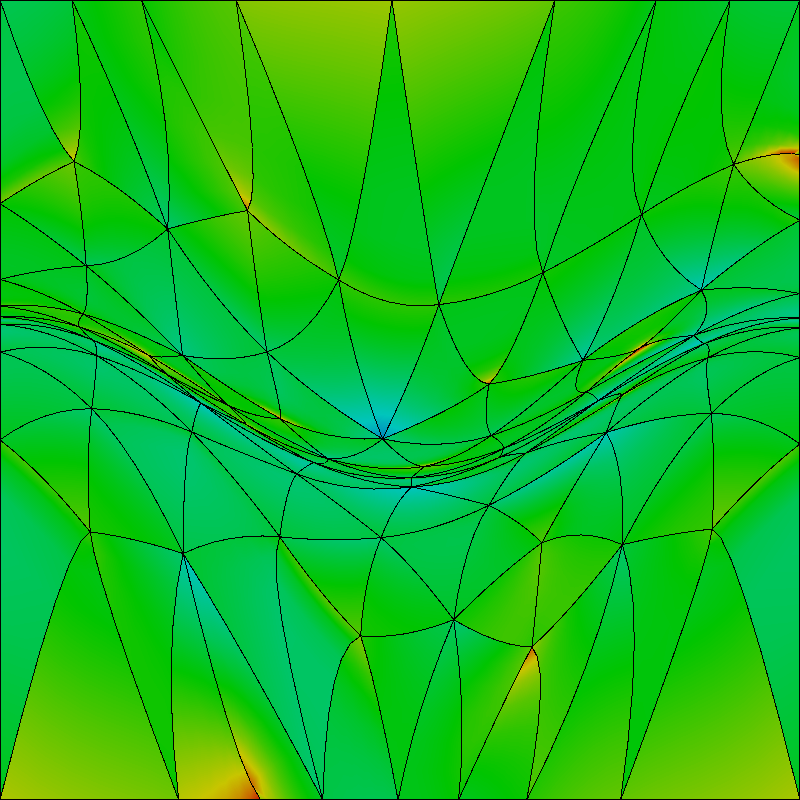}}
		&
		\subfloat[]{\label{fig:p4_1_a}
			\includegraphics[width=0.25\textwidth]{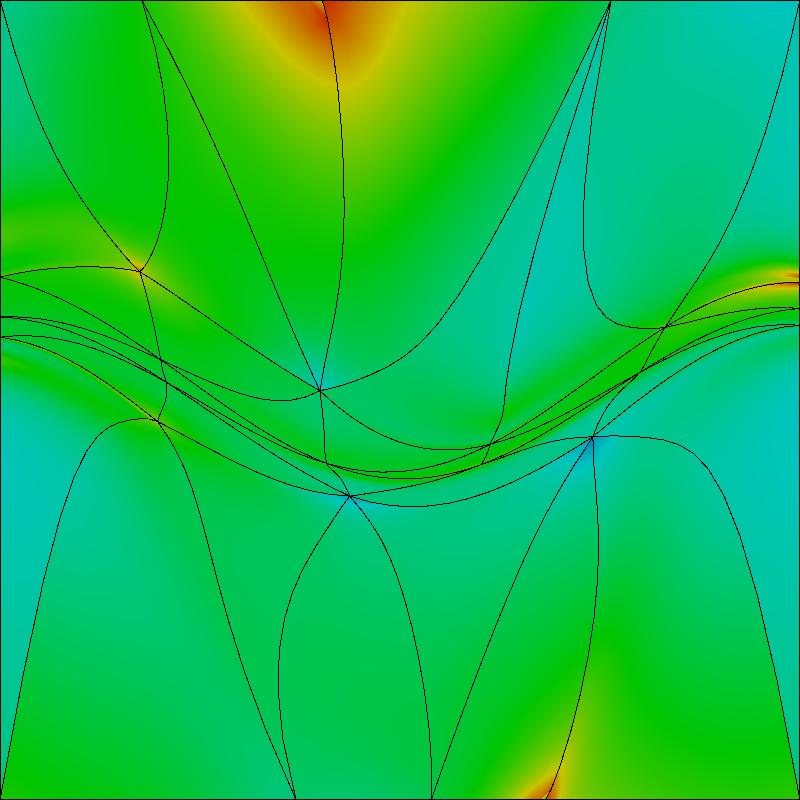}}
	\end{tabular}
	\\
	\includegraphics[width=0.35\textwidth]{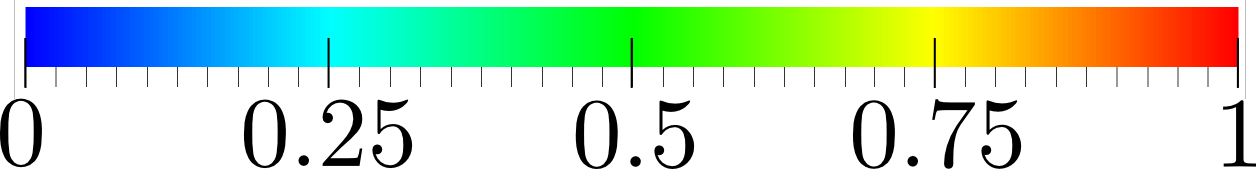}
	\caption{Point-wise distortion for triangular meshes of polynomial degree 1, 2, and 4 in columns. Initial straight-sided isotropic meshes, optimized meshes with discrete metric, and optimized meshes with analytic metric in rows.}
	\label{fig:ex}
\end{figure}
\begin{figure}[t!]
	\centering
	\begin{tabular}{ccc}
		\subfloat[]{\label{fig:p1_0_3}
			\includegraphics[width=0.3\textwidth]{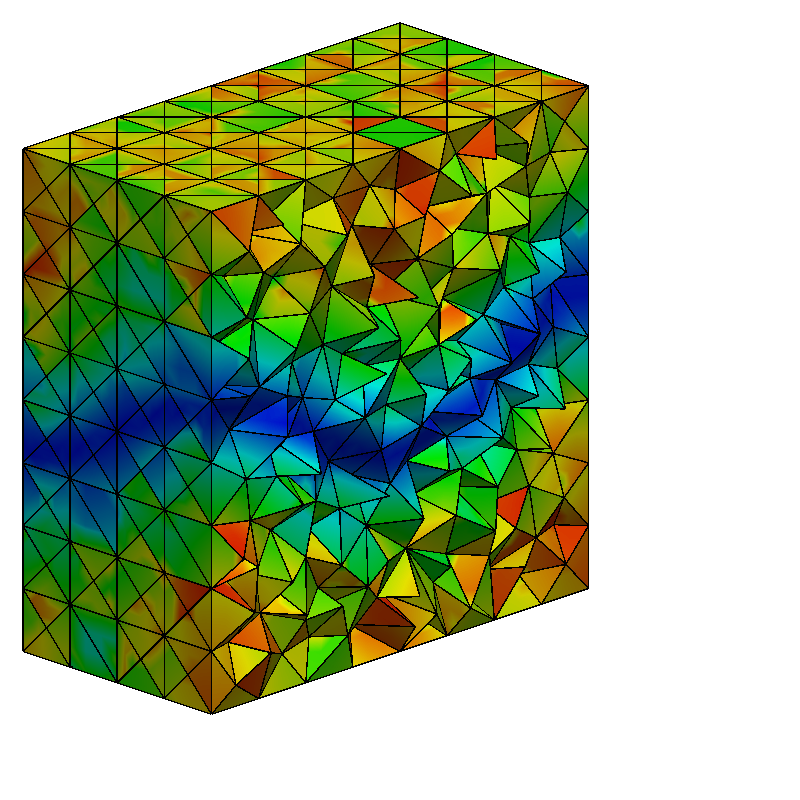}}
		&
		\subfloat[]{\label{fig:p2_0_3}
			\includegraphics[width=0.3\textwidth]{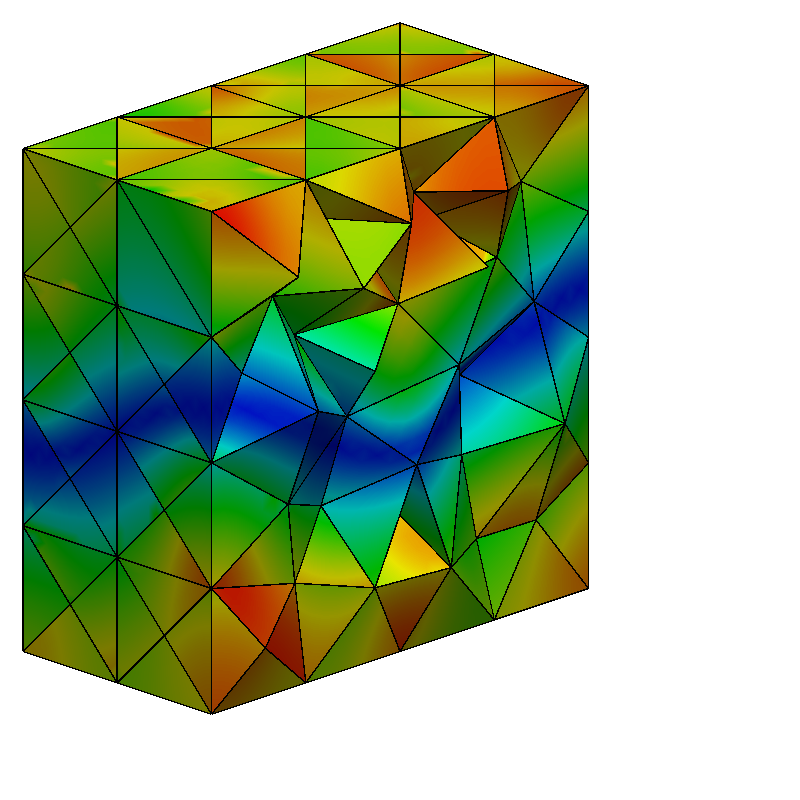}}
		&
		\subfloat[]{\label{fig:p4_0_3}
			\includegraphics[width=0.3\textwidth]{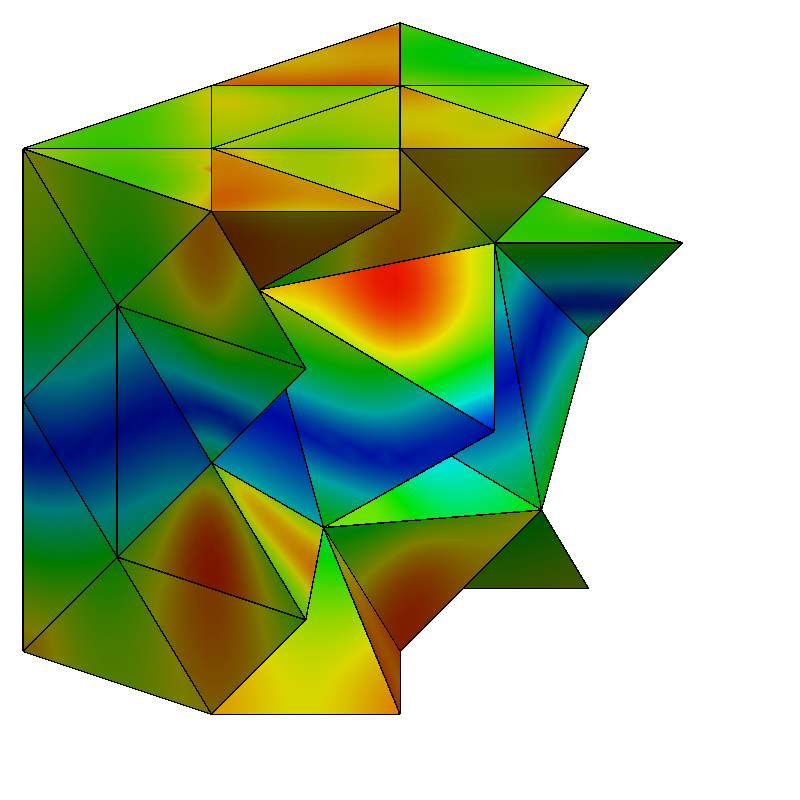}}
		\\
		\subfloat[]{\label{fig:p1_1_3}
			\includegraphics[width=0.3\textwidth]{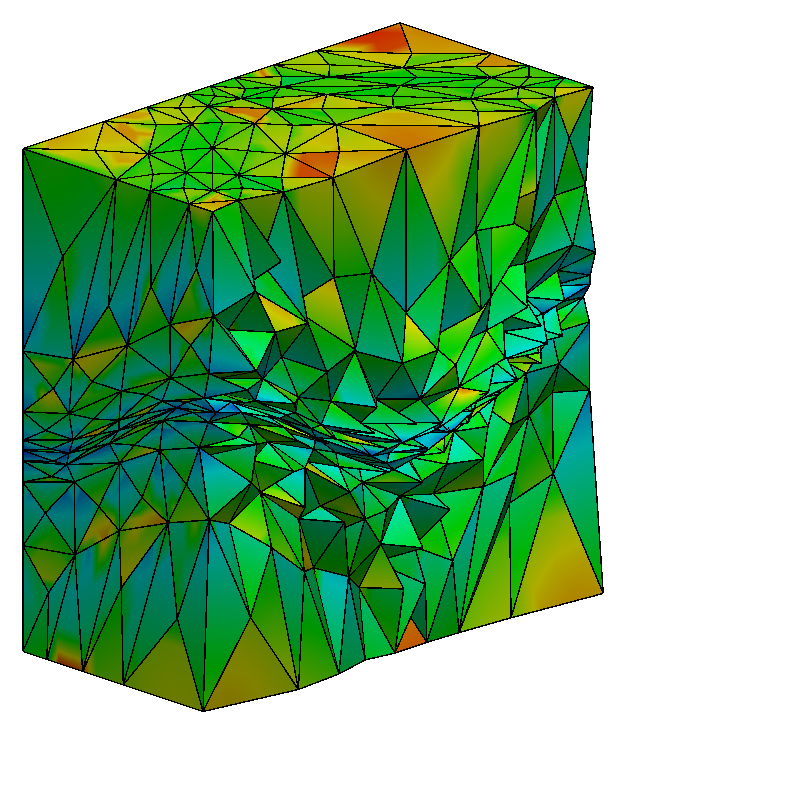}}
		&
		\subfloat[]{\label{fig:p2_1_3}
			\includegraphics[width=0.3\textwidth]{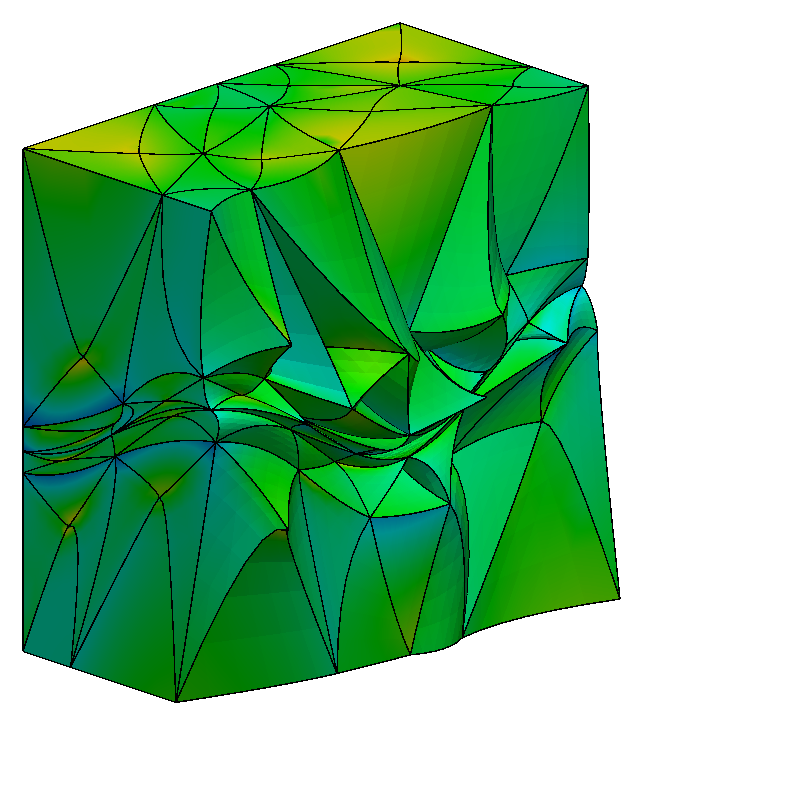}}
		&
		\subfloat[]{\label{fig:p4_1_3}
			\includegraphics[width=0.3\textwidth]{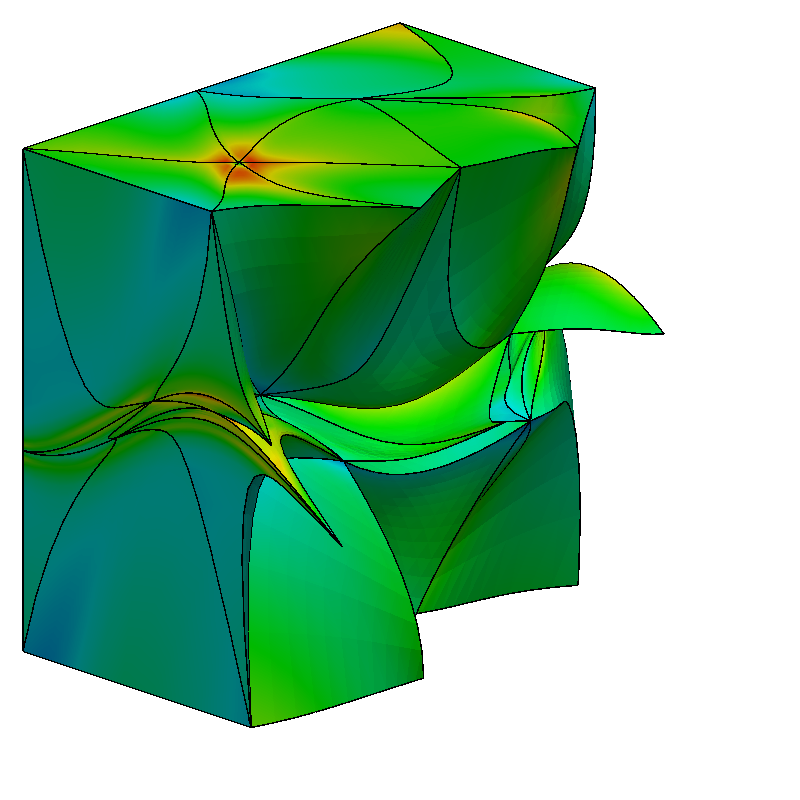}}
		\\
	\end{tabular}
	\\
	\includegraphics[width=0.35\textwidth]{/qualBarParaview_color} 
	\caption{Clipped tetrahedral meshes of polynomial degree 1, 2, and 4 in columns. Initial straight-sided isotropic meshes and optimized meshes from initial meshes in rows.}
	\label{fig:ex3D}
\end{figure}
\begin{figure}[t!]
	\centering
	\begin{tabular}{ccc}
		\subfloat[]{\label{fig:p1_0_3b}
			\includegraphics[width=0.3\textwidth]{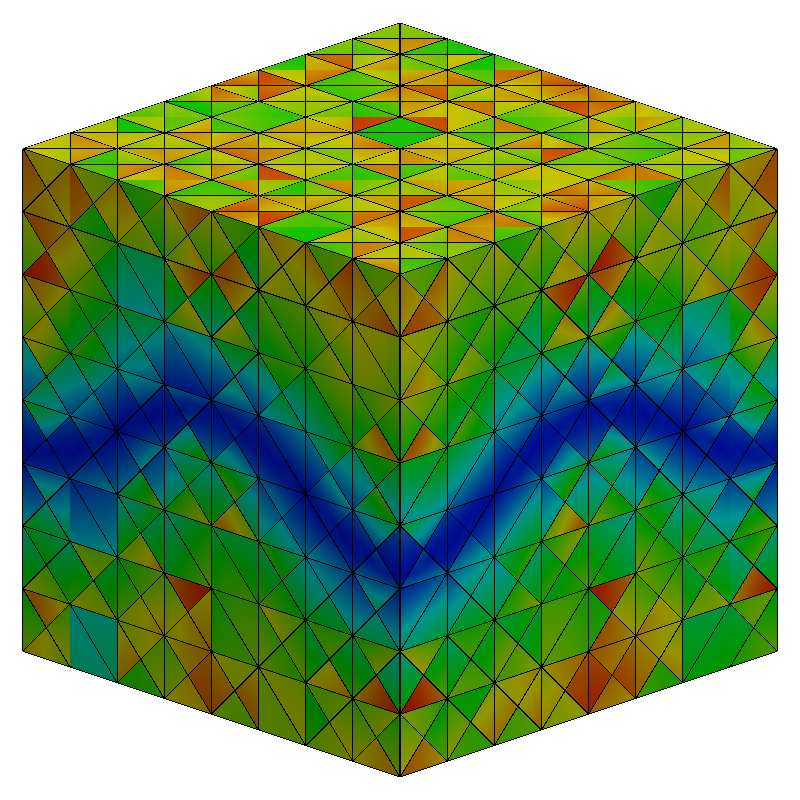}}
		&
		\subfloat[]{\label{fig:p2_0_3b}
			\includegraphics[width=0.3\textwidth]{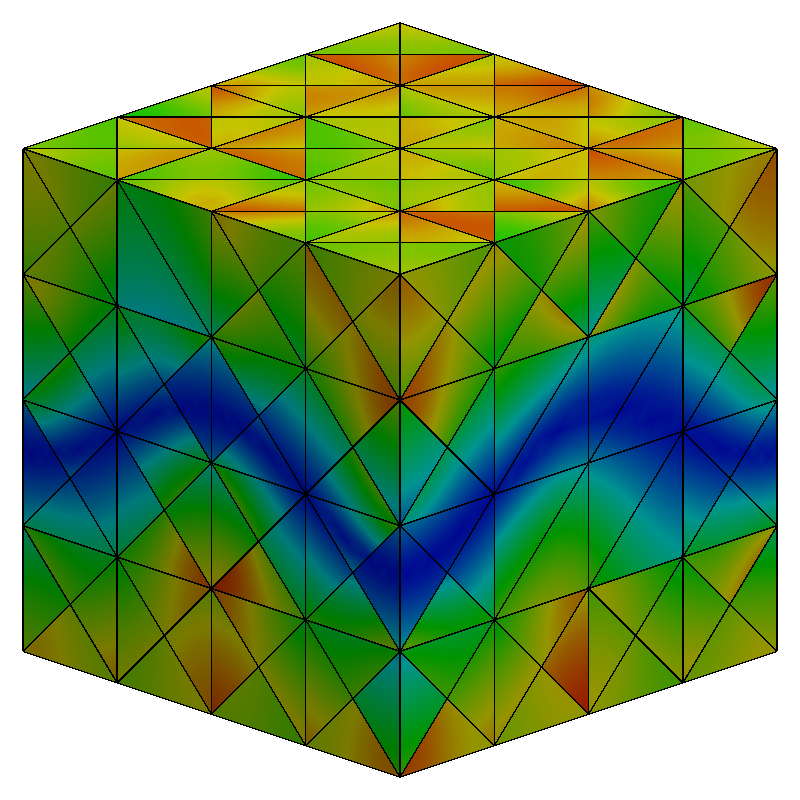}}
		&
		\subfloat[]{\label{fig:p4_0_3b}
			\includegraphics[width=0.3\textwidth]{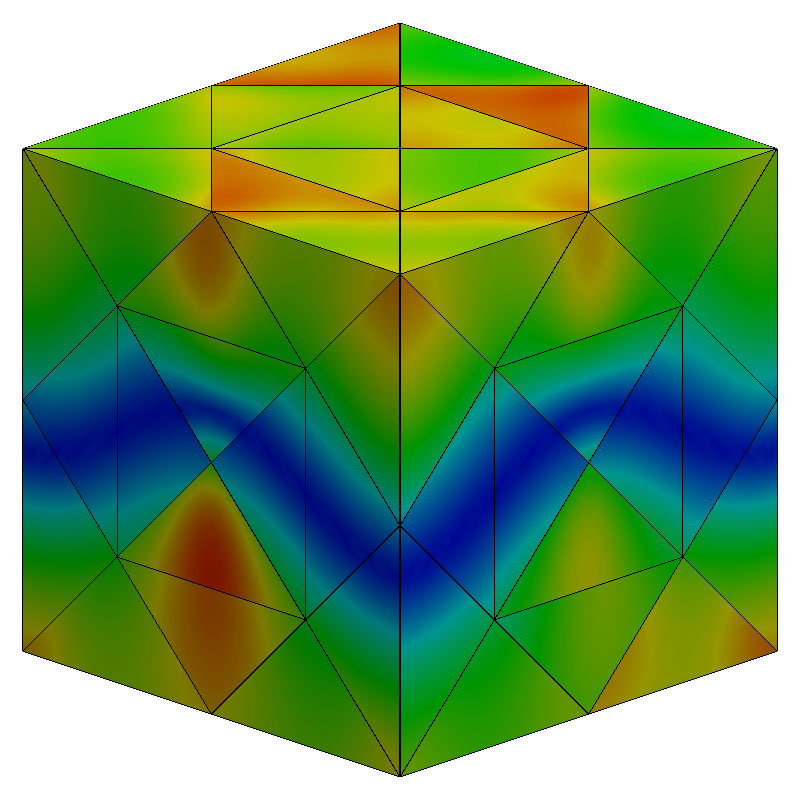}}
		\\
		\subfloat[]{\label{fig:p1_1_3b}
			\includegraphics[width=0.3\textwidth]{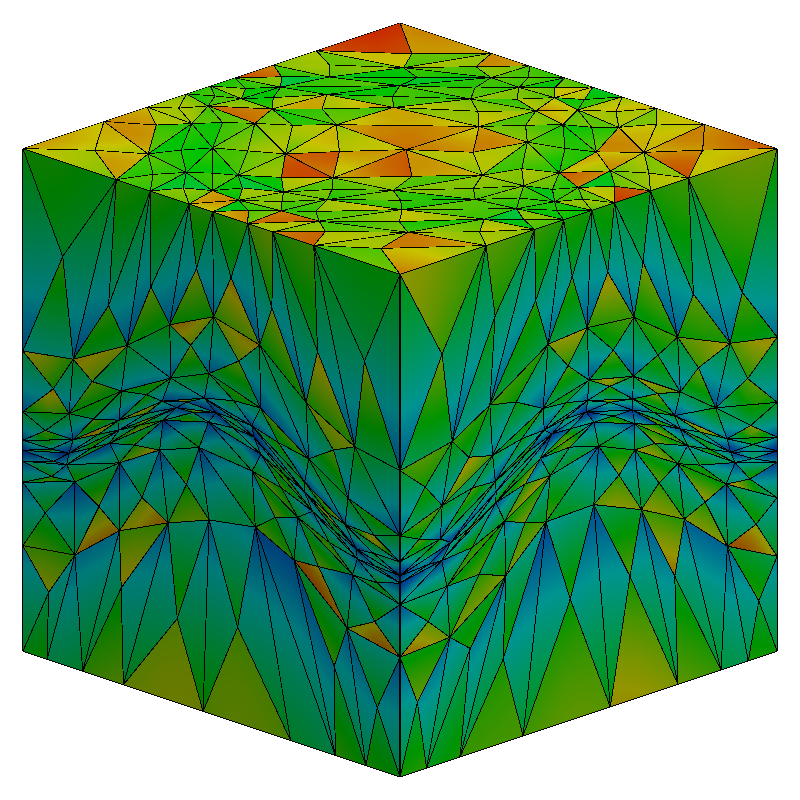}}
		&
		\subfloat[]{\label{fig:p2_1_3b}
			\includegraphics[width=0.3\textwidth]{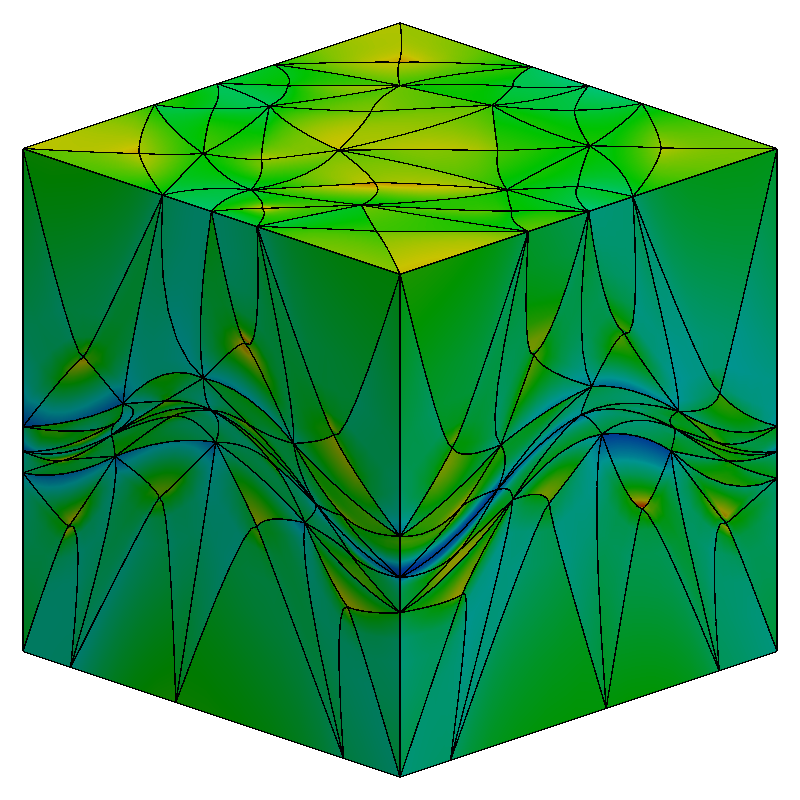}}
		&
		\subfloat[]{\label{fig:p4_1_3b}
			\includegraphics[width=0.3\textwidth]{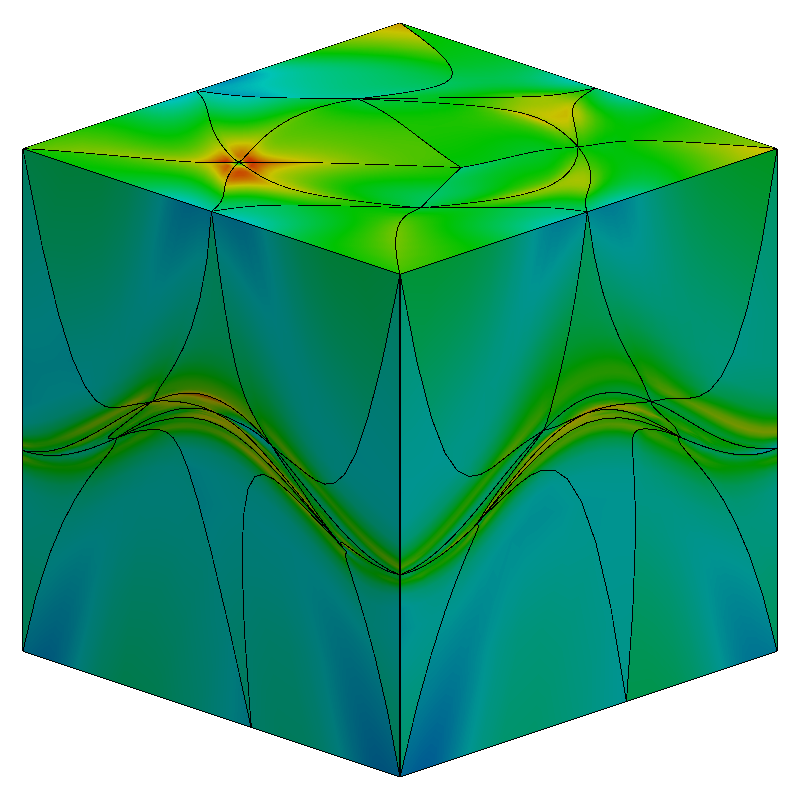}}
		\\
	\end{tabular}
	\\
	\includegraphics[width=0.35\textwidth]{/qualBarParaview_color} 
	\caption{Boundary of tetrahedral meshes of polynomial degree 1, 2, and 4 in columns. Initial straight-sided isotropic meshes and optimized meshes from initial meshes in rows.}
	\label{fig:3D}
\end{figure}
The initial meshes $\zmesh$ are of polynomial degree 1, 2, and 4.
The three meshes feature approximately the same number of nodes and they have approximately the same resolution over the domain.
In particular, in 2D the three initial meshes are respectively composed of 312, 321, and 337 nodes and 558, 144, and 38 triangles, see Figures \ref{fig:ex}\subref{fig:p1_0}, \ref{fig:ex}\subref{fig:p2_0}, and \ref{fig:ex}\subref{fig:p4_0}.
In 3D, they are respectively composed of $2\,356$, $2\,362$, and $2\,373$ nodes and $11\,699$, $1\,464$, and 184 tetrahedra.
Figures \ref{fig:ex3D}\subref{fig:p1_0_3}, \ref{fig:ex3D}\subref{fig:p2_0_3}, \ref{fig:ex3D}\subref{fig:p4_0_3}, and \ref{fig:3D}\subref{fig:p1_0_3b}, \ref{fig:3D}\subref{fig:p2_0_3b}, \ref{fig:3D}\subref{fig:p4_0_3b} show the clipped 3D meshes and the mesh boundary, respectively.
The meshes are colored according to the point-wise stretching and alignment quality measure, presented in Equation \eqref{eq:pointwisequality}.
Points in blue color have low quality and points with red color have high quality.
As we observe, the elements lying in the region of highest stretching ratio have less quality than the elements lying in the isotropic region.

We equip each mesh with the metric presented in Equation \eqref{eq:metricDeformation}.
We obtain the metric values from the log-Euclidean interpolation method presented in Section \ref{sec:interpolation}.
In particular, we interpolate the metrics from a background mesh $\zbmesh$.
The background meshes are of polynomial degree 1, 2, and 4 according to the polynomial degree of the initial meshes $\zmesh$. 
We impose the three background meshes to feature almost the same number of nodes and to have almost the same resolution over the domain, $h_{\min}/2$.
In particular, the resolution of the 2D background meshes is $h_{\min}/2 = 0.005$.
They are composed of $65\,170$, $64\,329$, and $62\,761$ nodes and $129\,318$, $31\,910$, and $7\,782$ triangles.
The resolution of the 3D background meshes is $h_{\min}/2 = 0.01$.
They are composed of $1\,773\,415$, $1\,798\,531$, and $1\,837\,851$ nodes and $10\,438\,221$, $1\,319\,008$, and $168\,441$ tetrahedra.

To obtain an optimal configuration $\zmesh^*$, we minimize the mesh distortion by relocating the mesh nodes while preserving their connectivity, as detailed in Section \ref{sec:preliminaries}.
The coordinates of the inner nodes, and the coordinates tangent to the boundary, are the design variables.
Thus, the inner nodes are free to move, the vertex nodes are fixed, while the rest of boundary nodes are enforced to slide along the boundary facets of the domain $\Omega$.
In Figures \ref{fig:ex}\subref{fig:p1_1}, \ref{fig:ex}\subref{fig:p2_1}, \ref{fig:ex}\subref{fig:p4_1} we illustrate the optimized 2D meshes.
In the 3D case, Figure \ref{fig:ex3D}\subref{fig:p1_1_3}, \ref{fig:ex3D}\subref{fig:p2_1_3}, \ref{fig:ex3D}\subref{fig:p4_1_3}, and \ref{fig:3D}\subref{fig:p1_1_3b}, \ref{fig:3D}\subref{fig:p2_1_3b}, \ref{fig:3D}\subref{fig:p4_1_3b} show the clipped 3D meshes and the mesh boundary, respectively.
We align the axes according to the ones of Figure \ref{fig:stretching}.
We observe that 
the elements lying in the anisotropic region are compressed to attain the stretching and alignment prescribed by the metric.

\begin{table*}[t!]
	\small
	\caption{Quality statistics for the initial and optimized meshes with interpolated 2D metric.}
	\label{table:interpolative}
	\centering
	\par\medskip
	\begin{tabular}{ c c c c c c c c c}
		\hline\noalign{\smallskip}
		Mesh & \multicolumn{2}{c}{Minimum}&\multicolumn{2}{c}{Maximum}&\multicolumn{2}{c}{Mean}&\multicolumn{2}{c}{Std dev.}\\
		deg. & Initial & Final & Initial & Final & Initial & Final & Initial & Final\\
		\noalign{\smallskip}\hline\noalign{\smallskip}
		1 & 0.0299 & 0.1724 & 0.9957 & 0.9551 & 0.6100 & 0.4462 & 0.2769 & 0.1039 \\
		2 & 0.0554 & 0.2878 & 0.9921 & 0.6268 & 0.5918 & 0.4545 & 0.2835 & 0.0638 \\
		4 & 0.0803 & 0.3072 & 0.9835 & 0.5806 & 0.5339 & 0.4439 & 0.2922 & 0.0760 \\
		\noalign{\smallskip}\hline\noalign{\smallskip}
	\end{tabular}
\end{table*}
\begin{table*}[t!]
	\small
	\caption{Quality statistics for the initial and optimized meshes with interpolated 3D metric.}
	\label{table:interpolative3D}
	\centering
	\par\medskip
	\begin{tabular}{ c c c c c c c c c}
		\hline\noalign{\smallskip}
		Mesh & \multicolumn{2}{c}{Minimum}&\multicolumn{2}{c}{Maximum}&\multicolumn{2}{c}{Mean}&\multicolumn{2}{c}{Std dev.}\\
		deg. & Initial & Final & Initial & Final & Initial & Final & Initial & Final\\
		\noalign{\smallskip}\hline\noalign{\smallskip}
		1 & 0.0175 & 0.1222 & 0.9905 & 0.9334 & 0.5550 & 0.4236 & 0.2660 & 0.1241 \\
		2 & 0.0320 & 0.2987 & 0.9695 & 0.7467 & 0.5194 & 0.4576 & 0.2735 & 0.0691\\
		4 & 0.0409 & 0.3231 & 0.8931 & 0.6737 &	0.4490 & 0.4702 & 0.2711 & 0.0749 \\
		\noalign{\smallskip}\hline\noalign{\smallskip}
	\end{tabular}
\end{table*}
Tables \ref{table:interpolative} and \ref{table:interpolative3D} show the quality statistics of both the initial and optimized meshes for the 2D and 3D cases, respectively.
In all the optimized meshes the minimum is improved and the standard deviation of the element qualities is reduced when compared with the initial configuration.
In addition, when comparing the curved meshes with the straight-edged ones, we observe that the curved meshes are more flexible.
That is, the curved meshes achieve a higher improvement of the minimum quality and the standard deviation.
This is because the curved elements can approximate the curved stretching of the metric in the point-wise sense and hence, more accurately.

\subsection{Validation: analytic versus discrete}\label{sec:avsd}
To validate the proposed method, we compare 2D curved $r$-adaption results for the high-order metric interpolation with the results corresponding to an analytic metric evaluation.
Considering the initial meshes presented in the previous section, we optimize the distortion measure by evaluating the analytical metric expression, instead of interpolating it in the background mesh.
In Figure \ref{fig:ex} we show the initial and optimized meshes.
They are colored according to the point-wise quality measure of Equation \eqref{eq:pointwisequality} using the analytical metric expression.

\begin{table*}[t!]
	\small
	\caption{Quality statistics for the initial and optimized meshes with analytic 2D metric.}
	\label{table:analytic}
	\centering
	\par\medskip
	\begin{tabular}{ c c c c c c c c c}
		\hline\noalign{\smallskip}
		Mesh & \multicolumn{2}{c}{Minimum}&\multicolumn{2}{c}{Maximum}&\multicolumn{2}{c}{Mean}&\multicolumn{2}{c}{Std dev.}\\
		deg. & Initial & Final & Initial & Final & Initial & Final & Initial & Final\\
		\noalign{\smallskip}\hline\noalign{\smallskip}
		1 & 0.0279 & 0.1684 & 0.9957 & 0.9581 & 0.6100 & 0.4484 & 0.2770 & 0.1088 \\
		2 & 0.0563 & 0.3358 & 0.9921 & 0.6432 & 0.5919 & 0.4569 & 0.2835 & 0.0623 \\
		4 & 0.0799 & 0.3096 & 0.9835 & 0.6318 & 0.5339 & 0.4473 & 0.2923 & 0.0634 \\
		\noalign{\smallskip}\hline\noalign{\smallskip}
	\end{tabular}
\end{table*}
To compare quantitatively both results, we compute the maximum distance of the node coordinates of the optimized configurations.
The maximum distances are around $2.2\cdot10^{-2}$, $7.6\cdot10^{-2}$, and $8.2\cdot10^{-2}$ for the linear, quadratic, and quartic cases, obtaining comparable nodal configurations, as it can be observed when comparing Figures \ref{fig:ex}\subref{fig:p1_1}, \ref{fig:ex}\subref{fig:p2_1}, and \ref{fig:ex}\subref{fig:p4_1} with Figures \ref{fig:ex}\subref{fig:p1_1_a}, \ref{fig:ex}\subref{fig:p2_1_a}, and \ref{fig:ex}\subref{fig:p4_1_a}, respectively.
In Table \ref{table:analytic}, we present the quality statistics of the initial and optimized meshes using the analytical metric evaluation.
To compare the quality improvement of both approaches, we compute the difference between the mean of the analyzed quality statistics, obtaining a value below $10^{-2}$.
Thus, the quality improvement driven by the optimization using the proposed metric interpolation procedure is analogous to the one given by the analytical metric, obtaining in all cases high-quality configurations with a minimum quality over 0.1.





\subsection{Distortion minimization: initial anisotropic straight-edged meshes}
\label{sec:iniAnisotropic}

The results presented in Section \ref{sec:meshes} show the application of the metric interpolation procedure to optimize isotropic meshes in a domain equipped with a metric.
However, in practice, anisotropic meshes are generated combining topological mesh operations that modify the mesh connectivity and mesh $r$-adaption procedures \cite{alauzet:AnisotropicMeshAdaptation}.
To illustrate a practical example, we consider an initial anisotropic straight-sided mesh.
Then, we apply the anisotropic $r$-adaption method presented in this work.

Although we generate meshes adapted to a target metric with MMG \cite{dobrzynski2012mmg3d}, our goal is not to compare the distortion minimization with the MMG package. Actually, we acknowledge MMG because it generates an initial straight-edged mesh that matches the stretching and alignment of the target metric.

\begin{figure}[t!]
	\centering
	\setlength{\tabcolsep}{10pt}
	\renewcommand{\arraystretch}{1.5}
	\begin{tabular}{ccc}
		\subfloat[]{\label{fig:p1_0mmg}
			\includegraphics[width=0.25\textwidth]{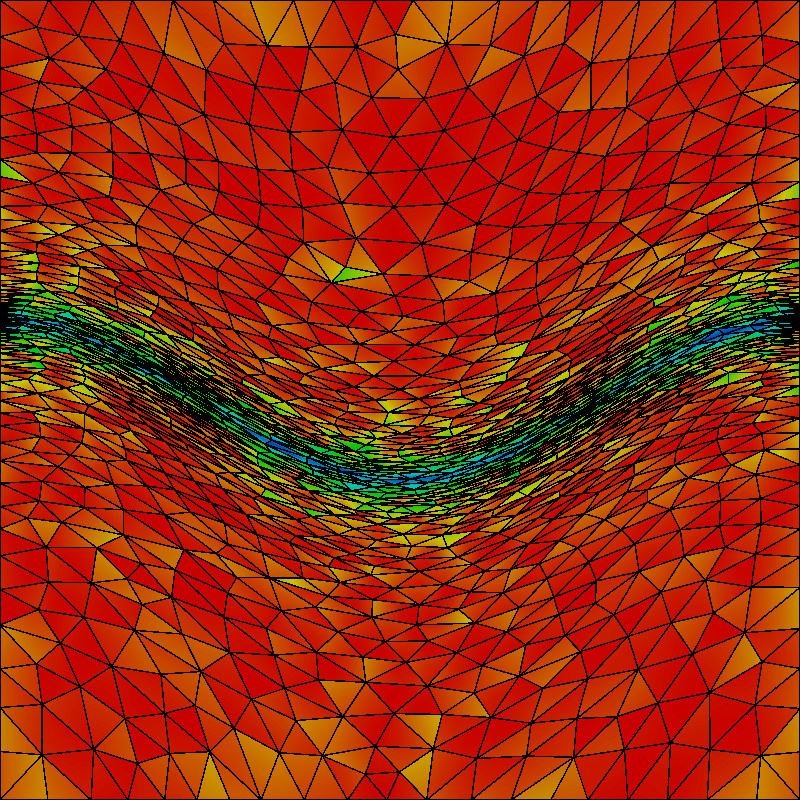}}
		&
		\subfloat[]{\label{fig:p2_0mmg}
			\includegraphics[width=0.25\textwidth]{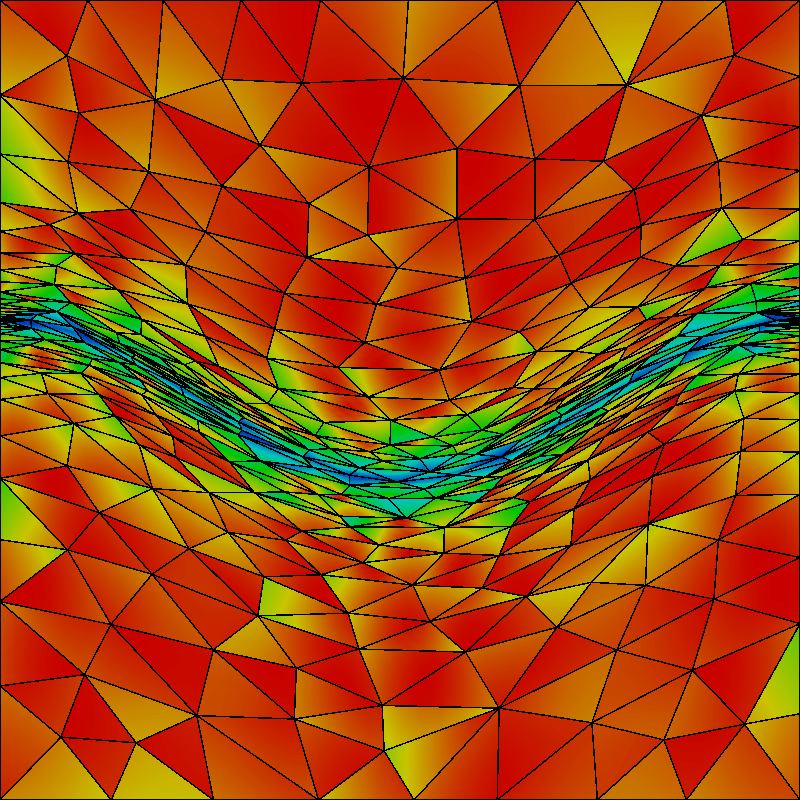}}
		&
		\subfloat[]{\label{fig:p4_0mmg}
			\includegraphics[width=0.25\textwidth]{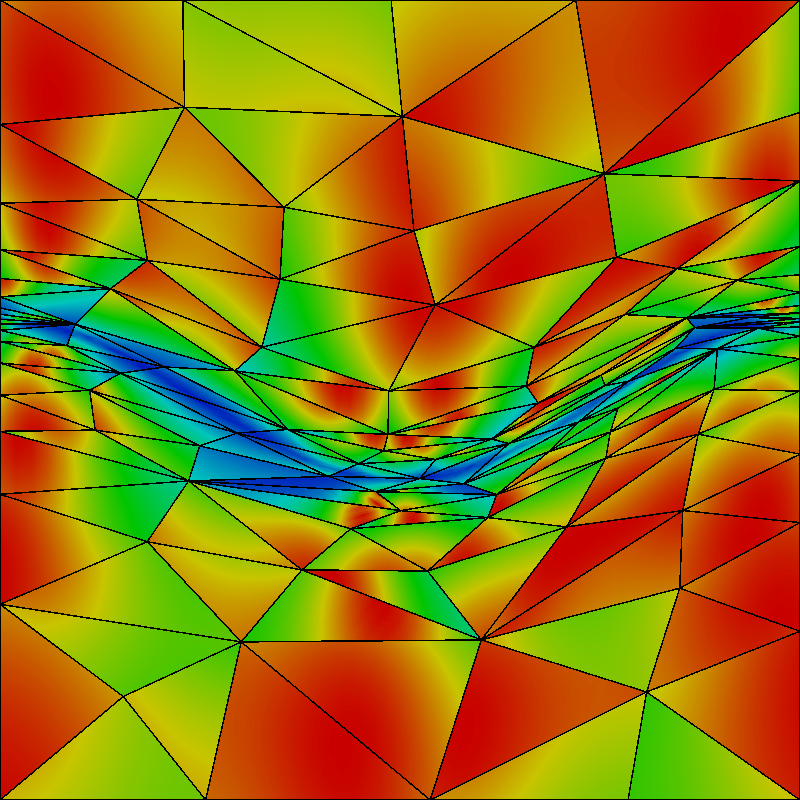}}
		\\
		\subfloat[]{\label{fig:p1_1mmg}
			\includegraphics[width=0.25\textwidth]{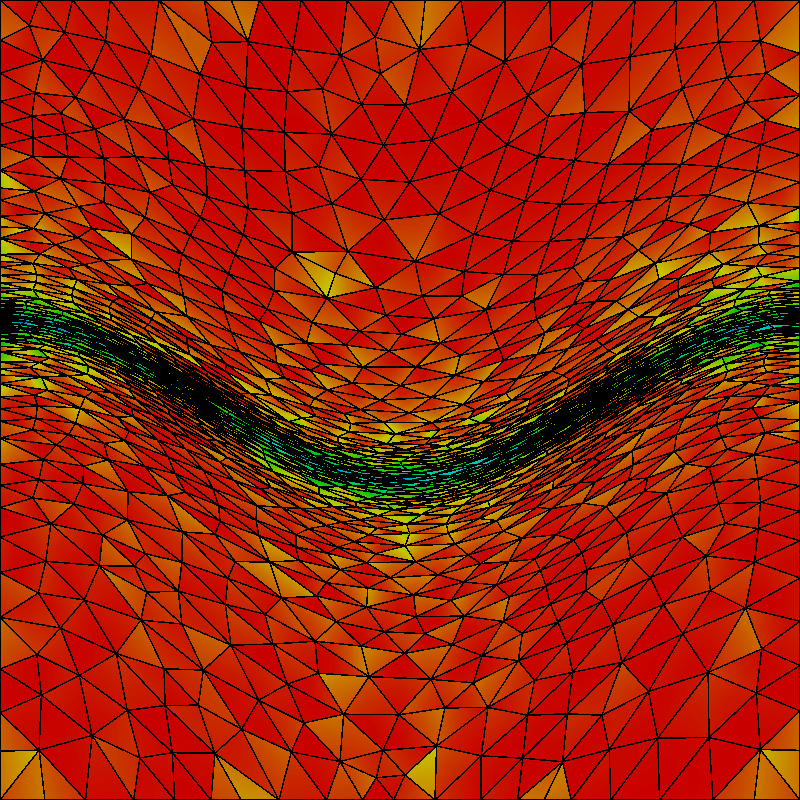}}
		&
		\subfloat[]{\label{fig:p2_1mmg}
			\includegraphics[width=0.25\textwidth]{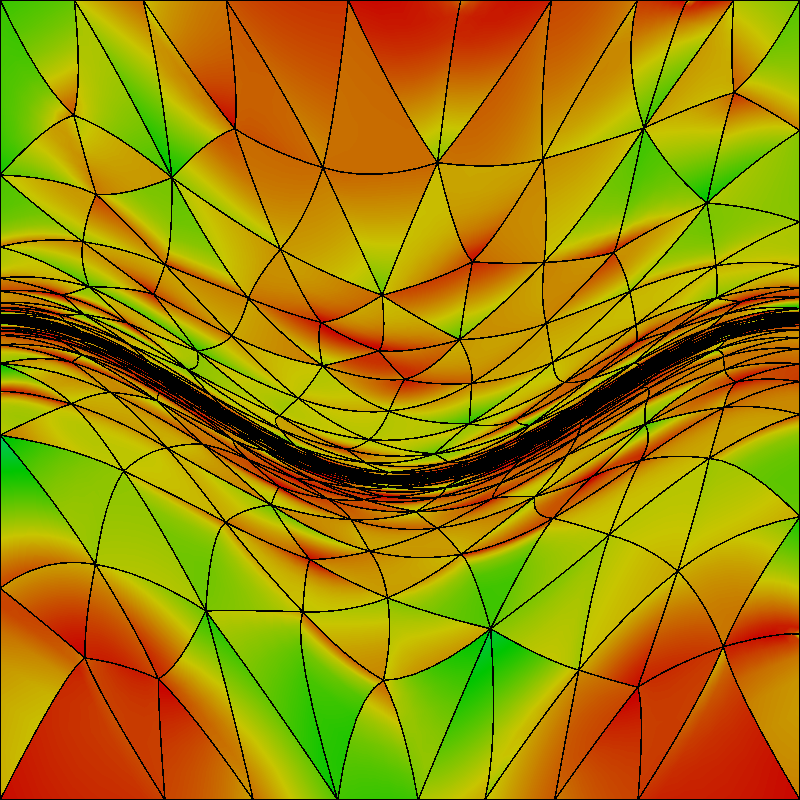}}
		&
		\subfloat[]{\label{fig:p4_1mmg}
			\includegraphics[width=0.25\textwidth]{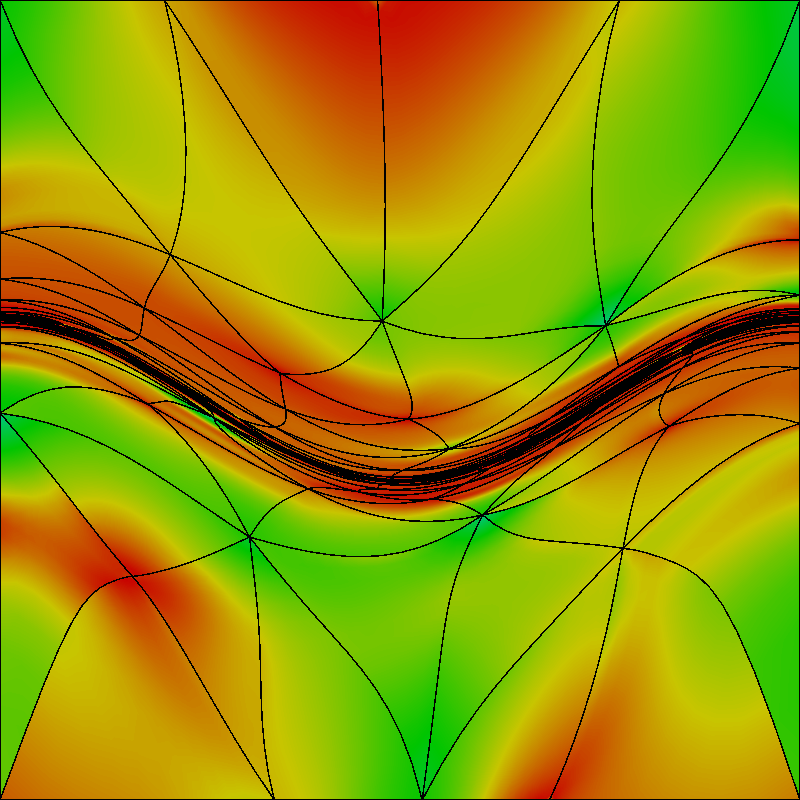}}
		\\
	\end{tabular}
	\\
	\includegraphics[width=0.35\textwidth]{/qualBarParaview_color} 
	\caption{Point-wise distortion for triangular meshes of polynomial degree 1, 2, and 4 in columns. Initial straight-sided anisotropic meshes and optimized meshes from initial meshes in rows.}
	\label{fig:exmmg}
\end{figure}
First, we consider the target metric presented in Equation \eqref{eq:metricDeformation} with $h_{\min} = 0.01$.
Second, we generate a linear isotropic triangular background mesh $\hat{\zmesh}$ of input size $h_{\min}/2 = 0.005$ with MATLAB.
We normalize the target metric according to the size of the physical meshes $\zmesh$ namely, $0.0625,\ 0.125, \text{ and } 0.25$ for the linear, quadratic, and quartic case, respectively.
These sizes are chosen in order to obtain a comparable mesh resolution according to the mesh polynomial degree.
Then, we couple each background mesh with the target metric evaluated at the background mesh vertices.
We apply the MMG algorithm to obtain an initial straight-sided anisotropic physical mesh $\zmesh$ of polynomial degree 1, 2, and 4,
see Figures \ref{fig:exmmg}\subref{fig:p1_0mmg}, \ref{fig:exmmg}\subref{fig:p2_0mmg}, and \ref{fig:exmmg}\subref{fig:p4_0mmg}.
In particular, the physical meshes are composed by $1\,161$ nodes and $2\,137$ triangles, $1\,333$ nodes and 624 triangles and, $1\,525$ nodes and 180 triangles, respectively.

The physical meshes $\zmesh$ are then optimized using the metric interpolation approach presented in this work.
In Figures \ref{fig:exmmg}\subref{fig:p1_1mmg}, \ref{fig:exmmg}\subref{fig:p2_1mmg}, and \ref{fig:exmmg}\subref{fig:p4_1mmg}, we illustrate the optimized meshes $\zmesh^*$.
We observe that the elements lying in the anisotropic region are compressed to attain the stretching and alignment prescribed by the metric.

\begin{table*}[t!]
	\small
	\caption{Quality statistics for the initial MMG and optimized meshes with interpolated 2D metric.}
	\label{table:interpolativemmg}
	\centering
	\par\medskip
	\begin{tabular}{ c c c c c c c c c}
		\hline\noalign{\smallskip}
		Mesh & \multicolumn{2}{c}{Minimum}&\multicolumn{2}{c}{Maximum}&\multicolumn{2}{c}{Mean}&\multicolumn{2}{c}{Std dev.}\\
		deg. & Initial & Final & Initial & Final & Initial & Final & Initial & Final\\
		\noalign{\smallskip}\hline\noalign{\smallskip}
		1 & 0.0365 & 0.1794 & 0.9988 & 0.9989 & 0.7806 & 0.7961 & 0.2273 & 0.2040 \\
		2 & 0.0624 & 0.6300 & 0.9982 & 0.9913 & 0.6966 & 0.8692 & 0.2558 & 0.0788 \\
		4 & 0.0424 & 0.6063 & 0.9774 & 0.9965 & 0.5677 & 0.9137 & 0.2681 & 0.0886 \\
		\noalign{\smallskip}\hline\noalign{\smallskip}
	\end{tabular}
\end{table*}
In Table \ref{table:interpolativemmg}, we show the quality statistics of both the initial and optimized meshes.
In all the optimized meshes the minimum is improved and the standard deviation of the element qualities is reduced when compared with the initial configuration.
We conclude that, with the same metric data and hence, the same inputs, the $r$-adaption mesh post-processing improves the quality of the meshes generated with the MMG algorithm.
In addition, for the straight-edged case, we have presented a global method to improve the stretching and alignment prescribed by the metric after applying an $h$-adaption approach.

For a fixed metric, usually the better the initial straight-edged mesh is, the better the optimized mesh is.
For instance, for different degrees, the mean quality statistics for the initial anisotropic meshes, Table \ref{table:interpolativemmg}, are better than for the isotropic meshes, Table \ref{table:interpolative}.
The anisotropic meshes have this advantage because their topology and geometry are adapted to match the corresponding scaling of the target metric.
This prior metric matching facilitates that the curved optimization reaches a better final quality.

As in the examples presented in Section \ref{sec:meshes}, when comparing the curved meshes with the straight-edged ones, we observe that the curved meshes are more flexible.
That is, the curved meshes achieve a higher improvement of the minimum quality and the standard deviation.
This is because the curved elements can approximate the curved stretching of the metric in the point-wise sense and hence, more accurately.

\subsection{Distortion minimization: curved boundaries}\label{sec:curvedboundaries}
We following illustrate that our approach is compatible with curved boundaries.
We consider a 2D example, in Section \ref{sec:curvedboundaries2D}, and a 3D example, in Section \ref{sec:curvedboundaries3D}.
To this end, we first construct the geometric model with FreeCAD \cite{riegel2016freecad}.
Next, we consider their implicit representation, see Section \ref{sec:implicit}.
Then, we generate the background and initial physical meshes coupled with a discrete metric, see Section \ref{sec:interpolation}.
Finally, we apply our $r$-adaption method, presented in Section \ref{sec:preliminaries}, by taking into account both the discrete metric and the implicit representation of the geometry.
This enables an optimized physical mesh that approximates the stretching and alignment of the metric while preserving the curvature of the boundary.

\begin{figure}[t!]
	\centering
	\begin{tabular}{cc}
		\subfloat[]{\label{fig:circleparam}
			\includegraphics[width=0.3\textwidth]{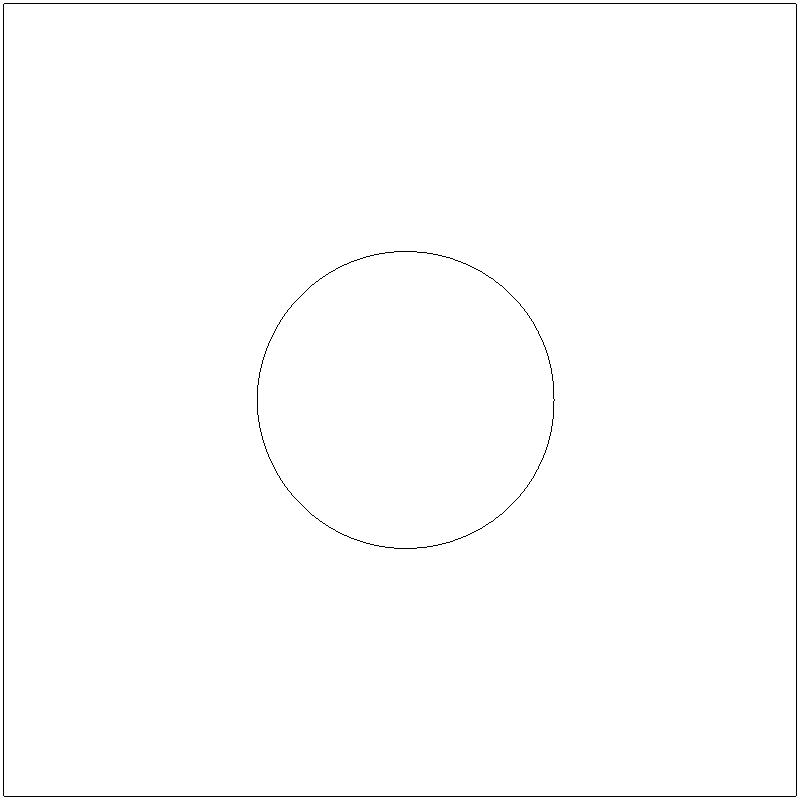}}
		&
		\subfloat[]{\label{fig:circle}
			\includegraphics[width=0.3\textwidth]{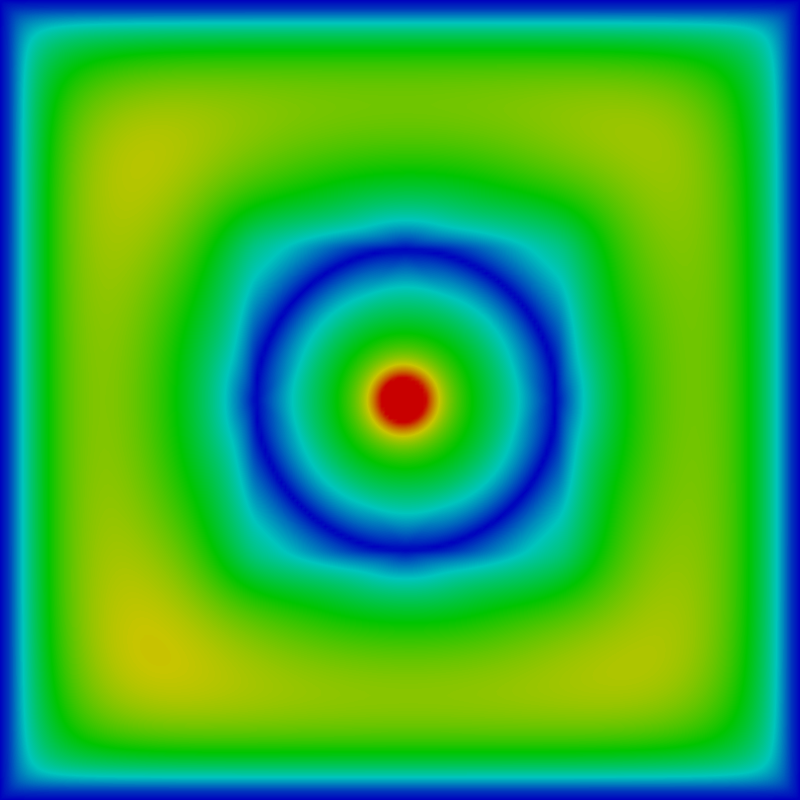}}
		\\
	\end{tabular}
	\caption{Parametric CAD and global implicit representation for the 2D model of a square with a circular hole.}
	\label{fig:2Dmodels}
\end{figure}
\begin{figure}[p!]
	\centering
	\vskip -50pt
	\setlength{\tabcolsep}{20pt}
	\renewcommand{\arraystretch}{1.5}
	\begin{tabular}{cc}
		\subfloat[]{\label{fig:p2_0mmg2}
			\includegraphics[width=0.35\textwidth]{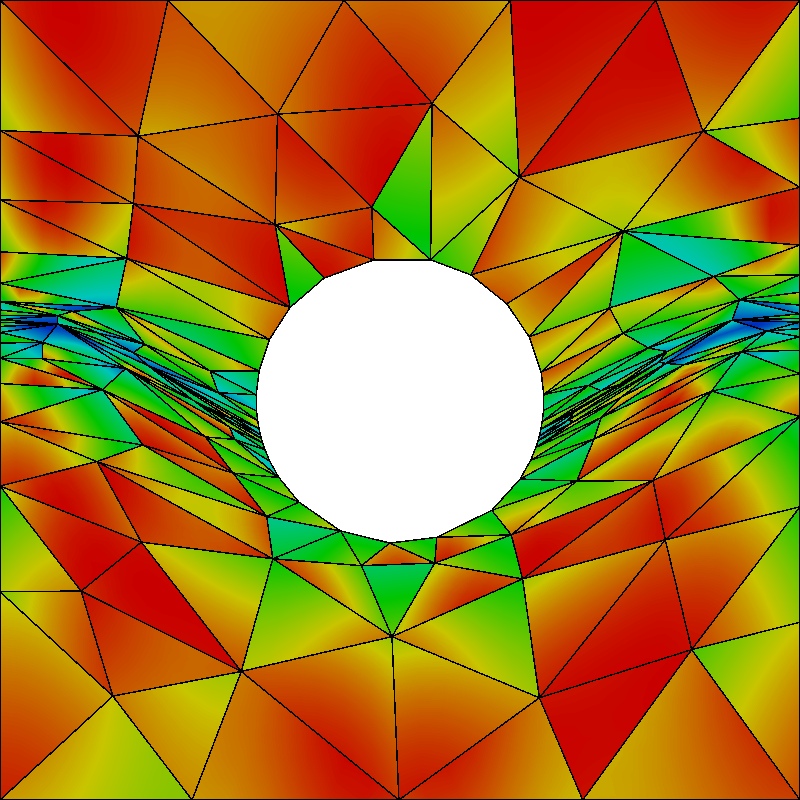}}
		&
		\subfloat[]{\label{fig:p2_1mmg2}
			\includegraphics[width=0.35\textwidth]{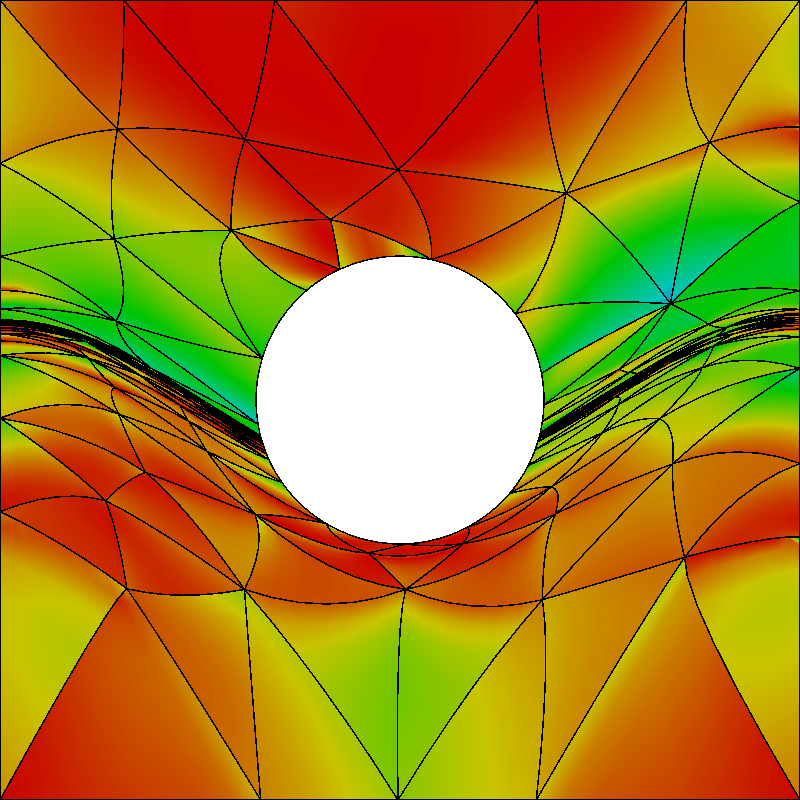}}
		\\
		\subfloat[]{\label{fig:p2_0mmg2zoom}
			\includegraphics[width=0.35\textwidth]{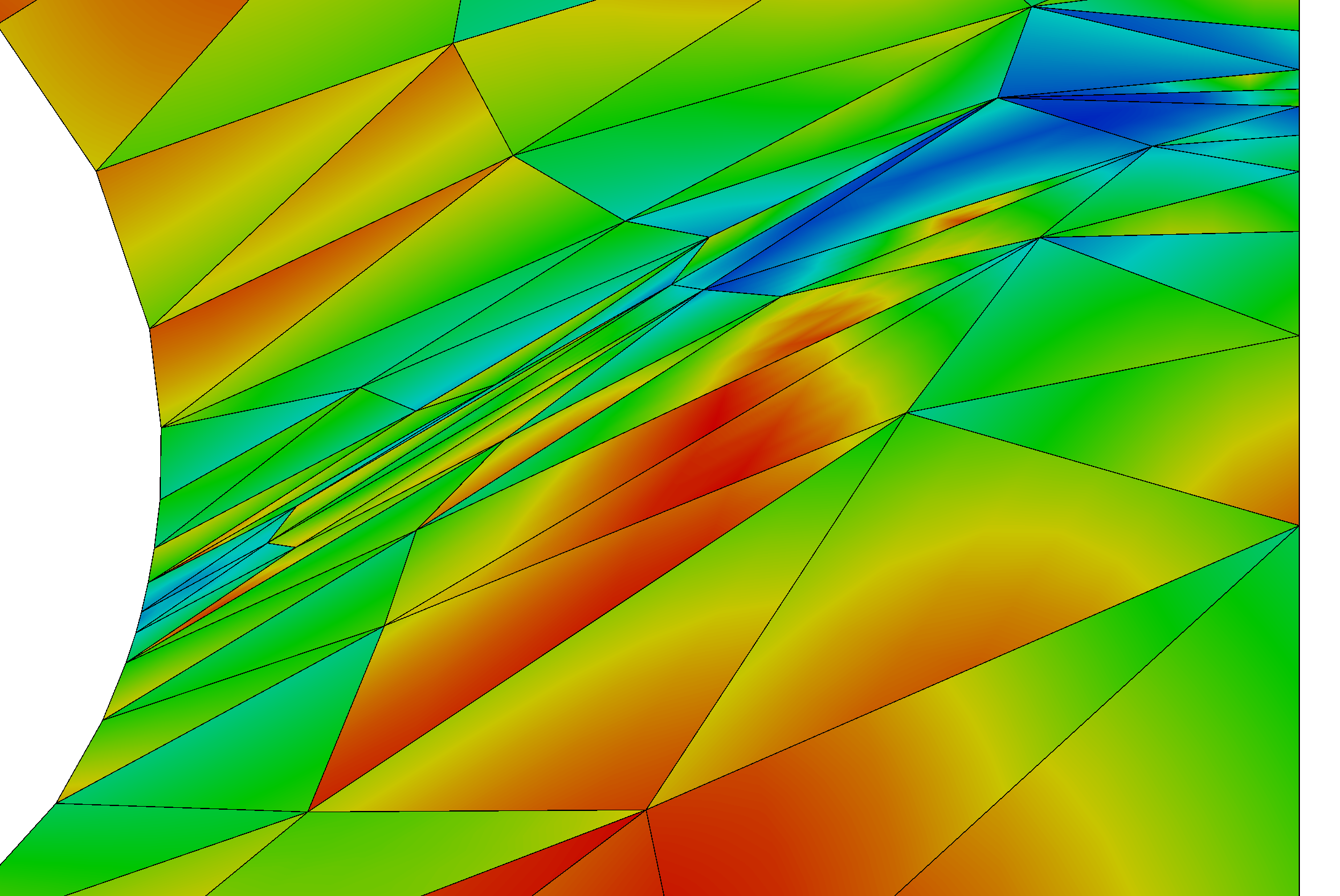}}
		&
		\subfloat[]{\label{fig:p2_1mmg2zoom}
			\includegraphics[width=0.35\textwidth]{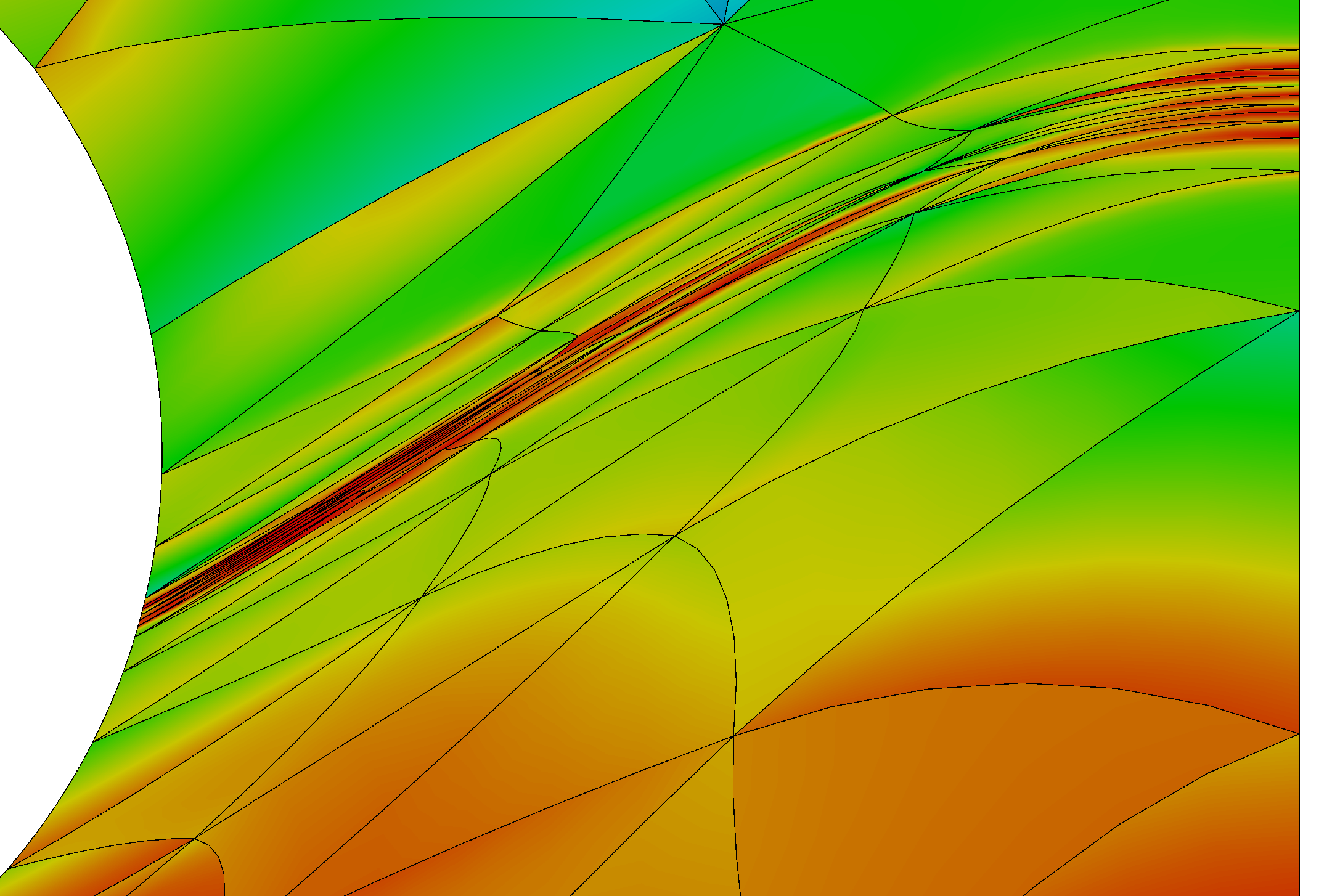}}
		\\
		\subfloat[]{\label{fig:p4_0mmg2}
			\includegraphics[width=0.35\textwidth]{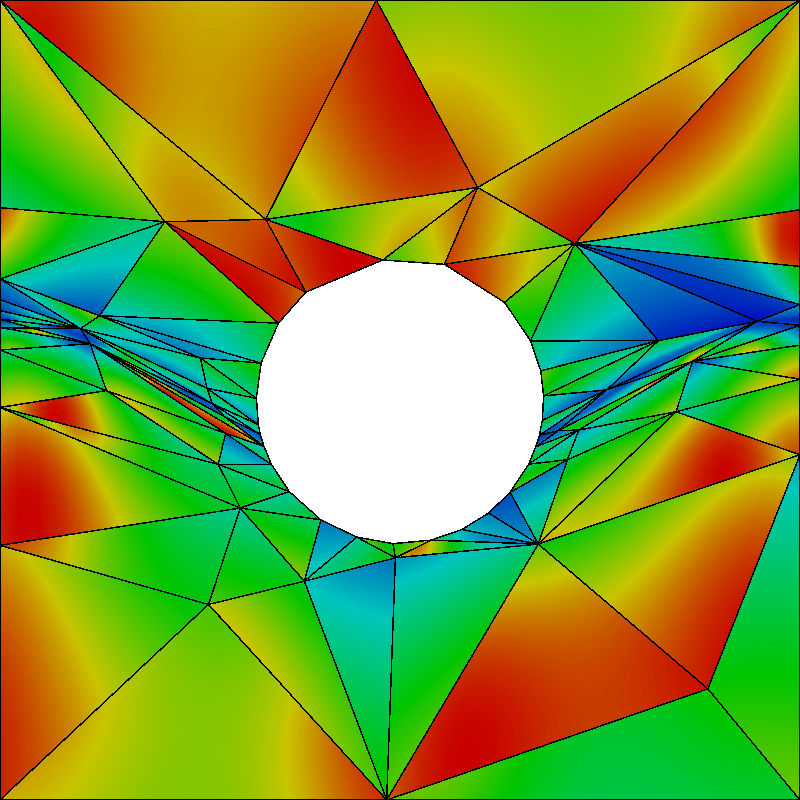}}
		&
		\subfloat[]{\label{fig:p4_1mmg2}
			\includegraphics[width=0.35\textwidth]{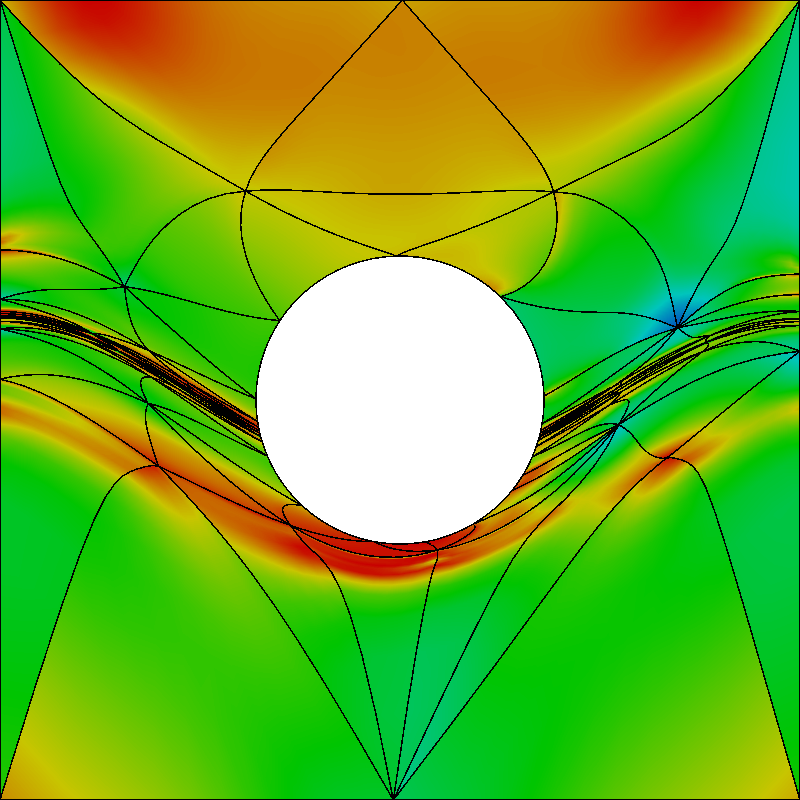}}
		\\
		\subfloat[]{\label{fig:p4_0mmg2zoom}
			\includegraphics[width=0.35\textwidth]{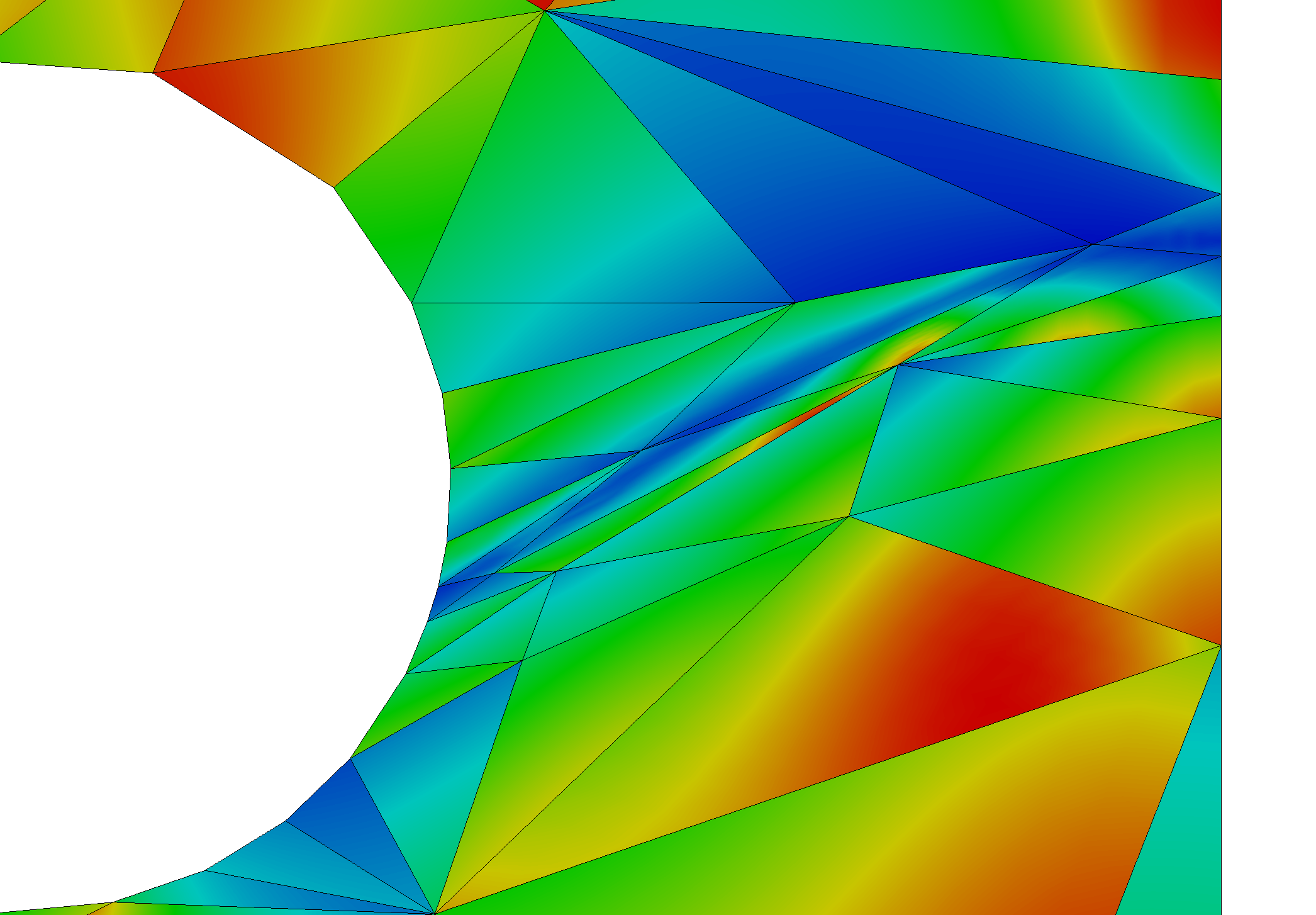}}
		&
		\subfloat[]{\label{fig:p4_1mmg2zoom}
			\includegraphics[width=0.35\textwidth]{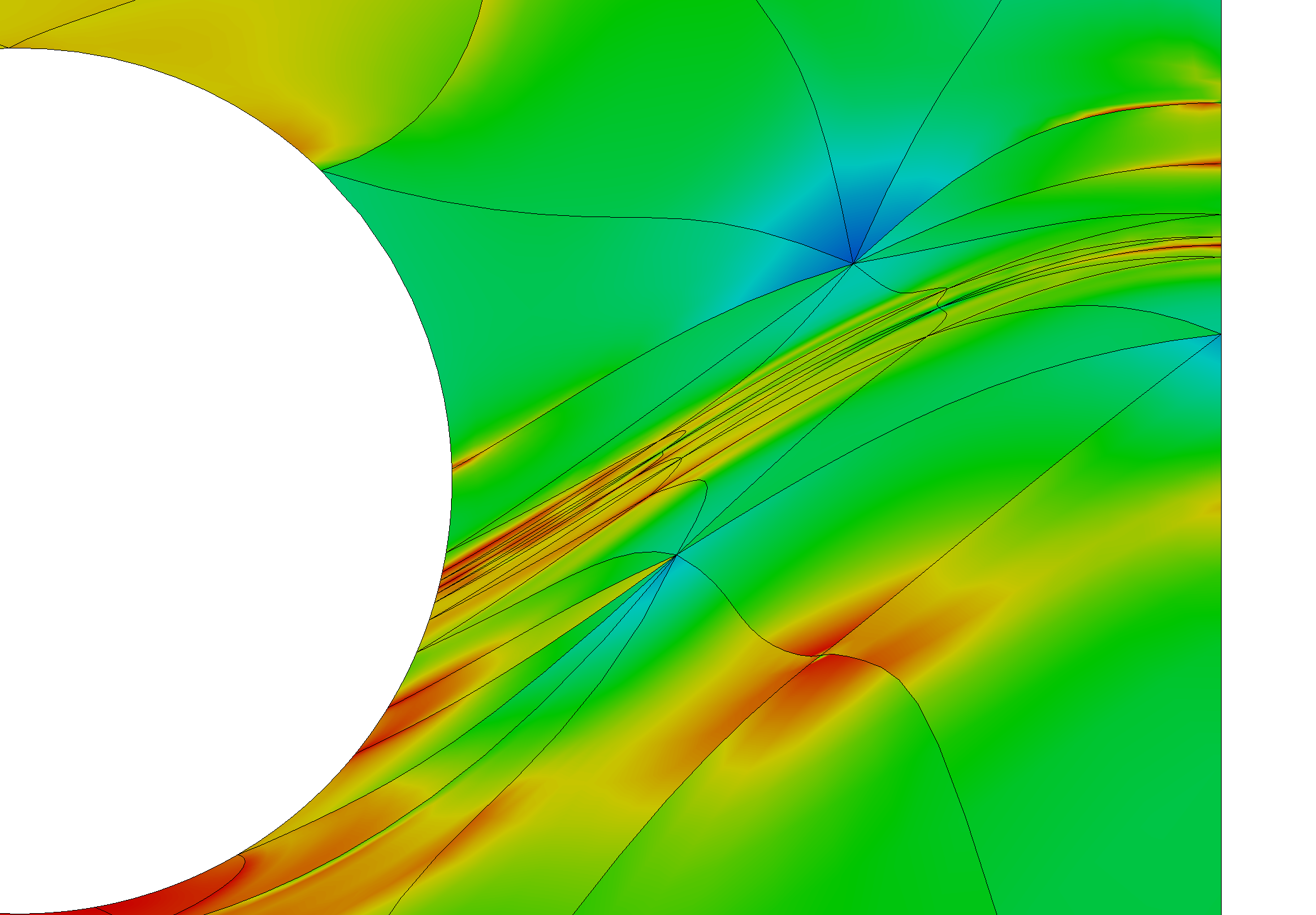}}
	\end{tabular}
	\\
	\includegraphics[width=0.35\textwidth]{/qualBarParaview_color} 
	\caption{Point-wise distortion for triangular meshes of polynomial degree 2 in first and second (zoom) rows, and 4 in third and fourth (zoom) rows. Initial straight-sided anisotropic mesh and optimized mesh in columns.}
	\label{fig:exmmg2}
\end{figure}
To accommodate the curved boundaries we include, to the presented functional, a boundary term that takes into account the mesh deviation to the boundaries of the domain, see Section \ref{sec:penalty}.
Specifically, we set the penalty parameter $\lambda := 10^4$ in all examples, see Equation \eqref{eq:curving}.
In addition, to approximate the metric stretching, we optimize the mesh using the metric interpolation approach presented in this work.
Finally, when optimizing the mesh functional all mesh nodes coordinates are free that is, each mesh node moves in $\zR^2$, in the 2D case, and in $\zR^3$, in the 3D case.
\subsubsection{2D curved model: square with a circular hole}\label{sec:curvedboundaries2D}
For the 2D model $\zmodel_1$, we consider a square with a circular hole.
Specifically, the domain is denoted by $\Omega_1 = K_1 \backslash C_1$, where $K_1 = [-0.5,0.5]^2$ is a square, and where $C_1$ is the circle with radius equal to $0.18$ and centered at the origin, see Figure \ref{fig:2Dmodels}\subref{fig:circleparam}.
The domain $\Omega_1$ has two boundaries, the one of the square $K_1$ and the one of the circle $C_1$.
We illustrate in Figure \ref{fig:2Dmodels}\subref{fig:circle} a global implicit representation of the boundary $\zmodel_1:=\partial\Omega_1$, using the method presented in Section \ref{sec:cadimplicit}.
Although the inner boundary is smooth, the outer boundary contains sharp features such as corners.

We equip the domain $\Omega_1$ with the target metric presented in Equation \eqref{eq:metricDeformation} with $h_{\min} = 0.01$.
Then, we generate with MATLAB two isotropic triangular background meshes $\hat{\zmesh}$ of polynomial degree 2 and 4.
They have an input resolution $h_{\min}/2 = 0.005$ over $\Omega_1$ that is, of input size $0.01$ and $0.02$, respectively.
We normalize the target metric according to size $h = 0.25$ in the quadratic case, and according to size $h = 0.5$ in the quartic case.
Then, we couple each background mesh with the target metric evaluated at the background mesh vertices.
From each background mesh $\hat{\zmesh}$, we obtain an initial straight-sided anisotropic physical mesh $\zmesh$ by applying the MMG algorithm, see Figures \ref{fig:exmmg2}\subref{fig:p2_0mmg2}, and \ref{fig:exmmg2}\subref{fig:p4_0mmg2}.
The quadratic and quartic physical meshes are respectively composed by 518 nodes and 220 triangles, and 944 nodes and 106 triangles.
Note that, since the MMG algorithm requires a linear background mesh, we subdivide the background meshes in order to preserve their resolution.
Specifically, our linear background meshes for the MMG algorithm are obtained by subdividing the quadratic background mesh once, and the quartic background mesh twice.

\begin{table*}[t!]
	\small
	\caption{Quality statistics for the initial MMG and optimized mesh with interpolated 2D metric at the square with a circular hole.}
	\label{table:interpolativemmg2}
	\centering
	\par\medskip
	\begin{tabular}{ c c c c c c c c c}
		\hline\noalign{\smallskip}
		Mesh & \multicolumn{2}{c}{Minimum}&\multicolumn{2}{c}{Maximum}&\multicolumn{2}{c}{Mean}&\multicolumn{2}{c}{Std dev.}\\
		deg. & Initial & Final & Initial & Final & Initial & Final & Initial & Final\\
		\noalign{\smallskip}\hline\noalign{\smallskip}
		2 & 0.0823 & 0.4140 & 0.9914 & 0.9943 & 0.5764 & 0.8224 & 0.2508 & 0.1281 \\
		4 & 0.0590 & 0.4045 & 0.9646 & 0.9850 & 0.4177 & 0.7321 & 0.2292 & 0.1461 \\
		\noalign{\smallskip}\hline\noalign{\smallskip}
	\end{tabular}
\end{table*}
In Figures \ref{fig:exmmg2}\subref{fig:p2_1mmg2}, and \ref{fig:exmmg2}\subref{fig:p4_1mmg2}, we illustrate the optimized meshes $\zmesh^*$.
We observe that the elements lying in the anisotropic region are compressed to attain the stretching and alignment prescribed by the metric.
Note that the boundary elements are curved to match both the metric and the curved domain boundaries.
In Table \ref{table:interpolativemmg2}, we show the quality statistics of both the initial and optimized mesh.
In the optimized mesh the minimum, the mean, and the standard deviation of the element qualities are improved when compared with the initial configuration.
From the results, we observe that, when compared with straight-sided elements, curved elements approximate more faithfully the metric while preserving the curved features of the boundary.
In this case, the stretching direction is almost aligned according to the tangent of the geometry.
When considering straight-edged elements, in Figures \ref{fig:exmmg2}\subref{fig:p2_0mmg2zoom} and \ref{fig:exmmg2}\subref{fig:p4_0mmg2zoom}, accumulating more degrees of freedom in the stretched regions may worsen the boundary representation at non-stretched regions.
In contrast, when considering curved elements, in Figures \ref{fig:exmmg2}\subref{fig:p2_1mmg2zoom} and \ref{fig:exmmg2}\subref{fig:p4_1mmg2zoom}, we observe that a single curved element represents the boundary more faithfully than several straight-sided elements.
This flexibility of curved elements allows the degrees of freedom to slide and accumulate, from non-stretched regions to the stretched regions, featuring high-quality elements.
For that reason, we observe how the elements are stretched, aligned, and curved according to the stretching and alignment of the metric.
Hence, curved elements allow an improved representation of the metric while preserving the curved features of the boundary.

We use a non-optimized prototype to demonstrate that the detailed derivatives enable Newton's method.
Nevertheless, to illustrate the computational cost, we next report the wall-clock time and the most expensive parts when matching a target metric and curved boundary.
The report is an initial reference for future improvements because the prototype is unoptimized.

For this two-dimensional example, the total wall-clock time is $2\ 194$ seconds for degree two and $17\ 911$ seconds for degree four. The wall-clock time is higher for the second case because of two main reasons: the number of mesh points and the polynomial degree.

First, the mesh features more points for degree four (944 points) than for degree two (518 points). Note that both cases are initialized with a straight-edged mesh adapted to the corresponding scaling of the metric. This scaling accounts for the difference of points between an element of degree two and an element of degree four. Unfortunately, the resulting adapted straight-edged mesh features 220 and 106 elements for degrees two and four, respectively. Thus, the initial meshes do not feature a comparable number of points, a difference that computationally benefits the example of degree two.

Second, the higher the order, the higher the computational cost is. For higher orders, the Hessians of the objective function densify, and the initial approximations worsen. Regarding density, note that the elemental contributions to the Hessian have around six times more non-zero entries for degree four than for degree two. In this example, computing each elemental contribution to the Hessian needs 0.15 seconds for degree four and 0.03 seconds for degree two. Regarding initial approximations, they are worse because the initial straight-edged mesh is of degree one, and thus, the difference of degrees is higher for degree four. In this example, the non-linear problem needs 693 iterations for degree four and 229 iterations for degree two.

Finally, for both degrees, the most expensive part is to compute the elemental contributions to the gradient and the Hessian, a computation that needs the derivatives of the metric interpolation and the geometry implicitation. For the metric interpolation, the percentage of the total wall-clock time computing the derivatives is 45

\subsubsection{3D curved model: a cube trimmed by a cylinder}\label{sec:curvedboundaries3D}
\begin{figure}[t!]
	\centering
	\setlength{\tabcolsep}{-40pt}
	\begin{tabular}{cc}
		\subfloat[]{\label{fig:cylinderparam}
			\includegraphics[width=0.6\textwidth]{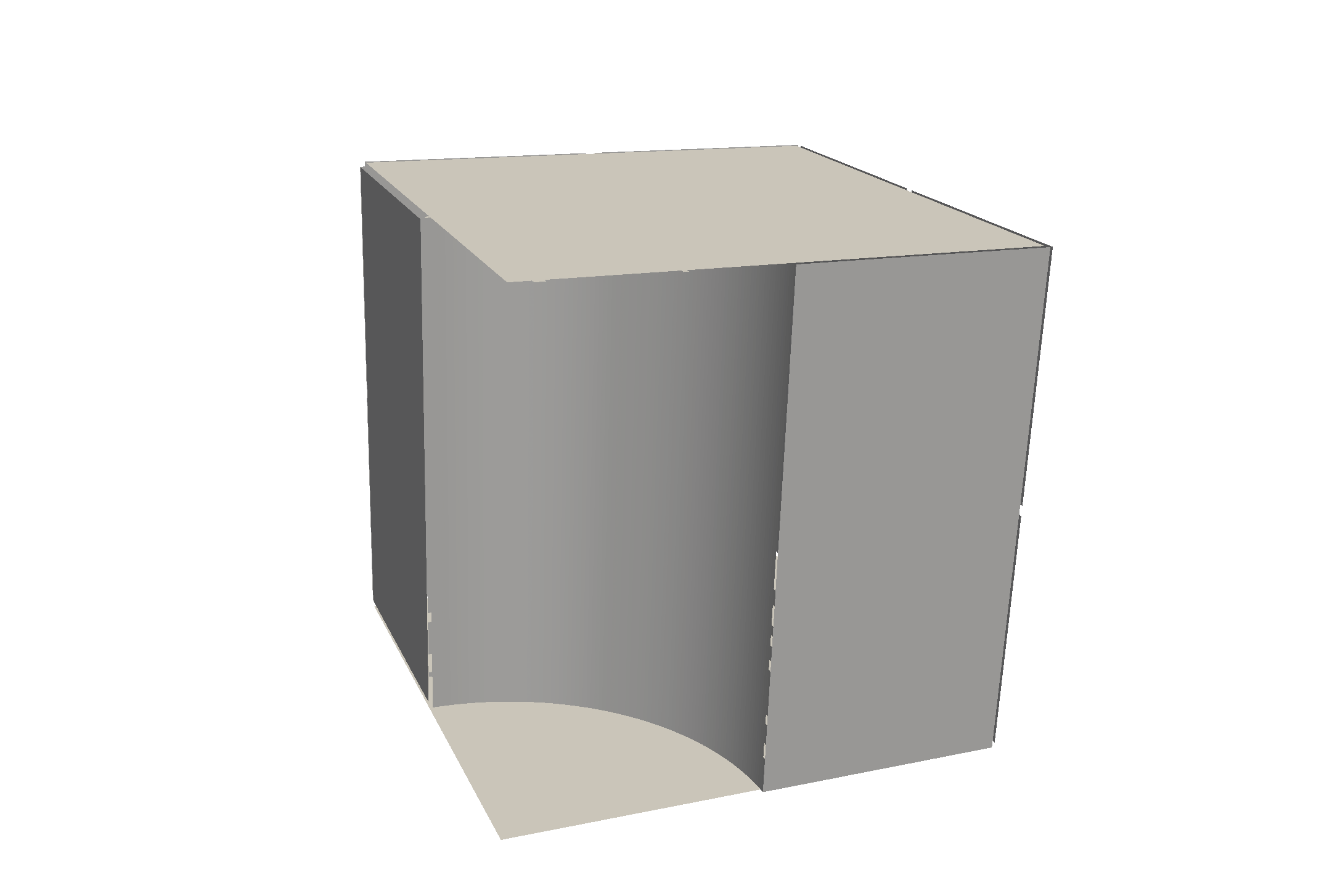}}
		&
		\subfloat[]{\label{fig:cylinderimplicit}
			\includegraphics[width=0.6\textwidth]{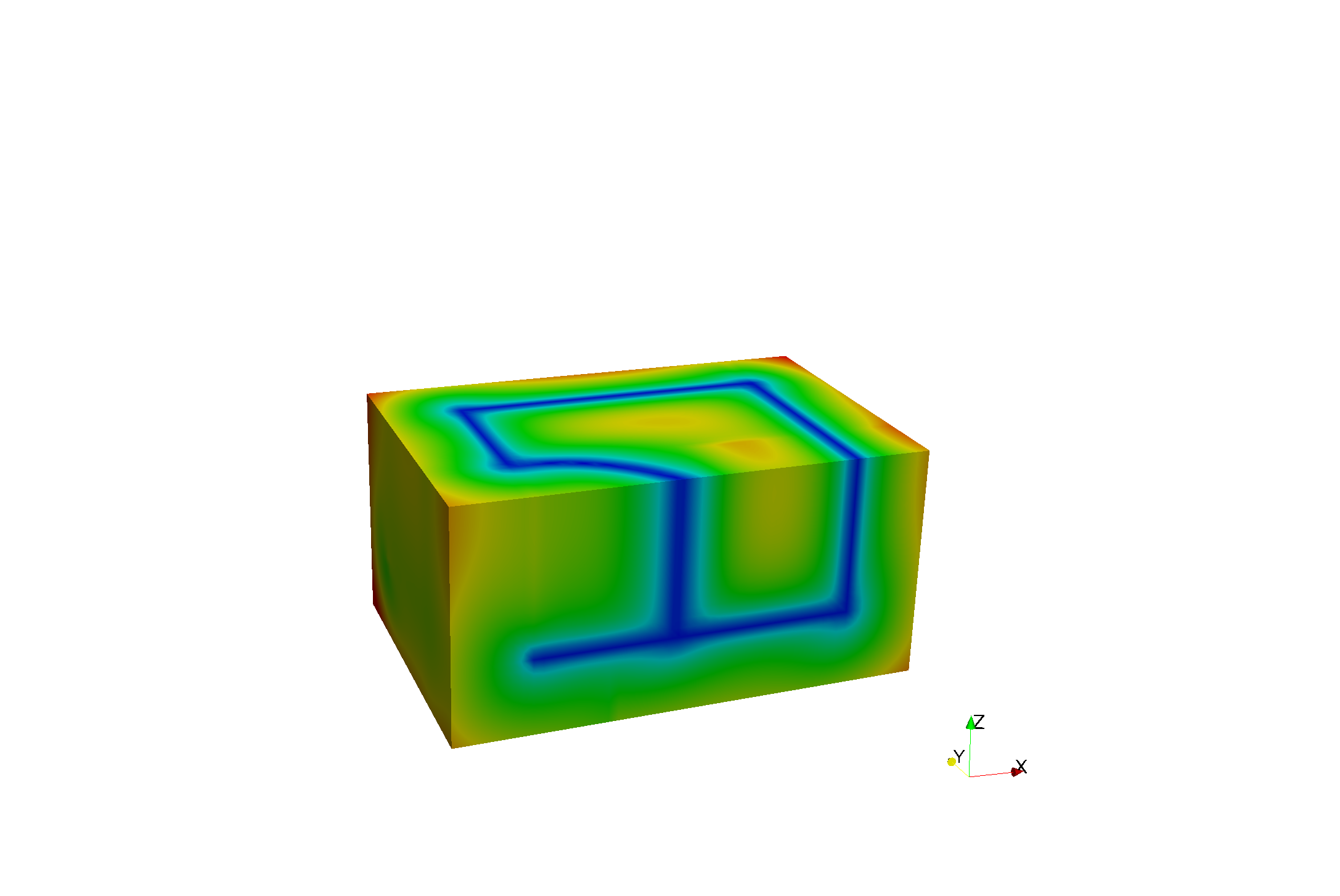}}
		\\
	\end{tabular}
	\caption{Parametric CAD and sliced global implicit representation for the 3D model of a cube trimmed by a cylinder.}
	\label{fig:3Dmodels}
\end{figure}
For the 3D model $\zmodel_2$, we consider a cube trimmed by a cylinder.
Specifically, our domain is denoted by $\Omega_2 = K_2 \backslash C_2$ where $K_2 = [-0.5,0]^2\times[-0.25,0.25]$ is a box, and where $C_2$ is the cylinder with radius equal to $0.25$, height equal to $1/2$, and centered at the origin, see Figure \ref{fig:3Dmodels}\subref{fig:cylinderparam}.
The boundary of the domain $\Omega_2$ is composed of seven curves and seven surfaces.
Six surfaces correspond to the cube $K_2$ and one correspond to the cylinder $C_2$.
Six curves correspond to the boundary curves of each surface boundary of the cube, and one curve correspond to the intersection of the surface boundary of the cylinder $C_2$ with the cube.
We illustrate in Figure \ref{fig:3Dmodels}\subref{fig:cylinderimplicit} a global implicit representation of the boundary $\zmodel_2:=\partial\Omega_2$, using the method presented in Section \ref{sec:cadimplicit}.
Although the inner boundary is smooth, the outer boundary contains sharp features such as corners and sharp edges.

\begin{figure}[p!]
	\centering
	\vskip-70pt
	\setlength{\tabcolsep}{-30pt}
	\begin{tabular}{ c c c }
		\subfloat[]{\label{fig:cylinder0_0}
			\includegraphics[width=0.7\textwidth]{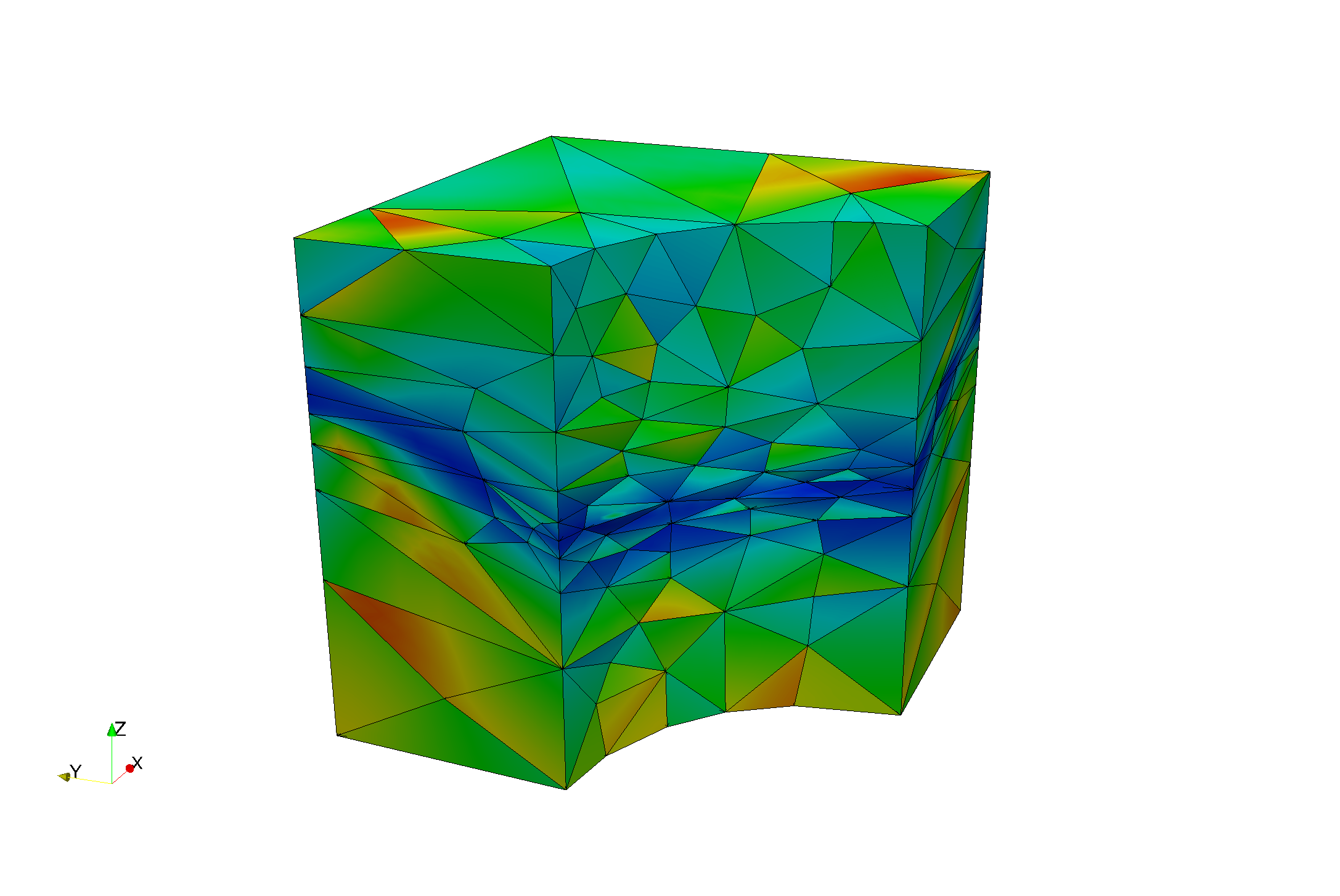}}
		&
		\subfloat[]{\label{fig:cylinder0_1}
			\includegraphics[width=0.7\textwidth]{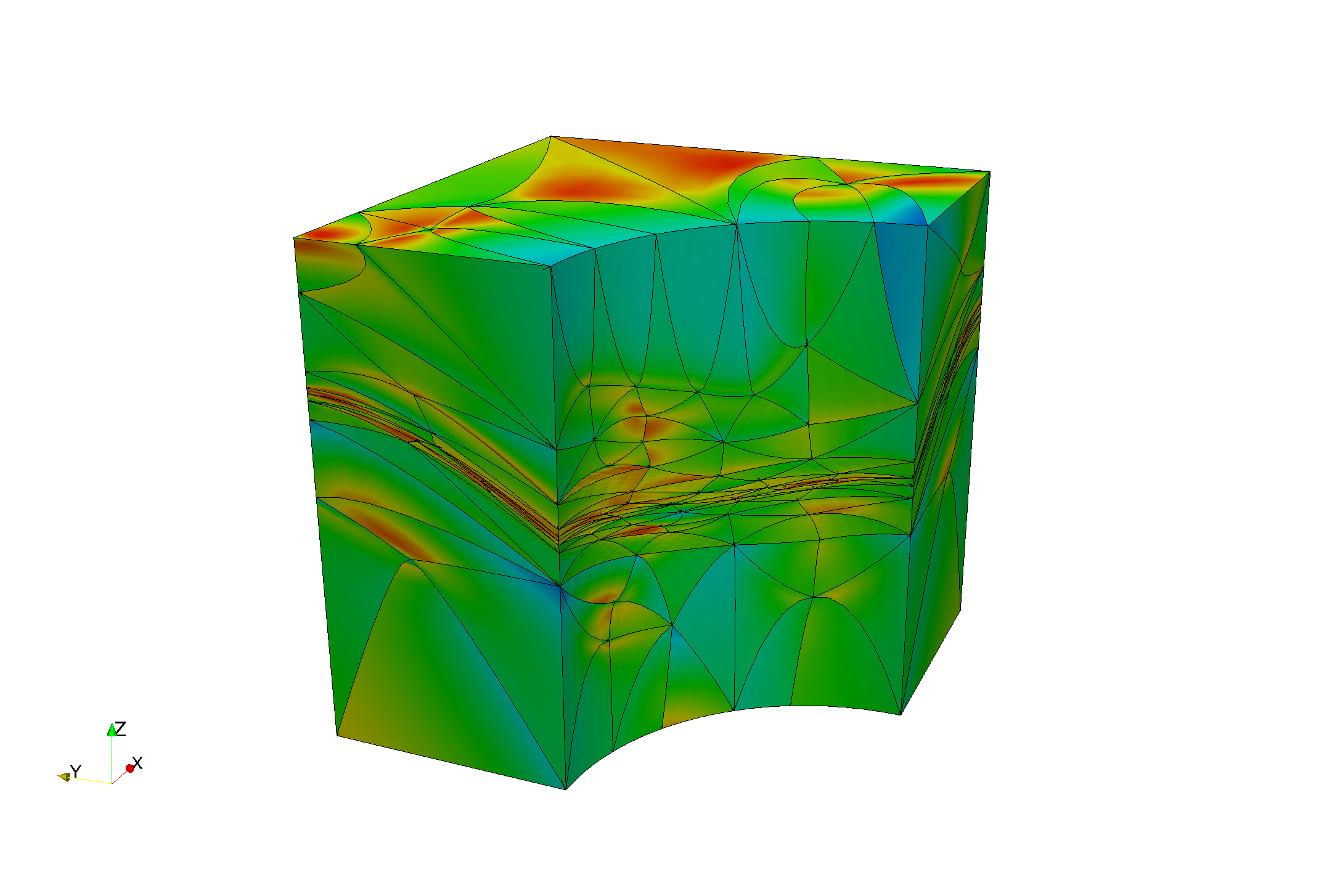}}
		\\
		\subfloat[]{\label{fig:cylinder1_0}
			\includegraphics[width=0.7\textwidth]{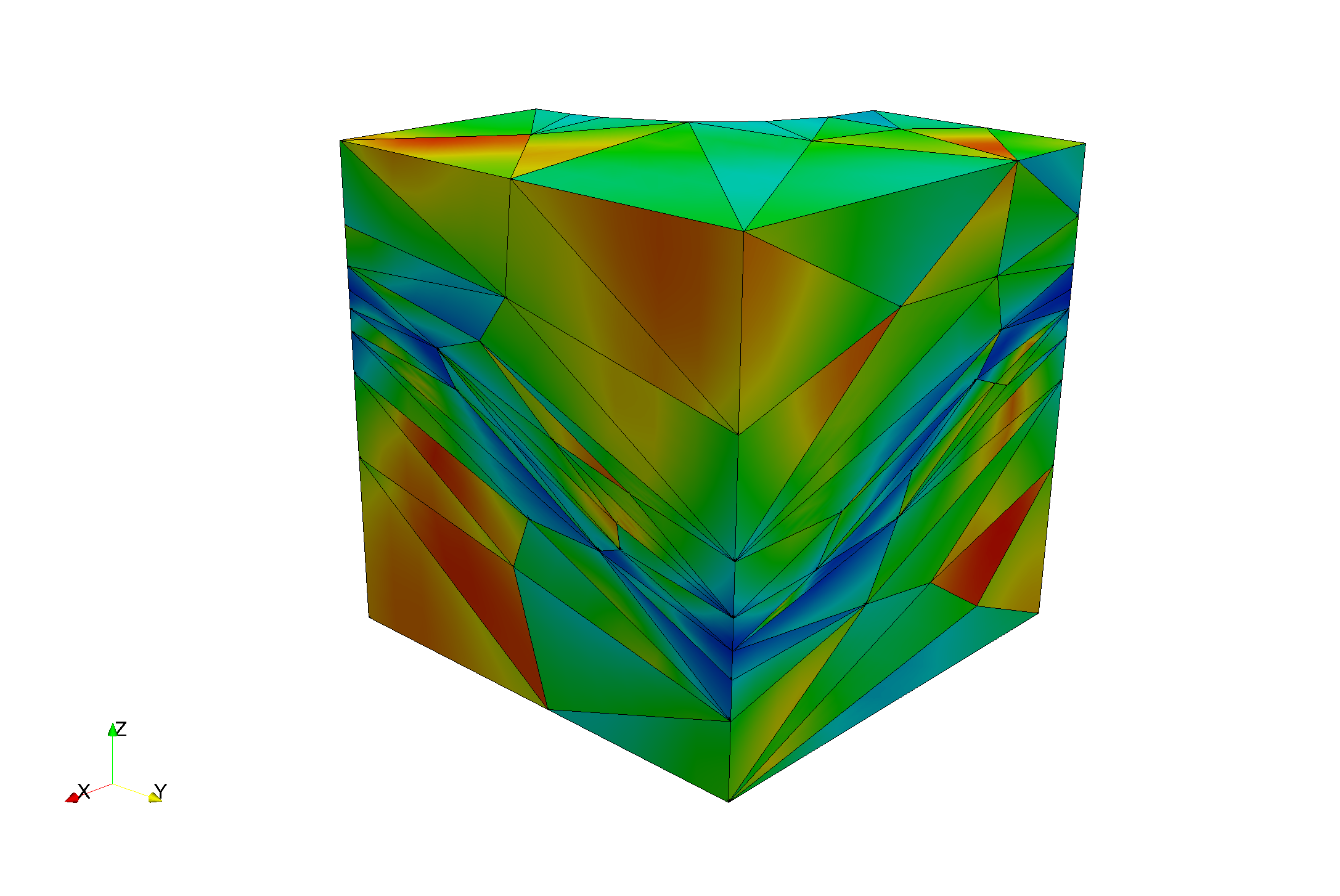}}
		&
		\subfloat[]{\label{fig:cylinder1_1}
			\includegraphics[width=0.7\textwidth]{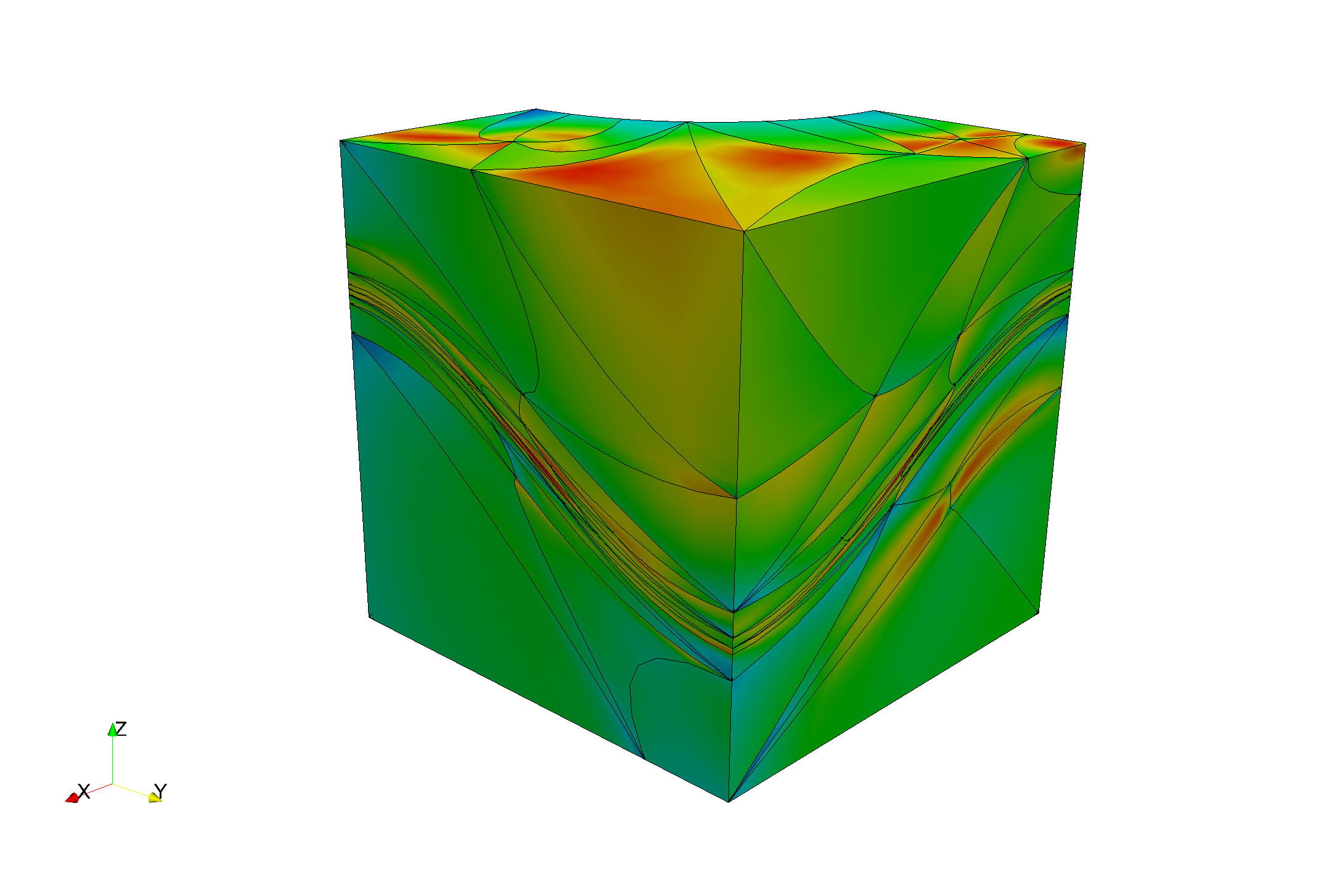}}
		\\
		\subfloat[]{\label{fig:cylinder2_0}
			\includegraphics[width=0.7\textwidth]{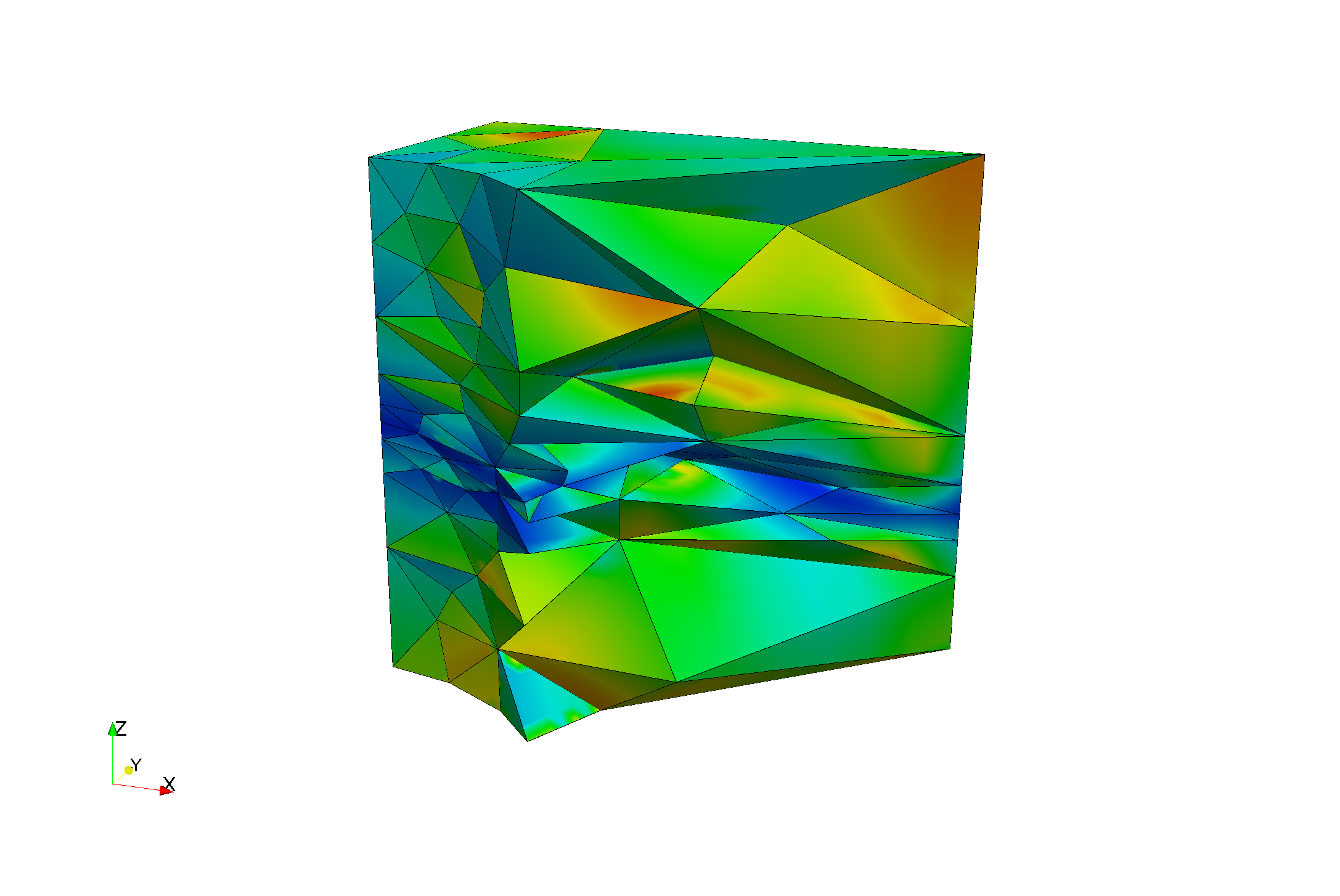}}
		&
		\subfloat[]{\label{fig:cylinder2_1}
			\includegraphics[width=0.7\textwidth]{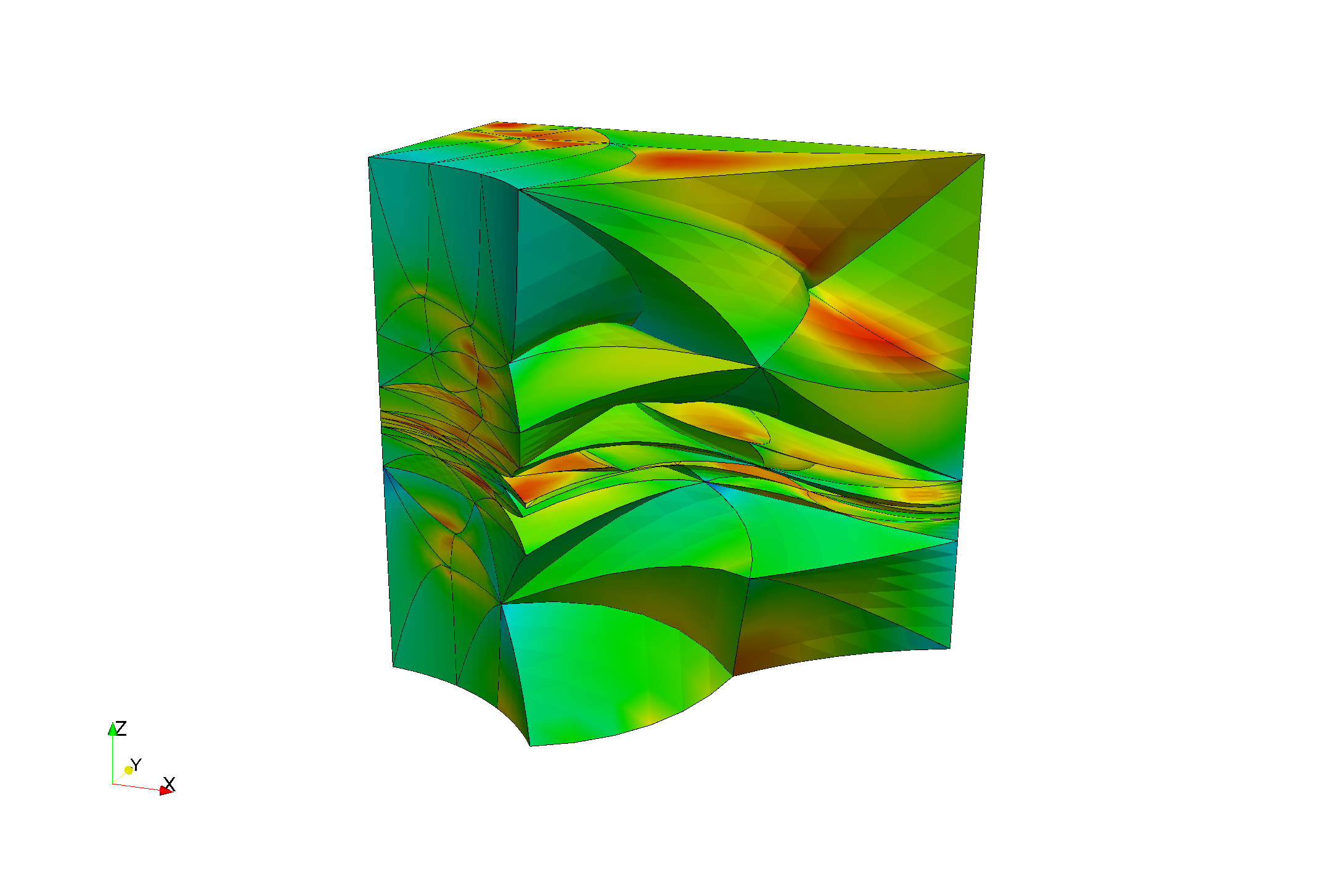}}
		\\
	\end{tabular}
	\\
	\includegraphics[width=0.35\textwidth]{/qualBarParaview_color} 
	\caption{Point-wise distortion for quadratic tetrahedral meshes. Initial straight-sided anisotropic mesh and optimized mesh in columns.}
	\label{fig:cylinder}
\end{figure}
We equip the domain $\Omega_2$ with the target metric presented in Equation \eqref{eq:metricDeformation} with $h_{\min} = 0.02$.
Then, we generate with MATLAB a quadratic isotropic tetrahedral background mesh $\hat{\zmesh}$ of input resolution $h_{\min} = 0.02$ over $\Omega_2$ that is, of input size $0.04$.
We normalize the target metric according to size $h = 0.5$.
Then, we couple each background mesh with the target metric evaluated at the background mesh vertices.
From this background mesh $\hat{\zmesh}$, we obtain an initial quadratic straight-sided anisotropic physical mesh $\zmesh$ by applying the MMG algorithm, see Figures \ref{fig:cylinder}\subref{fig:cylinder0_0}, \ref{fig:cylinder}\subref{fig:cylinder1_0}, and \ref{fig:cylinder}\subref{fig:cylinder2_0}.
The physical mesh is composed by $1\,261$ nodes and 695 tetrahedra.
Note that, since the MMG algorithm requires a linear background mesh, we subdivide once our quadratic background mesh in order to preserve its resolution.

\begin{table*}[t!]
	\caption{Quality statistics for the initial MMG and optimized mesh with interpolated 3D metric at the cube trimmed by a cylinder.}
	\label{table:cylinder}
	\centering
	\par\medskip
	\begin{tabular}{ c c c c c }
		\hline\noalign{\smallskip}
		Mesh&Minimum&Maximum&Mean&Standard deviation\\
		\noalign{\smallskip}\hline\noalign{\smallskip}
		Initial & 0.0506 & 0.9489 & 0.3519 & 0.1874 \\
		Optimized & 0.3315 & 0.9198 & 0.6661 & 0.1144 \\
		\noalign{\smallskip}\hline\noalign{\smallskip}
	\end{tabular}
\end{table*}
In Figures \ref{fig:cylinder}\subref{fig:cylinder0_1}, \ref{fig:cylinder}\subref{fig:cylinder1_1}, and \ref{fig:cylinder}\subref{fig:cylinder2_1}, we illustrate the optimized meshes $\zmesh^*$.
We observe that the elements lying in the anisotropic region are compressed to attain the stretching and alignment prescribed by the metric.
Note that the boundary elements are curved to match both the metric and the curved domain boundaries.
In Table \ref{table:cylinder}, we show the quality statistics of both the initial and optimized mesh.
In the optimized mesh the minimum, the mean, and the standard deviation of the element qualities are improved when compared with the initial configuration.

From the results, we observe that, when compared with straight-sided elements, curved elements approximate more faithfully the metric while preserving the curved features of the boundary.
In this case, the stretching direction and the curvature of the geometry are independent.
Accordingly, when considering straight-edged elements, in Figure \ref{fig:cylinder}\subref{fig:cylinder2_0}, more stretched elements may enable a lower resolution of the boundary.
That is, the achieved resolution of the boundary limits the achieved stretching, and vice-versa.
In contrast, when considering curved elements, in Figure \ref{fig:cylinder}\subref{fig:cylinder2_1}, we observe that more degrees of freedom can be accumulated at the stretched directions while preserving the curved features of the boundary.
As before, we conclude that curved elements allow an improved representation of the metric while preserving the curved features of the boundary.

\section{Concluding remarks}
\label{sec:conclusions}

In conclusion, we have obtained unprecedented second-order optimization results in curved $r$-adaption to a metric and geometry targets. We have represented the discrete metric in a curved background mesh as a high-order log-Euclidean metric interpolation. For this metric interpolation, we have detailed the first and second derivatives in terms of the physical coordinates. Moreover, we have  considered the geometry model as an implicit representation of the NURBS entities. For this implicit representation, we have detailed the first and second derivatives.

The derivatives of the metric interpolation and the implicit representation have allowed minimizing the objective function with Newton's method, an objective function that accounts for the metric and geometry deviations. The discrete metric results compare well with the analytic metric results. In all the results, the method exploits the non-constant Jacobian of curved high-order elements. This mechanism allows the technique to simultaneously match curved features of the metric and the geometry.

To meet our goal, we have enabled Newton's method for curved $r$-adaption.
Nevertheless, we have planned new directions and improvements for the near future.
First, to demonstrate the applications of our method and the advantages of adapted curved meshes, we have planned to $r$-adapt the curved meshes to the steady state of inviscid flows.
At this point, we cannot obtain the required discrete metrics because we need to implement existing goal-oriented error estimators for high-order methods \cite{yano2012,COULAUD2016353}.
Second, we have demonstrated a key ingredient for curved $r$-adaption.
Nevertheless, combining curved $r$-adaption with curved $h$-adaption might be more efficient.
To illustrate this combination, we have used an external straight-edged adaptive mesher.
However, to properly match the requirements of high-order methods in $h$-adaption, it is mandatory to use local cavity operators for curved meshes.
Regarding these curved operators, we have planned to combine existing approaches \cite{zhang2018curvilinear,zahr2020implicit,rochery2021p2,FEUILLET2020102846} with our approaches.
Specifically, our distortion minimization for high-order metric and curved boundaries can also optimize a local cavity.
To this end, we will match the cavity interior to the target high-order metric while the old cavity boundaries represent the target geometry.

In perspective, this capability to match metric and geometry features might be an attractive ingredient for curved high-order adaption. Specifically, in goal-oriented or indicator-based adaptive processes, one would have a target high-order metric field in a current mesh approximating a target geometry. The combined approach would drive the curved $r$-adaption to globally (locally) relocate the current curved mesh (re-meshed cavity) according to the curved features of the solution and the geometry (cavity) boundary.

\section{Acknowledgements}
This project has received funding from the European Research Council (ERC) under the European Union's Horizon 2020 research and innovation programme under grant agreement No 715546. This work has also received funding from the Generalitat de Catalunya under grant number 2017 SGR 1731. The work of X. Roca has been partially supported by the Spanish Ministerio de Econom\'ia y Competitividad under the personal grant agreement RYC-2015-01633.


\appendix
\section{Derivatives of the eigenvalue decomposition}\label{sec:appendix}

In this Appendix, we detail the first and second-order spatial derivatives of the eigenvalue decomposition (eigenvalues and eigenvectors), first presented in \cite{andrew1993derivatives} and rewritten herein using our notation.

Let us consider, for $\ell = 1,...,d$, the eigenvalue equation for the eigenvector $\zu_{\ell}$ with eigenvalue $\lambda_\ell$
\begin{equation*}
\textbf{L}_{\ell}\zu_{\ell} := \left(\textbf{L} - \lambda_{\ell}\textbf{I}\right)\zu_{\ell} = 0,
\end{equation*}
where $\textbf{L}$ is a symmetric matrix and $\textbf{I}$ is the identity matrix.
Then, by taking its first-order and second-order derivatives we respectively obtain
\begin{equation}\label{eq:eigendiff1}
0 = \partial_{j}\left(\textbf{L}_{\ell}\zu_{\ell}\right) = \left(\partial_{j}\textbf{L}_{\ell}\right)\cdot \zu_{\ell} + \textbf{L}_{\ell}\cdot \partial_{j}\zu_{\ell},
\end{equation}
\begin{eqnarray}\label{eq:eigendiff2}
0 = \partial_{jk}\left(\textbf{L}_{\ell}\zu_{\ell}\right) = \left(\partial_{jk}\textbf{L}_{\ell}\right)\cdot \zu_{\ell}  + \textbf{L}_{\ell}\cdot \partial_{jk}\zu_{\ell} + \\
\notag \left(\partial_{j}\textbf{L}_{\ell}\right)\cdot \partial_{k}\zu_{\ell} + \left(\partial_{k}\textbf{L}_{\ell}\right)\cdot \partial_{j}\zu_{\ell}.
\end{eqnarray}

For each $\ell$ one first computes the first-order derivative of the eigenvalue $\lambda_{\ell}$ by left-multiplying by $\zu_{\ell}$ to Equation \eqref{eq:eigendiff1}. Then, by solving the remaining unknown term of Equation \eqref{eq:eigendiff1} one obtains the first-order derivatives of the eigenvector $\zu_{\ell}$. In particular, the first-order derivatives of the eigenvalues and the eigenvectors are given by
\begin{equation*}
\partial_j \lambda_{\ell} = \zu_{\ell}^{\text{T}} \cdot \partial_j \textbf{L} \cdot \zu_{\ell},\ \ \partial_j \zu_\ell = -\textbf{L}_{\ell}^+ \cdot \partial_j \textbf{L}_{\ell} \cdot \zu_{\ell},
\end{equation*}
where the operation $\textbf{L}_{\ell}^+$ is the Moore-Penrose pseudo-inverse matrix for the matrix $\textbf{L}_{\ell}$. We use the Moore-Penrose pseudo-inverse matrix instead of the inverse matrix because the matrix $\textbf{L}_{\ell}$ is singular. In addition, the redundant equations are satisfied automatically.

The second-order derivatives are obtained by applying a similar procedure. For each $\ell$ one first computes the second-order derivative of the eigenvalue $\lambda_{\ell}$ by left-multiplying by $\zu_{\ell}$ to Equation \eqref{eq:eigendiff2}. Then, by solving the remaining unknown term of Equation \eqref{eq:eigendiff2} one obtains the second-order derivatives of the eigenvector $\zu_{\ell}$. In particular, the second-order derivatives of the eigenvalues are given by
\begin{equation*}
\partial_{jk} \lambda_{\ell} = \zu_{\ell}^{\text{T}} \cdot \left(\partial_k \textbf{L}_{\ell} \cdot \partial_j\zu_{\ell} + \partial_j \textbf{L}_{\ell} \cdot \partial_k\zu_{\ell} + \partial_{jk}\textbf{L} \cdot \zu_{\ell}\right),
\end{equation*}
\begin{equation*}
\partial_{jk} \zu_\ell = -\textbf{L}_{\ell}^+ \cdot \left(\partial_k\textbf{L}_{\ell}\cdot \partial_{j}\zu_{\ell} + \partial_j\textbf{L}_{\ell}\cdot \partial_{k}\zu_{\ell} + \partial_{jk} \textbf{L}_{\ell} \cdot \zu_{\ell} \right) - \left(\partial_j \zu_{\ell}\cdot\partial_k \zu_{\ell}\right) \zu_{\ell},
\end{equation*}
where the last term of the second-order derivative of the eigenvector is obtained by imposing the second-order derivative of the imposed normalization condition $\zu_{\ell}^{\text{T}}\cdot \zu_{\ell} = 1$
\begin{equation*}
0 = \partial_{jk}\left(\zu_{\ell}^{\text{T}}\cdot \zu_{\ell}\right) = 2\partial_{jk}\zu_{\ell}^{\text{T}} \cdot \zu_{\ell} + 2\partial_{j}\zu_{\ell}^{\text{T}} \cdot \partial_{k}\zu_{\ell}.
\end{equation*}

\section{Derivatives of the implicit representation}\label{sec:appendix2}
In this Appendix, we detail the first and second-order derivatives of the normalized representation, the convex-hull representation, and the implicit representation of a B\'{e}zier patch.
They are used in the computation of the gradient and Hessian for the implicit representation, see Section \ref{sec:implicitders}.

Herein, we consider the gradient and Hessian of the normalized representation $\gnfun$, presented in Equation \eqref{eq:normalized}.
As before, we denote by $\nabla f * \nabla g$ the matrix with coefficients $\partial_{j} f \partial_{k}g$ for $j,\ k = 1,...,d$.
In addition, we consider the symmetric term $\nabla f \otimes \nabla g := \nabla f * \nabla g + \nabla g * \nabla f$ given by the matrix with coefficients $\partial_{j} f \partial_{k}g + \partial_{k} f \partial_{j}g$ for $j,\ k = 1,...,d$.
Then, the derivatives of the normalized representation are given by
\begin{equation}\label{eq:gradnormalization}
\nabla\gnfun = \frac{\nabla\gfun - \gnfun\nabla\|\nabla\gfun\|}{\|\nabla\gfun\|},
\end{equation}
and
\begin{equation}\label{eq:Hessnormalization}
\gnfun\nabla^2\gnfun = \frac{\gnfun\nabla^2\gfun - \gnfun^2\nabla^2\|\nabla\gfun\| - \nabla\gnfun \otimes \gnfun\nabla \| \nabla\gfun\|}{\|\nabla\gfun\|},
\end{equation}
where
\begin{equation*}
\gnfun\nabla\|\nabla\gfun\| = \frac{\gnfun\nabla^2\gfun\cdot\nabla\gfun}{\|\nabla\gfun\|},
\end{equation*}
\begin{equation*}
\gnfun^2\nabla^2\|\nabla\gfun\| = \frac{\gnfun^2\nabla^3\gfun\cdot\nabla\gfun + \gnfun\nabla^2\gfun\cdot\gnfun\nabla^2\gfun - \gnfun\nabla\|\nabla\gfun\|* \gnfun\nabla\|\nabla\gfun\|}{\|\nabla\gfun\|}.
\end{equation*}
We observe that they require the first, second, and third derivatives of $\gfun$.
In addition, we consider these terms when differentiating the trimming operation, see Equations \eqref{eq:dertrim} and \eqref{eq:der2trim} for $h = \gnfun$.
In particular, the chain rule involves the terms $\nabla\gnfun$ and $\gnfun\nabla^2\gnfun$, and the terms $\nabla\gfun$, $\gfun\nabla^2\gfun$, and $\gfun^2\nabla^3\gfun$.
As we can see, this observation is advantageous because a straight-forward computation of the second and third derivatives, $\nabla^2\gfun$, and $\nabla^3\gfun$, involves a singularity at the corresponding zero level-set of $\gfun$.
For this reason, instead of computing directly the derivatives we consider them multiplied by the representation $\gfun$.

Next, we compute the derivatives of the convex-hull representation $\gnfun_{\zchull\left(\znurbs\right)}$.
In particular, note that these derivatives are trivial since the representation of each hyperplane entity is linear.
Then, we differentiate the $r$-conjunction between the hyperplane representations $\gfun_{\zchull\left(\znurbs\right)}$, see Equations \eqref{eq:gradrconj} and \eqref{eq:Hessrconj}.
Finally, we differentiate the normalization of the convex-hull representation $\gnfun_{\zchull\left(\znurbs\right)}$, see Equations \eqref{eq:gradnormalization} and \eqref{eq:Hessnormalization}.

Now, we compute the derivatives for the determinant $\gfun$ of Equation \eqref{eq:implicit}.
That is, $\nabla\gfun$, $\gfun\nabla^2\gfun$, and $\gfun^2\nabla^3\gfun$.
First, compute the gradient of the determinant by using the Jacobi's formula
\begin{equation}\label{eq:jacobi}
\nabla \gfun\left( \zx \right) = \tr\left(\zadj\cdot \nabla\mathbb{N}\left(\zx\right)\right),
\end{equation}
where $\mathbb{N}\left(\zx \right) := \mathbb{M}\left(\zx\right)\cdot \mathbb{M}\left(\zx\right)^{\text{T}}$.
We consider the adjugate matrix $\zadj$, instead of the inverse matrix, to avoid the singularity issues at the patch $\znurbs$.
In particular, the adjugate matrix of $\mathbb{N}\left(\zx \right)$ is defined by the transposed cofactor matrix, and satisfying the relation $\zadj = \gfun\left(\zx\right)\mathbb{N}\left(\zx\right)^{-1}$ \cite{upreti2014algebraic}.
Secondly, we compute the higher-order derivatives $\gfun\nabla^2\gfun$, and $\gfun^2\nabla^3\gfun$ by differentiating the terms inside the trace function $\nabla \gfun$, see Equation \eqref{eq:jacobi}.
In particular, using the same notation as in Section \ref{sec:interpDeriv}, we compute the second derivatives for each $j$ and $k$ as
\begin{equation*}
\gfun\left(\zx\right)\partial_{jk} \gfun\left( \zx \right) = \tr\left(\gfun\left(\zx\right)\partial_{k}\zadj\cdot \partial_{j}\mathbb{N}\left(\zx\right) + \gfun\left(\zx\right)\zadj\cdot \partial_{jk}\mathbb{N}\left(\zx\right)\right).
\end{equation*}
In addition, the third derivatives are given by
\begin{eqnarray*}
\gfun\left(\zx\right)^2\partial_{j k \ell}\gfun\left( \zx \right) = \tr\left(\gfun\left(\zx\right)^2\partial_{k \ell}\zadj \cdot \partial_j \mathbb{N}\left(\zx\right) +\right.\\
\left(\gfun\left(\zx\right)^2\partial_k\zadj \cdot \partial_{j \ell} \mathbb{N}\left(\zx\right) + \gfun\left(\zx\right)^2\partial_\ell\zadj \cdot \partial_{jk} \mathbb{N}\left(\zx\right)  \right),
\end{eqnarray*}
for each $j$, $k$, and $\ell$. Note that, there is no third order term $\partial_{jk\ell} \mathbb{N}\left(\zx\right)$ because $\mathbb{N}\left(\zx\right)$ is a quadratic function on $\zx$, see Equation \eqref{eq:implicit}.

Finally, we provide the derivatives of the adjugate matrix $\zadj$.
In particular, we present them in terms of the derivatives of the inverse matrix multiplied by the determinant.
Then, to rewrite the obtained expression in terms of the adjugate matrix, we multiply both expressions by the determinant $\gfun$.
Specifically, the gradient is given by
\begin{eqnarray*}
\gfun\left(\zx\right)\nabla\zadj = \gfun\left(\zx\right)\nabla\left(\gfun\left(\zx\right)\mathbb{N}\left(\zx\right)^{-1}\right) =\\
\zadj\nabla\gfun\left(\zx\right) - \mathbb{N}\left(\zx\right)\cdot\zadj\cdot\mathbb{N}\left(\zx\right).
\end{eqnarray*}
We apply the same reasoning for the Hessian by computing
\begin{equation*}
\gfun\left(\zx\right)^2\nabla^2\zadj = \gfun\left(\zx\right)^2 \nabla\left(\frac{1}{\gfun\left(\zx\right)}\gfun\left(\zx\right)\nabla\zadj\right).
\end{equation*}
In particular, using the same notation as in Section \ref{sec:interpDeriv}, for each $j$ and $k$ we have
\begin{eqnarray*}
	\gfun^2\partial_{jk}\zadjj = \left(\gfun \partial_{jk}\gfun\right)\zadjj + 
	\left(\partial_{j}\gfun\right)\gfun\partial_k\zadjj - \left(\partial_{k}\gfun\right)\gfun\partial_j\zadjj-\\
	\gfun\zadjj\cdot\partial_{jk}\zmat\cdot\zadjj -
	\gfun \partial_k\zadjj \cdot \partial_j\zmat \cdot\zadjj -
	\zadjj \cdot \partial_j\zmat\cdot\gfun \partial_k\zadjj,
\end{eqnarray*}
where, for the sake of brevity, we omit the dependence on the $\zx$ variable of the functions $\gfun$ and $\zmatrix$.

\bibliography{globalreferences}

\end{document}